\documentclass[prd,aps,
longbibliography,
preprintnumbers,
nofootinbib,
floatfix,
superscriptaddress]{revtex4}
\usepackage{graphicx}
\usepackage{epsfig}
\usepackage{rotating}
\usepackage{amssymb}
\usepackage{dsfont}
\usepackage{psfrag}
\usepackage{amsmath,euscript,array,mathrsfs}
\usepackage{bbold}
\usepackage{bm}
\usepackage{multirow}
\usepackage{dcolumn}
\RequirePackage{doi}
\usepackage{hyperref}
\usepackage{caption}
\usepackage{subcaption}
\usepackage[skip=10pt plus1pt, indent=20pt]{parskip}
\usepackage{hyperref}
\usepackage{float}
\floatstyle{plaintop}
\restylefloat{table}
\usepackage{bbding}
\usepackage{orcidlink}
\usepackage{verbatim}
\usepackage{xcolor}

\newcommand{\beq}{\begin{equation}}
\newcommand{\eeq}{\end{equation}}
\newcommand{\beqs}{\begin{eqnarray}}
\newcommand{\eeqs}{\end{eqnarray}}
\newcommand{\Tr}{{\rm Tr}}

\newcommand{\be}{\begin{equation}}
\newcommand{\ee}{\end{equation}}
\newcommand{\ba}{\begin{array}}
\newcommand{\ea}{\end{array}}

\newcommand{\orcidauthorBENNETT}{0000-0002-1678-6701}
\newcommand{\orcidauthorLUCINI}{0000-0001-8974-8266}
\newcommand{\orcidauthorPIAI}{0000-0002-2251-0111} 
\newcommand{\orcidauthorFORZANO}{0000-0003-0985-8858}
\newcommand{\orcidauthorVADACCHINO}{0000-0002-5783-5602}
\newcommand{\orcidauthorHILL}{0000-0003-2383-940X}
\newcommand{\orcidauthorHONG}{0000-0002-3923-4184}
\newcommand{\orcidauthorDELDEBBIO}{0000-0003-4246-3305}
\newcommand{\orcidauthorLUPO}{0000-0001-9661-7811}
\newcommand{\orcidauthorLIN}{0000-0003-3743-0840}
\newcommand{\orcidauthorLEE}{0000-0002-4616-2422}
\newcommand{\orcidauthorZIERLER}{0000-0002-8670-4054}
\newcommand{\orcidauthorHSIAO}{0000-0002-8522-5190}

\begin{document}

\title{Meson spectroscopy from spectral densities in lattice gauge theories }

\author{Ed Bennett\,\orcidlink{\orcidauthorBENNETT}}
\email{E.J.Bennett@swansea.ac.uk}
\affiliation{Swansea Academy of Advanced Computing, Swansea University (Bay Campus), Fabian Way, Swansea SA1 8EN, United Kingdom}

\author{Luigi Del Debbio\,\orcidlink{\orcidauthorDELDEBBIO}}
\email{luigi.del.debbio@ed.ac.uk}
\affiliation{Higgs Centre for Theoretical Physics, School of Physics and Astronomy, 
The University of Edinburgh, Peter Guthrie Tait Road, Edinburgh EH9 3FD, United Kingdom}

\author{Niccolò Forzano\,\orcidlink{\orcidauthorFORZANO}}
\email{2227764@swansea.ac.uk}
\affiliation{Department of Physics, Faculty  of Science and Engineering, Swansea University, Singleton Park, SA2 8PP, Swansea, United Kingdom}

\author{Ryan~C. Hill\,\orcidlink{\orcidauthorHILL}}
\email{ryan.hill@ed.ac.uk}
\affiliation{School of Physics and Astronomy, The University of Edinburgh, Edinburgh EH9 3FD, United Kingdom}

\author{Deog~Ki Hong\,\orcidlink{\orcidauthorHONG}}
\email{dkhong@pusan.ac.kr}
\affiliation{Department of Physics, Pusan National University, Busan 46241, Korea}
\affiliation{Extreme Physics Institute, Pusan National University, Busan 46241, Korea}

\author{Ho Hsiao\,\orcidlink{\orcidauthorHSIAO}}
\email{thepaulxiao.sc09@nycu.edu.tw}
\affiliation{Institute of Physics, National Yang Ming Chiao Tung University, 1001 Ta-Hsueh Road, Hsinchu 30010, Taiwan}

\author{Jong-Wan Lee\,\orcidlink{\orcidauthorLEE}}
\email{j.w.lee@ibs.re.kr}
\affiliation{ Particle Theory  and Cosmology Group, Center for Theoretical Physics of the Universe, Institute for Basic Science (IBS), Daejeon, 34126, Korea }

\author{C.-J. David Lin\,\orcidlink{\orcidauthorLIN}}
\email{dlin@nycu.edu.tw}
\affiliation{Institute of Physics, National Yang Ming Chiao Tung University, 1001 Ta-Hsueh Road, Hsinchu 30010, Taiwan}
\affiliation{Centre for High Energy Physics, Chung-Yuan Christian University, Chung-Li 32023, Taiwan}
\affiliation{ Centre for Theoretical and Computational Physics, National Yang Ming Chiao Tung University, 1001 Ta-Hsueh Road, Hsinchu 30010, Taiwan }

\author{Biagio Lucini\,\orcidlink{\orcidauthorLUCINI}}
\email{B.Lucini@Swansea.ac.uk}
\affiliation{Swansea Academy of Advanced Computing, Swansea University (Bay Campus), Fabian Way, Swansea SA1 8EN, United Kingdom}
\affiliation{Department of Mathematics, Faculty of Science and Engineering, Swansea University (Bay Campus), Fabian Way, SA1 8EN Swansea, United Kingdom}

\author{Alessandro Lupo\,\orcidlink{\orcidauthorLUPO}}
\email{alessandro.lupo@cpt.univ-mrs.fr}
\affiliation{Aix-Marseille Université, Université de Toulon, CNRS, CPT, Marseille, France}

\author{Maurizio Piai\,\orcidlink{\orcidauthorPIAI}}
\email{m.piai@swansea.ac.uk}
\affiliation{Department of Physics, Faculty  of Science and Engineering, Swansea University, Singleton Park, SA2 8PP, Swansea, United Kingdom}

\author{Davide Vadacchino\,\orcidlink{\orcidauthorVADACCHINO}}
\email{davide.vadacchino@plymouth.ac.uk}
\affiliation{Centre for Mathematical Sciences, University of Plymouth, Plymouth, PL4 8AA, United Kingdom}

\author{Fabian Zierler\,\orcidlink{\orcidauthorZIERLER}}
\email{fabian.zierler@swansea.ac.uk}
\affiliation{Department of Physics, Faculty  of Science and Engineering, Swansea University, Singleton Park, SA2 8PP, Swansea, United Kingdom}

\date{\today}

\begin{abstract}

Spectral densities encode non-perturbative information that
enters the calculation of a plethora of physical observables
in strongly coupled field theories. Phenomenological applications encompass  aspects of
standard-model hadronic physics, observable at current colliders, as well as 
 correlation functions characterizing
new physics proposals, testable in future experiments.
By making use of numerical data produced in a $Sp(4)$ lattice gauge theory with matter transforming in an admixture of 
fundamental and 2-index antisymmetric representations of the gauge group, we perform a systematic study to 
demonstrate the effectiveness of recent technological progress in the reconstruction of spectral densities.

To this purpose, we write and test new software packages that use energy-smeared spectral densities
to analyze the mass spectrum of mesons. We assess the effectiveness of different smearing kernels
and optimize the smearing parameters to the characteristics of available lattice ensembles.
We generate new ensembles for the theory in consideration, with lattices 
that have a longer extent in the time direction with respect to the spatial ones.
We run our tests on these ensembles, obtaining new results about the spectrum of light mesons
and their excitations. We make available our algorithm and software for the extraction of spectral densities,
that can be applied to theories with other gauge groups, 
including the theory of strong interactions (QCD)  governing
 hadronic physics in the standard model.

\end{abstract}
\preprint{CTPU-PTC-24-11}
\preprint{PNUTP-24/A02}
\maketitle

\tableofcontents

\section{Introduction} 
\label{Sec:introduction}

Recent years have seen the development of new technology aimed at extracting
spectral densities from numerical lattice data
obtained in the non-perturbative study of strongly coupled field theories---see for instance Refs.~\cite{Hansen:2019idp,
Hansen:2017mnd,Bulava:2019kbi,Bailas:2020qmv,Gambino:2020crt,Bruno:2020kyl,Lupo:2021nzv,
Gambino:2022dvu,DelDebbio:2022qgu,Bulava:2021fre,Lupo:2022nuj,Bulava:2023brj,DeSantis:2023rjl,Lupo:2023qna,Kades:2019wtd,Pawlowski:2022zhh,DelDebbio:2021whr,Bergamaschi:2023xzx,Bonanno:2023thi,Frezzotti:2023nun,Frezzotti:2024kqk}.
Spectral densities are inverse Laplace transforms of space-averaged two-point functions involving time-separated operators. 
They can be used to compute high-precision spectral observables that are otherwise difficult to access with conventional methodologies. 
Additionally, spectral densities encode information about off-shell physics, providing an alternative framework to compute scattering amplitudes~\cite{Bulava:2019kbi}, inclusive rates~\cite{Gambino:2020crt, Bulava:2021fre}, finite-volume energies and matrix elements. In QCD, spectral densities enter the calculation of  hadronic observables,
 such as the $R$-ratio~\cite{ExtendedTwistedMassCollaborationETMC:2022sta} and 
inclusive decays of the $\tau$ lepton~\cite{Evangelista:2023fmt,Alexandrou:2024gpl}. They can also be used to access the properties of glueballs~\cite{Smecca:2023hgr}. In finite temperature QCD, spectral functions enable to investigate transport properties of the quark gluon plasma (see, e.g., the reviews~\cite{Asakawa:2000tr,Meyer:2011gj, Ratti:2017qgq} and the more recent Ref.~\cite{Aarts:2023vsf}), including the electrical conductivity, as discussed, for instance, in Refs~\cite{Aarts:2007wj, Aarts:2014nba}. In 
analogy to  the derivation of the Weinberg sum rules from the properties of the spectral functions~\cite{Weinberg:1967kj,Dash:1967fq,Bernard:1975cd},  the spectral representation 
of two-point functions involving both mesons and baryons enters the effective potential of new physics models~\cite{Contino:2010rs,Banerjee:2023ipb}, that implement Higgs compositeness~\cite{Kaplan:1983fs,Georgi:1984af,Dugan:1984hq}
and top partial compositeness~\cite{Kaplan:1991dc} (see also the reviews in Refs.~\cite{Panico:2015jxa,Witzel:2019jbe,Cacciapaglia:2020kgq,Bennett:2023wjw},
and the summary tables in Refs.~\cite{Ferretti:2013kya,Ferretti:2016upr,Cacciapaglia:2019bqz}), and may trigger
electroweak symmetry breaking  via 
vacuum misalignment~\cite{Das:1967it,Peskin:1980gc,Preskill:1980mz}.

This paper has two main objectives.
The first is to develop, test, tune, optimize, and benchmark the effectiveness of a new software package that allows the computation of spectral densities from correlation functions measured on the lattice.
The spectral densities of interest are smeared in energy, and thus have a well-defined infinite-volume limit. After such limit is taken, the energy smearing can in principle be removed. This is however not necessary in this work, where we perform correlated fits of smeared spectral densities to extract the finite-volume spectrum of mesons, as proposed in Ref.~\cite{DelDebbio:2022qgu}. The computation of smeared spectral densities can be systematically improved by reducing the statistical noise, and by increasing the number of time slices in the lattice~\cite{Hansen:2019idp}.
To this purpose, we generate and analyze ensembles with long time extent.
This work sets the stage for future applications, both in the context of QCD and of new physics models,
by demonstrating its viability as an analysis tool for the aforementioned ambitious endeavors.
To validate our techniques., we apply these analysis tools to observables in theories for which 
alternative ways exist to gain access to the relevant
non-perturbative information.

The second objective is to make significant progress toward understanding the spectrum of 
a specific gauge theory that is a prototypical candidate for UV-completion of Composite Higgs Models~\cite{Barnard:2013zea}.
To this purpose, we generate new ensembles for the $Sp(4)$ theory coupled to two Dirac fermions
transforming in the fundamental and three in the two-index antisymmetric representation of the gauge group. 
This theory and its variations have been recently studied in the literature on $Sp(2N)$ theories~\cite{Bennett:2017kga,Lee:2018ztv,Bennett:2019jzz,Bennett:2019cxd,Bennett:2020hqd,
Bennett:2020qtj,Lucini:2021xke,Bennett:2021mbw,Bennett:2022yfa,Bennett:2022gdz,Bennett:2022ftz,Lee:2022elf,
Hsiao:2022kxf,Maas:2021gbf,Zierler:2021cfa,Kulkarni:2022bvh,Bennett:2023wjw,
Bennett:2023rsl,Bennett:2023gbe,Bennett:2023mhh,Bennett:2023qwx,Pomper:2024otb}.
We use the Grid software suite~\cite{Boyle:2015tjk,Boyle:2016lbp,Yamaguchi:2022feu}, including adaptations~\cite{Bennett:2023gbe} previously made to implement symplectic groups.
The data are analyzed with the HiRep code~\cite{DelDebbio:2008zf}.
We write and make available new software that reconstructs the smeared energy spectral density.
We focus our analysis on the two-point correlation functions of meson operators, 
with a general basis of Dirac structures, but restrict our attention to flavor non-singlet states.
We will analyze the spectra of flavor singlet mesons and of chimera baryons in separate publications.

Our results on the spectrum of mesons show improved statistical accuracy, 
with better control over systematics, 
with respect to existing published results, and extend over a larger range of parameter space.
We also detect
 new excited states, not accessible with earlier existing ensembles.
 We make publicly available both the software we developed and the data analysis flow~\cite{analysis_release},
 as well as the new data generated for this study~\cite{data_release}.
In the present paper, we defer to future studies of the extraction of the couplings (decay constants) of the associated mesons.

The paper is organized as follows.
We define the lattice theories of interest in Sec.~\ref{Sec:lattice}. We present the properties of the ensembles we study in the main body of the paper.
We apply the gradient flow  
as scale-setting procedure, 
as well as a smoothening process for topological observables. 
We monitor the topological charge, and use it to 
estimate autocorrelations.
In Sec.~\ref{Sec:correlation_functions} we introduce the flavored, mesonic operators of interest and their correlation functions.
We also describe our implementation of Wuppertal and APE smearing, and exemplify how the spectra can be extracted from a variational 
approach based upon the generalized eigenvalue problem (GEVP).
Spectral densities and (energy) smearing kernels are introduced in Sec.~\ref{Sec:spectral_density}.
We also discuss how the signal and its statistical significance depend on the choices of 
smearing parameters.
We devote Sec.~\ref{Sec:varyingT} to a systematic investigation of how the length of the time extent of the lattice affects the
spectral density reconstruction.
Our (new) results on the spectrum of mesons are summarized and critically discussed in Sec.~\ref{Sec:spectra}.
We conclude by 
highlighting avenues for further investigation in Sec.~\ref{Sec:outlook}.
Technical details about the spectral density reconstruction are relegated to the Appendix.

\section{Lattice field theory} 
\label{Sec:lattice}

In this section, we present the quantum field theory of interest and the discretized lattice action 
we adopt for its study. 
By doing so, we fix the notation so that the presentation is self-contained. 
We also tabulate and characterize the lattice ensembles generated for the purposes of this paper,
in which we report our results on the spectrum of flavored mesons. We postpone to future publications
the measurement, on these same ensembles,  of the spectra of other bound states---flavor singlet mesons and chimera baryons.

\subsection{Lattice discretization and bare parameters}
\label{Sec:lattice_setup}

The $Sp(4)$ gauge theory of interest has a continuum Lagrangian density that,
in the presence of   $N_{\rm f}$ Dirac fermions, $Q^I$, transforming  in the fundamental representation, together
with  $N_{\rm as}$ Dirac fermions, $\Psi^k$, transforming in the two-index antisymmetric representation, is given by
\begin{align}\label{eq:continuum_lagrangian}
    \mathcal L = -\frac{1}{2} \Tr\, G_{\mu\nu} G^{\mu\nu} +  
    \sum_{I,J=1}^{N_{\rm f}} \bar Q^I \left( i\delta_{IJ}\,\gamma^\mu D_\mu - m_{IJ}^{\rm f} \right) Q^J 
    + \sum_{k,\ell=1}^{N_{\rm as}} \bar \Psi^k \left( i\delta_{k\ell}\,\gamma^\mu D_\mu - m_{k\ell}^{\rm as} \right) \Psi^{\ell}\,,
\end{align}
where $\gamma^{\mu}$ are Dirac gamma matrices, 
$\mu=0,\,1,\,2,\,3$ refers to the coordinates in Minkowski space, while  $I,\,J=1,\,\cdots,\,N_{\rm f}$,  
and $k,\,\ell=1,\,\cdots,\, N_{\rm as}$, are flavor indexes---color and spin indexes are understood.
Throughout this paper we assume mass-degenerate fermions, for which $m_{IJ}^{\rm f} = m^{\rm f}\delta_{IJ}$ and $m_{k\ell}^{\rm as} = m^{\rm as}\delta_{k\ell}$. 
The field-strength tensor, $G_{\mu\nu}$, and the covariant derivative, $D_\mu$, acting on the fermions are defined following the conventions of Ref.~\cite{Bennett:2019cxd}: 
\begin{align}\label{eq:covariant_derivative}
      G_{\mu\nu} &= \partial_\mu A_\nu - \partial_\nu A_\mu + ig \left[ A_\mu, A_\nu  \right]\,, \\
      D_\mu Q &= \partial_\mu Q + ig A_\mu Q\,, \\
      D_\mu \Psi &= \partial_\mu \Psi + ig A_\mu \Psi + ig \Psi A_\mu^T\,,
\end{align}
where $g$ is the gauge coupling.

An element, $M \in Sp(4)$, of the gauge group  acts on the fermion fields with the  gauge transformation
$Q \to M Q$ and $\Psi \to M \Psi M^T$. Because the fundamental representation is pseudo-real, while the  2-index antisymmetric one is real, the global symmetries are enhanced in comparison with a complex representation (such as that of QCD). The global symmetry of the Lagrangian is $U(1)_{\rm f} \times U(1)_{\rm as} \times SU(2N_{\rm f}) \times SU(2N_{\rm as})$. One combination of the $U(1)$ factors is broken by the axial anomaly.\footnote{Due to the multi-representation nature of this theory, both $U(1)$ symmetries are expected to be spontaneously broken which would lead to two additional (pseudo-) Nambu-Goldstone bosons (PNGBs): one combination of them is broken by the anomaly, while the other combination is non-anomalous and may have an implication on the phenomenological studies of composite Higgs models~\cite{Belyaev:2015hgo}.} The bilinear condensate of fundamental fermions breaks spontaneously the associated $SU(2N_{\rm f})$ symmetry to its $Sp(2N_{\rm f})$ subgroup,  while the condensate made of antisymmetric fermions gives rise to the breaking pattern $SU(2N_{\rm as}) \to SO(2N_{\rm as})$~\cite{Kosower:1984aw}. The mass terms break explicitly the symmetry along the same pattern, providing
masses for the PNGBs. From hereon, we ignore the $U(1)$ symmetries and specify the numbers of flavors to $N_{\rm f}=2$ and $N_{\rm as}=3$. We hence have $5+20$ PNGBs associated with the two non-Abelian cosets.

For the $Sp(2N)$ Euclidean action on the lattice---see Refs.~\cite{Bennett:2022yfa, Bennett:2023wjw, Bennett:2023gbe} for technical details---we adopt the standard Wilson plaquette action. We write it in terms of the gauge links, $U_\mu(x)$, as
\begin{align}
    S_g = \beta \sum_x \sum_{\mu<\nu} \left(1-\frac{1}{2N} {\rm Re}\, {\Tr}\, \left( U_\mu (x) U_\nu (x+\hat{\mu}) U_\mu^\dagger(x+\hat{\nu}) U_\nu^\dagger(x) \right)  \right)\,,
\end{align}
independently on the representation of the gauge links, and $\mu,\,\nu$ denote the direction of the link, starting from lattice site $x$, while $\hat{\mu},\,\hat{\nu}$ are unit displacement on the lattice.
The fermions are described by the Wilson fermion action~\cite{Wilson:1974sk}, for both representations:\begin{align}    
    S_f =  a^4 \sum_{J=1}^{N_{\rm f}}\sum_{x,y} \overline{Q}^J(x) D^{\mathrm{(f)}}(x,y) Q^J(y)+ a^4 \sum_{k=1}^{N_{\rm as}}\sum_{x,y} \overline{\Psi}^k(x) D^{\mathrm{(as)}}(x,y) \Psi^k(y)\,.
\end{align}
The lattice spacing is denoted by $a$, while the fundamental and antisymmetric Dirac operators,  $D^{\mathrm{(f)}}$ and $D^{\mathrm{(as)}}$, depending on the gauge links in the respective representation. 
We borrow their definitions from
Refs.~\cite{Bennett:2023gbe}:
\begin{align}
    D^{(R)}(x,y) \equiv \left( \frac{4}{a}+m^{R}_0 \right) \delta(x,y)\label{eq:Dirac_full} 
    -\frac{1}{2a}\sum_{\mu=1}^{4}
    \left\{(1-\gamma_\mu)U^{(R)}_\mu(x)\delta(x+\hat{\mu},y)
    +(1+\gamma_\mu)U^{(R)\dagger}_\mu(x)\delta(x-\hat{\mu},y)\right\}\,,    
\end{align}
where $R={\rm f}, {\rm as}$ denotes the different representations. Specifically, $m_0^{\mathrm{f}}$ and $m_0^{\mathrm{as}}$ are the bare masses 
of fermions of species ${\rm (f)}$ and ${\rm (as)}$, respectively. For the link variable, $U^{\mathrm{(f)}}_\mu(x)=U_\mu(x)$,
while a choice of parameterisation for $U^{\mathrm{(as)}}_\mu(x)$ can be found  in Ref.~\cite{Bennett:2022yfa}.

\begin{table}
    \begin{tabular}{|c|c|c|c|c|c|c|c|c|c|c|c|c|}
    \hline \hline
    Label & $\beta$ & $am_0^{\rm as}$ & $am_0^{\rm f}$ & $N_t$ & $N_s$ & $N_{\rm therm}$ & $n_{\rm skip}$ & $N_{\rm conf}$ & $\langle P \rangle$ & $w_0 / a$ & $\tau_{\rm int}^Q$ &  $\bar{Q}$ \\ 
    \hline
        M1 & 6.5 & -1.01 & -0.71 & 48 & 20 & 3006 & 14 & 479 & 0.585172(16) & 2.5200(50) & 6.9(2.4) & 0.38(12)\\ 
        M2 & 6.5 & -1.01 & -0.71 & 64 & 20 & 1000 & 28 & 698 & 0.585172(12) & 2.5300(40) & 7.1(2.1) & 0.58(14)\\ 
        M3 & 6.5 & -1.01 & -0.71 & 96 & 20 & 4000 & 26 & 436 & 0.585156(13) & 2.5170(40) & 6.4(3.3) & -0.60(19)\\ 
        M4 & 6.5 & -1.01 & -0.70 & 64 & 20 & 1000 & 20 & 709 & 0.584228(12) & 2.3557(31) & 10.6(4.8) & -0.31(19)\\ 
        M5 & 6.5 & -1.01 & -0.72 & 64 & 32 & 3020 & 20 & 295 & 0.5860810(93) & 2.6927(31) & 12.9(8.2) & 0.80(33)\\ 
    \hline \hline
    \end{tabular}
    \caption{Ensembles generated for this study of the $Sp(4)$ theory coupled to fermions in multiple representations. The inverse coupling is denoted as $\beta$ and the bare masses of the (Wilson-Dirac) fermions transforming according to the fundamental and antisymmetric representation of the gauge group are denoted by $am_0^{\mathrm{f}}$ and $am_0^{\mathrm{as}}$, respectively. The lattice volume is $N_t N_s^3 a^4$. For each ensemble, we further report the number of gauge configurations, $N_{\rm conf}$, as well as the average plaquette, $\langle P \rangle$, the gradient flow scale  in lattice units, $w_0/a$, the mean of the 
    topological charge, $\bar{Q}$, and the correlation length of the topological charge, $\tau_{\rm int}^Q$.
    We also report, for reproducibility purposes, the number of thermalization steps, $N_{\rm therm}$, discarded from the analysis, as well as the number of complete sweeps between configurations $n_{\textrm{skip}}$, which have been discarded to avoid larger autocorrelation. Notice however that $\tau_{\rm int}^Q$ has been computed on the remaining $N_{\textrm{conf}}$ configurations only, and yet $\tau_{\rm int}^Q>1$.
    \label{tab:ensembles}}
\end{table}

Gauge configurations are generated using Grid~\cite{Boyle:2015tjk,Boyle:2016lbp,Yamaguchi:2022feu}, which has
the functionality to work with $Sp(2N)$ gauge groups~\cite{Bennett:2023gbe}. We include dynamical fermions using the Hybrid Monte-Carlo (HMC) algorithm~\cite{Duane:1987de} for the two ${\rm (f)}$ fermions and the rational HMC (RHMC)~\cite{Clark:2006fx} algorithm for the three ${\rm (as)}$ Dirac fermions. In principle, the inclusion of an odd number of degenerate fermions might give rise to a sign problem, but for an odd number of fermions in the antisymmetric representation the determinant of the Dirac operator remains positive and real~\cite{Hands:2000ei,Bennett:2022yfa}. Acceptance rates were tuned to be around $70\%$ to $80\%$, which corresponds to $27-36$ integrator steps for a single unit of Monte-Carlo time. Resymplecticization was performed after every gauge configuration update. We use even-odd preconditioning as in Ref.~\cite{Bennett:2022yfa}. It was also shown in Ref.~\cite{Bennett:2023gbe} that the HMC and RHMC implementations in $Sp(4)$ yield compatible results for even number of fermions, and that one could equivalently adopt the HMC algorithm for two of the ${\rm (as)}$ fermions and RHMC for the third, without visible changes to the results.

Our hypercubic lattices have $N_s$ lattice sites in the spatial and $N_t > N_s$ lattice sites in the temporal direction, hence the volume is  ${V}_4 = (N_s a)^3 \times (N_t a) = L^3 \times T$. We impose periodic boundary conditions for the gauge fields. For the fermion fields, we impose periodic boundary conditions in the spatial dimensions and anti-periodic boundary conditions along the temporal direction. 
The lattice action has three free parameters: the inverse gauge coupling, $\beta = 2 N_c / g^2 = 4 N / g^2 =8/g^2$, and the masses, $m_0^{\mathrm{f}}$ and $m_0^{\mathrm{as}}$, of the two types of fermions. At strong coupling (small $\beta$) a transition into an unphysical bulk phase occurs. For this action it was found that a value of $\beta \gtrsim 6.3$ is sufficient to avoid this lattice phase~\cite{Bennett:2022yfa}. We choose $\beta=6.5$ for the ensembles discussed in this paper, and we keep the bare mass of the ${\rm (as)}$ fermions fixed, to be  $am_0^{as}=-1.01$. We allow the mass of the ${\rm (f)}$ fermions to vary over a modest range of values, as listed in Tab.~\ref{tab:ensembles}.
 For each ensemble, we compute the average plaquette, $\langle P \rangle$, defined as
\beqs
\langle P \rangle &\equiv&\frac{1}{6 N_t N_s^3}\sum_{x}\sum_{\mu>\nu}
{\rm Re}\, {\Tr}\, \frac{1}{2N} \left[\frac{}{} U_\mu (x) U_\nu (x+\hat{\mu}) U_\mu^\dagger(x+\hat{\nu}) U_\nu^\dagger(x) \frac{}{}\right]\,,
\eeqs
as this quantity enters the tadpole-improved gauge coupling $\tilde{g}^2\equiv g^2/\langle P \rangle $~\cite{Martinelli:1982mw,Lepage:1992xa}.

\subsection{Gradient flow, topological charge and autocorrelations}
\label{sec:flow}

In this study, we adopt the gradient flow~\cite{Luscher:2011bx,Luscher:2013vga}, 
and its lattice implementation, the Wilson flow~\cite{Luscher:2010iy}. This finds two applications: on one hand, it allows us to set the physical scale for our ensembles~\cite{BMW:2012hcm},
on the other hand, it will be used as a smoothening process in the extraction of topological properties.
We follow the convention and processes described in Ref.~\cite{Bennett:2022ftz}.
We define a new observable, $\mathcal W(t)$, as a function of a new gradient flow time, $t$, as 
\begin{align} \label{eq:gradient_flow_definitions}
    \mathcal W(t) &\equiv  \frac{\rm d }{\rm d \ln t}\left\{ t^2 \langle E(t) \rangle\right\}\,,
    \end{align}
    where in turn\footnote{Here we fix a typo in Eq.~(23) Ref.~\cite{Bennett:2022ftz}, 
in which a minus sign is missing, and which should read as our current Eq.~(\ref{eq:gradient_flow_energy_density}),
and in Eq.~(25) of the same reference, the right hand side of which should read as our current Eq.~(\ref{eq:gradient_flow_definitions}).}
\begin{align} \label{eq:gradient_flow_energy_density}
  E(t,x)  \equiv - \frac{1}{2} ~\Tr \, G_{\mu\nu}(t,x) G_{\mu\nu}(t,x).
\end{align}
The field strength tensor in Euclidean space-time, $G_{\mu\nu}(t,x)$, evaluated at non-vanishing flow time $t$, is defined in terms of the five-dimensional gauge field $A_\mu(t,x)$ as
\begin{align} \label{eq:gradient_flow_differential_equation}
    \frac{{\rm d} A_\mu(t, x)}{{\rm d} t} = D_\nu G_{\nu \mu} (t,x), ~~{\rm with}~~ A_\mu(t=0,x) = A_\mu(x).
\end{align}
We define the gradient flow scale, $w_0$, as the square root of the flow time, $t$, for which $\mathcal W(t)  \vert_{t = w_0^2}  = \mathcal W_0 = 0.2815$ and report all dimensionful quantities in units of $w_0$~\cite{Fodor:2012td}. Dimensionful quantities expressed in units of $w_0$ are denoted by a hat, for example masses are denoted as
$\hat m\equiv (m a) (w_0/a)$. 
 The measured values of $w_0/a$,
obtained on the lattice by discretizing Eq.~\eqref{eq:gradient_flow_differential_equation}, using the clover discretization of the field-strength tensor, are tabulated in Tab.~\ref{tab:ensembles}.

We monitor the topological charge, $Q$, of each configuration, which in the continuum is defined as
\begin{align}
    Q = \frac{1}{32\pi^2} \int {\rm d}^4x\, \epsilon^{\mu \nu \rho \sigma} \, \Tr \,G_{\mu\nu}G_{\rho\sigma}.
\end{align}
On the lattice, we measure $Q$ by following the same process as in Ref.~\cite{Bennett:2022ftz}, to which we refer the reader for details.
By applying the gradient flow to smoothen the gauge fields, UV fluctuations are removed in the measurement of $Q$. 
None of the ensembles used in this study show hard evidence of topological freezing, with the Monte Carlo algorithm sampling configurations with multiple values of $Q$. In order to quantify this statement we study the autocorrelations in 
Monte-Carlo time of both the topological charge, $Q$, 
as well as the average plaquette, $\langle P \rangle$.
We perform the measurements of the gradient flow scale, the topological charge and the hadron correlators using the HiRep code~\cite{DelDebbio:2008zf,HiRepSUN}, which has been extended to symplectic gauge groups~\cite{HiRepSpN}. To this purpose, we convert the configurations created with the Grid code using the Gauge Link Utility (GLU) library~\cite{GLU}.

\begin{figure}
    \includegraphics[width=0.8\linewidth]{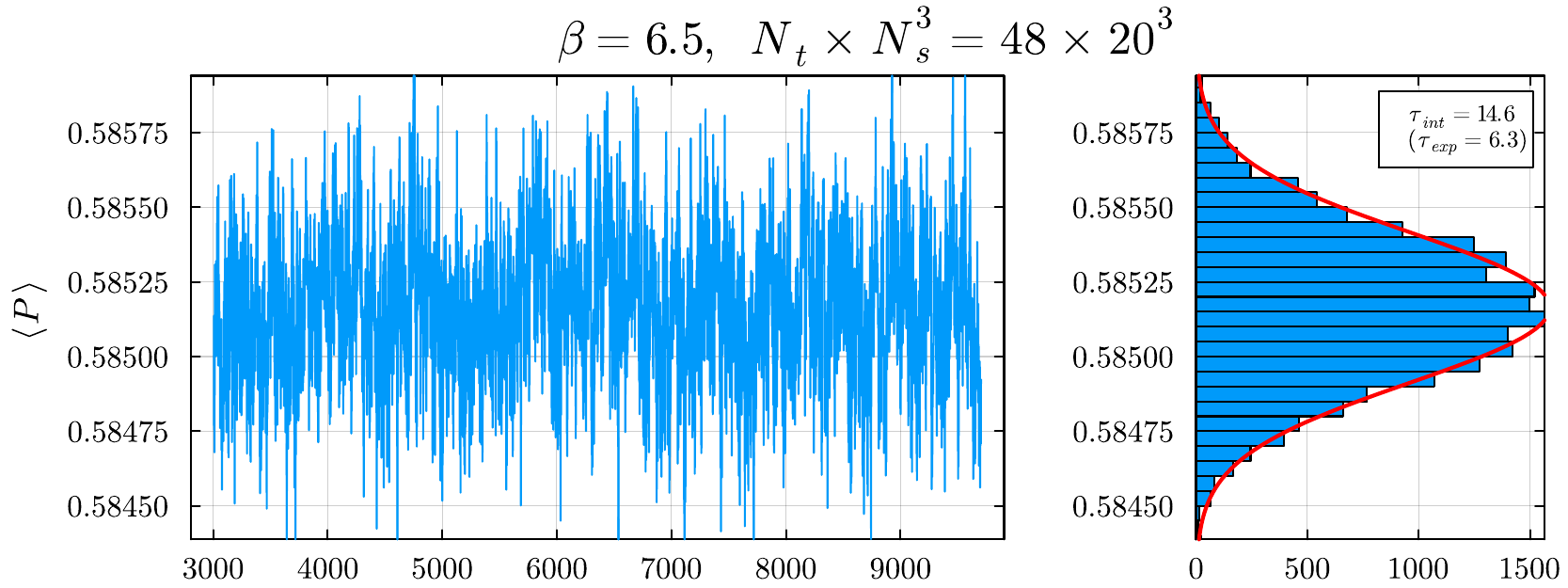}
    \caption{Thermalised trajectory of the average plaquette, $\langle P \rangle$, in Monte-Carlo time (left) and histogram of the same data (right), for the lattice ensemble denoted as M$1$ in Table~\ref{tab:ensembles}. All other ensembles show qualitatively similar behaviors. We fitted the histogram to a Gaussian distribution (red line on the right). Configuration sampling and binning of the raw data are discussed in the main text.}
        \label{fig:plaquette_autocor}
\end{figure}

For a generic  observable of interest, $X$, we denote as $\tau=1,\,\dots,\,N$,  the Monte-Carlo time, as  $X_i$
the individual measurement of the observable, 
 and as $\bar X$ the arithmetic mean of $X$. The Madras-Sokal integrated autocorrelation time, $\tau_{\rm int}$, is defined as follows~\cite{Madras:1988ei,Wolff:2003sm,Luscher:2004pav}:
\beqs
    \tau_{\rm int} &=& \max_{\tau_{\rm max}} \tau_{\rm int}(\tau_{\rm max})\,,
    \eeqs
    where
    \beqs
    \tau_{\rm int}(\tau_{\rm max}) &=& \frac{1}{2} + \sum_{t=1}^{\tau_{\rm max}} \Gamma(\tau)\,,
    \eeqs
    and
    \beqs
    \Gamma(\tau) &=& \sum_{i=1}^{N-\tau} \frac{\left( X_{i} - \bar X \right)\left(  X_{i+\tau} - \bar X  \right)}{N-\tau}\,.
\eeqs
In applying these definitions, we assume that the Monte-Carlo time series, $X_i$, is fully thermalized. In practice, we vary also the thermalization cut-off and choose the thermalization time so that histograms of the plaquette and topological charge show a Gaussian behavior. We use this definition by treating two values of $X_i$ as uncorrelated if they are separated by at least $2 \tau_{\rm int}$ in Monte-Carlo time $\tau_i$. For comparison, we additionally compute the exponential autocorrelation time, $\tau_{\rm exp}$, by fitting the autocorrelation function, $C_{X}(\tau)$,
defined as
    \beqs
    C_X(\tau) &=& \sum_{i=1}^{N-\tau} \left( X_{i} - \bar X \right)\left(  X_{i+\tau} - \bar X  \right)\,,
\eeqs
 to an exponential decay:
\beqs
    \frac{C_X(\tau)}{C_X(0)} &\sim& \exp \left( - \frac{\tau}{\tau_{\rm exp}} \right)\,.
    \eeqs
    
\begin{figure}
    \includegraphics[width=0.8\linewidth]{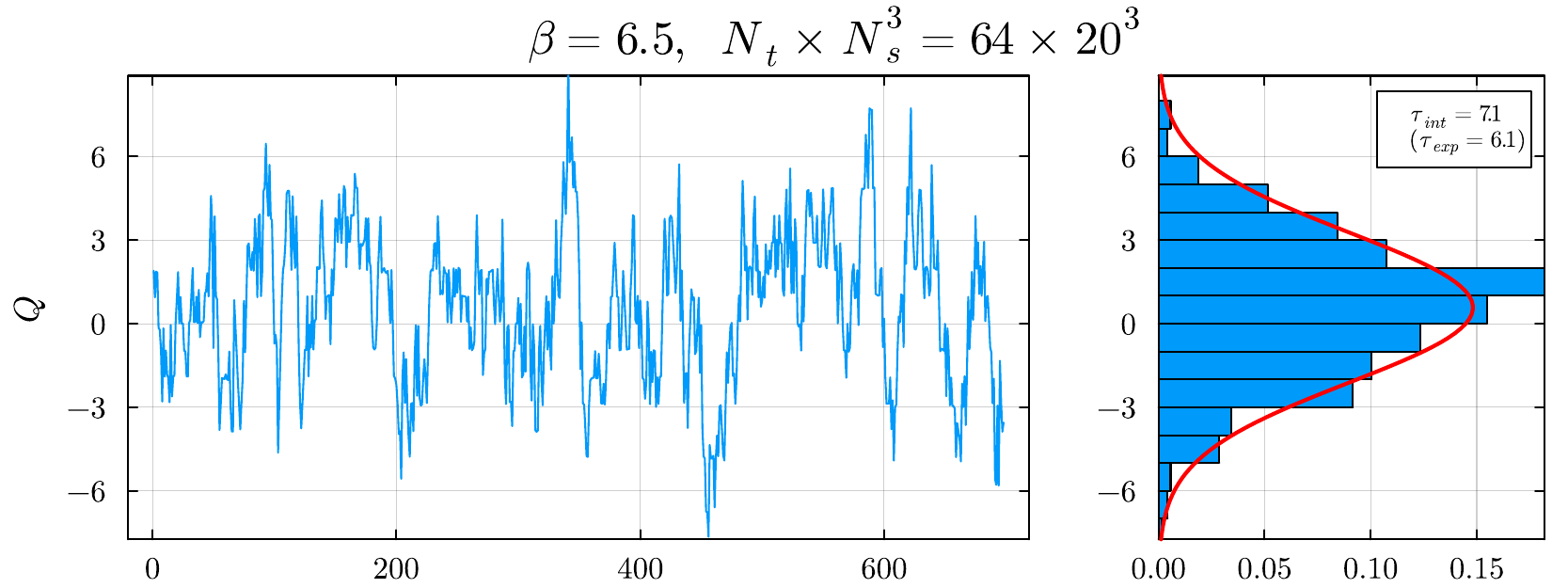}
    \caption{Thermalised trajectory of the topological charge, $Q$, in ensemble M2, restricted to configurations used in the remainder of this paper, after removing the effect of autocorrelation in the plaquette. We fitted the histogram to a Gaussian distribution (red line on the right). Gauge configurations show a modest level of residual autocorrelation, with respect to the topological charge. The other ensembles exhibit qualitatively similar behaviors.
        \label{fig:topology_autocor}}
\end{figure}

In Fig.~\ref{fig:plaquette_autocor} we display one example of average plaquette trajectory in Monte-Carlo time. We find the exponential autocorrelation to be always smaller than the integrated Madras-Sokal autocorrelation time. We keep one gauge configuration every $n_{\rm skip}$ Monte-Carlo time units, such that $n_{\rm skip} \gtrsim \tau_{\rm int}^{\langle P \rangle}$ for the autocorrelation time of the average plaquette. We measure our observables on these gauge configurations and bin the resulting dataset with bin size of $2$. We retain $N_{\rm conf}$ measurements, or  $N_{\rm conf}/2$ independent samples. We tabulate $N_{\rm conf}$, in Tab.~\ref{tab:ensembles}.

 \begin{table}[t!]
    \begin{tabular}{ |c|c|c|c|c| }
     \multicolumn{3}{c}{} \\
     \hline  \hline
     Label & Interpolating operator $\mathcal{O}^A$ & $J^P$ & $Sp(4)$ & $SO(6)$ \\
     \hline
     PS  & $\bar{Q}^I \gamma_5 Q^J$ & $0^{-}$ & $5$ & $1$ \\
     V  & $\bar{Q}^I \gamma_i Q^J$ & $1^{-}$ & $10$ & $1$  \\
     T  & $\bar{Q}^I \gamma_0 \gamma_i Q^J$ & $1^{-}$  & $10$ & $1$\\ 
     AV  & $\bar{Q}^I \gamma_5 \gamma_i Q^J$ & $1^{+}$ & $5$ & $1$  \\ 
     AT  & $\bar{Q}^I \gamma_0 \gamma_5 \gamma_i Q^J$ & $1^{+}$ & $10$ & $1$  \\ 
     S  & $\bar{Q}^I Q^J$ & $0^{+}$ & $5$ & $1$  \\
     \hline
    ps & $\bar{\Psi}^k \gamma_5 \Psi^{\ell}$ & $0^{-}$  & $1$ & $20$ \\
 v & $\bar{\Psi}^k \gamma_i \Psi^{\ell}$ & $1^{-}$& $1$ & $15$   \\
 t& $\bar{\Psi}^k \gamma_0 \gamma_i \Psi^{\ell}$ & $1^{-}$ & $1$ & $15$  \\ 
 av& $\bar{\Psi}^k \gamma_5 \gamma_i \Psi^{\ell}$ & $1^{+}$  & $1$ & $20$ \\
     at& $\bar{\Psi}^k \gamma_0 \gamma_5 \gamma_i \Psi^{\ell}$ & $1^{+}$ & $1$ & $15$  \\ 
    s & $\bar{\Psi}^k \Psi^{\ell}$ & $0^{+}$  & $1$ & $20$ \\
     \hline  \hline
    \end{tabular}
    \caption{\label{table:fermionic_operators} Meson interpolating operators, $\mathcal{O}^A(x)$, built using Dirac fermions transforming in the fundamental, 
    $Q^{I}$, and antisymmetric, $\Psi^{k }$, representations. Here,  $I,\,J = 1, \, 2$ and $k,\,\ell = 1, \, 2, \, 3$ are flavor indices, and we restrict our attention to the case $I\neq J$ and $k \neq \ell$. The index $i = 1, \, 2, \, 3$ refers to spatial directions. Color and spinor indices are implicit and summed over. $J^{P}$ are space-time quantum numbers. We report also the representation under the unbroken $Sp(4)$ and $SO(6)$ global symmetries.}
    \end{table}

We then compute the topological charge, $Q$, for the $N_{\rm conf}$ gauge configurations used in the remainder of this paper. As for the plaquette, we plot the trajectory, compute the autocorrelation time(s), and display the measurements in a histogram. One example of the results is shown in Fig.~\ref{fig:topology_autocor}. We report the autocorrelation of the topology, $\tau_{\rm int}^Q$, and the average topological charge, $\bar Q$, in Tab.~\ref{tab:ensembles}.
Qualitatively, all resulting distributions are Gaussian. Yet, $\tau_{\rm int}^Q > 1$  and a non-vanishing average value of $Q$ is measured for all ensembles.
 We conclude that the configurations used in this paper are affected by a moderate amount of residual
autocorrelation in the topological charge.\footnote{In pure gauge theory, no sizeable effect 
on the glueball spectrum was detected even in ensembles with complete topological freezing~\cite{Bonanno:2022yjr}.}

\section{Correlation functions }
\label{Sec:correlation_functions}

   Table~\ref{table:fermionic_operators} summarises the properties of the meson operators, $\mathcal{O}^A$, of interest in this paper.
They are constituted by two  fermions in the fundamental, ${\rm (f)}$, representation $Q^I$, where $I = 1,\, 2$, 
or two in the antisymmetric, ${\rm (as)}$,  representation, denoted $\Psi^k$, where $k = 1, \, 2, \, 3$. 
We label them as pseudoscalar, vector, tensor, axial-vector, axial-tensor and scalar,
both for mesons made of ${\rm (f)}$ fermions and ${\rm (as)}$ ones---in the latter case, we conventionally label them with lower case acronyms.
 We also display the quantum numbers, $J^P$, and the irreducible representation in the unbroken $Sp(4)\times SO(6)$ global symmetry.

Zero-momentum, two-point correlation functions are defined by averaging over lattice sites as:
\begin{equation}
\label{eq:correlators_average}
    C^{AB}(t) = \sum_{\vec{x}} \, \langle \mathcal{O}^A (t, \vec{x}) \, \bar{\mathcal{O}}^B (0) \rangle \equiv \langle \mathcal{O}^A(t) \bar{\mathcal{O}}^B(0) \rangle \,.
\end{equation}
For $A\neq B$, Equation~(\ref{eq:correlators_average}) can be recast for a generic $C(t)$ as a summation over a complete basis, $|n\rangle\langle n|$, with associated energies $E_n$, as follows:
\begin{equation}
\label{eq:correlator_matrix_elements}
C^{AB}(t) = \sum_n  \dfrac{\langle 0 | \mathcal{O}^{A}(0) | n \rangle \langle n |  \bar{\mathcal{O}}^{B}(0) | 0 \rangle }{2E_n} \, b(t,E_n)\,,
\end{equation}
where $b(t, E_n)$ becomes an exponential function in the limit of infinite temporal lattice extent, 
$
b(t, E_n) \xrightarrow[N_t \to \infty]{} e^{-t\,E_n}
$.
For a finite,  periodic lattice in the time extent, one expects the following functional form
\begin{equation}
\label{eq:basis_functions}
b(t, E_n) = b_+(E_n) e^{-t \,E_n} + b_-(E_n) e^{-(N_t - t)E_n}\,,
\end{equation}
which suggests identifying the interval for which the ground states dominate by examining the effective mass, defined as
\begin{equation} \label{Eq:effmass}
    a\,m_{\rm eff} (t) \equiv \cosh^{-1} \left[ \dfrac{C^{AB}(t+a) + C^{AB}(t-a)}{2C^{AB}(t)} \right]\,.
\end{equation}
At large Euclidean times, one expects $ a\,m_{\rm eff} (t)  \xrightarrow[t \to \infty]{} a\,E_0$.
To extract the ground state energy $E_0$, therefore, one has the freedom of fitting the plateau according to Eq.~\eqref{Eq:effmass} or the correlators as in Eq.~\eqref{eq:correlator_matrix_elements}. In the present paper, we fitted correlators setting coefficients $b_+(E_n) = b_-(E_n) = 1$ when $A = B$, and $b_+(E_n) = 1, \, b_-(E_n) = 0$ when $A \neq B$ over the interval that shows a plateau in the effective mass.

In the remainder of the section, we describe two techniques,  APE and Wuppertal smearing, that
 we adopt in order to optimize the extraction of physical information in the fitting procedure. 
Moreover, we introduce an additional, well established methodology in lattice field theory, the GEVP, that reduces the contributions from the excited states to ground states at large $t$. 

\subsection{ APE  and Wuppertal smearing algorithms}
\label{Sec:corr_smearings}

From Eq.~\eqref{eq:correlator_matrix_elements}, one expects contamination from excited states to affect meson two-point functions at moderate-to-small time separation.
Conversely, for any finite lattice, there is an intrinsic limitation on the maximum length of the time separation in the two-point functions. 
The combination of these two factors results in a limitation on the length of the plateau displaying (approximately) constant effective mass---see Eq.~(\ref{Eq:effmass})---that can be used for spectroscopy. 
As visible in Eq.~(\ref{eq:correlator_matrix_elements}), the contribution of the $n^{th}$ excited state, relative to the ground state, is determined by two factors: the exponential decays proportional to $\exp \left[ -(E_{n} - E_{0}) t \right]$, and the overlap functions
encoded in the matrix elements $\langle 0 | \mathcal{O}(0) | n \rangle$.
Increasing the overlap of the interpolating operators, $\mathcal{O}$, with the ground state, $| 0 \rangle$,  relative to the excited states,
 $|n \rangle$,  results in suppressed contamination from excitations,  longer plateaux in effective mass plots, and more precise spectroscopy.

Our implementation of these ideas combines 
 APE smearing~\cite{APE:1987ehd,Falcioni:1984ei}  of the gauge configurations with 
 Wuppertal smearing~\cite{Gusken:1989qx,Roberts:2012tp,Alexandrou:1990dq} 
  of the sink/source operators.
 APE smearing acts on gauge links, smoothening their ultraviolet fluctuations, and hence improving the statistical control over the plateaux being analyzed. 
Wuppertal smearing modifies the  fermion fields used to source the two-point functions, and increases the ground state overlap by using extended interpolators, instead of point-like ones. 

APE smearing consists of applying an iterative procedure involving the staple operator around each gauge link, $S_\mu (x) \equiv \sum_{\pm \nu \neq \mu} U_\nu(x)U_\mu(x+\hat{\nu})U^\dagger_\nu(x+\hat{\mu})$, as follows
\begin{equation}
\label{eq:APE_smearing_formula}
    U^{(m)}_\mu(x) = \mathcal{P} \left\{ (1-\alpha_{\textrm{APE}})U^{(m-1)}_\mu(x) + \frac{\alpha_\textrm{APE}}{6} S^{(m-1)}_\mu(x) \right \}\,,
\end{equation}
with the initial conditions $U_{\mu}^0=U_{\mu}$ and $S^0_{\mu}=S_{\mu}$.
The iteration number is $m = 1,\, \dots,\, N_{\rm APE}$ and $\alpha_\textrm{APE}$ is called \textit{APE-smearing step size}. In the case of interest in this study, we use
 the same process for gauge links transforming in the fundamental and antisymmetric representations.
As the gauge links at each iteration are summed over their neighboring staples according to Eq.~(\ref{eq:APE_smearing_formula}), they do not necessarily lie within the group manifold.
Therefore, we use a projection operator, $\mathcal{P}$, to project the smeared link variable into the group manifold.
The explicit form of the projector depends on the gauge group and representation considered. 

In order to illustrate the Wuppertal smearing of source and sink operators, one starts at first with the Dirac equation for point-like source and sink:
\begin{equation}
\label{eq:Wuppertal_dirac_eq}
    \sum_{y, \, \beta, \, b} D^{R}_{a \alpha \, b \beta} (x,y) \, S^{b \beta}_{R \, c \gamma} (y,0) = \delta_{x, 0}  \, \delta_{\alpha \gamma} \, \delta_{ac}\,,
\end{equation}
where $D^{R}_{a \alpha \, b \beta}$ is the Wilson-Dirac operator in the representation, $R$,
spinor indices are denoted as $\alpha,\,\beta,\,\gamma$,
and (generalised) color indexes as  $a,\,b,\, c$, respectively.
The solution, 
$S^{b \beta}_{R \, c \gamma}$, is the hyperquark propagator in such representation, $R$. The two-point correlation function is then written as
\begin{equation}
\label{eq:correlator_propagator}
    C(t) = \langle \mathcal{O}_{R} (t) \, \bar{\mathcal{O}}_{R} (0) \rangle = \Big\langle \sum_{\vec{x}} \Tr\, \left[ \left( \Gamma \,  S_R(x,0) \, \bar{\Gamma} \, S_R (x,0) \right)  \right] \Big\rangle\,,
\end{equation}
where $\Gamma$ and $\bar{\Gamma}$ depend on the spin structure of the interpolating operator, $\mathcal{O}_{R}$, listed in Tab.~\ref{table:fermionic_operators}.  

Wuppertal smearing consists of replacing $\delta_{x,0}$, in the right-hand side of Eq.~(\ref{eq:Wuppertal_dirac_eq}), with a function $q_R^{(n+1)}(x)$, defined through an iterative procedure based on a diffusion process:
\beqs
\label{eq:Wuppertal_smearing_formula}
    q_{\rm R}^{(n)}(x) &=& \frac{1}{1+6\varepsilon_{\rm R}} \left [ q_{\rm R}^{(n-1)}(x) + \varepsilon_{\rm R} \sum_{\mu= \pm1}^{\pm 3} U^{\rm R}_\mu(x)q_{\rm R}^{(n-1)}(x+\hat{\mu}) \right ] \,,\qquad (n > 0)\,,\\
    q_{\rm R}^{(0)}(x)&=&\delta_{x, 0}\,.
\eeqs
Here, $\varepsilon_{\rm R}$ is the representation-dependent Wuppertal-smearing step size. Each source smearing requires an inversion of the Dirac operator, and the result is a new, source-smeared, propagator denoted as $S^{(n)}_{R}(y,\,0)$. Sink smearing is obtained by applying the smearing iteration of Eq.~\eqref{eq:Wuppertal_smearing_formula} to the source-smeared propagator. Note, that this does not require further inversions. We denote the propagator with $N_{\rm source}$ iterations of source smearing and $N_{\rm sink}$ iterations of sink-smearing as $S^R_{(N_{\rm source},N_{\rm sink})} (x,\,0)$. The smeared two-point functions read
\begin{equation}
    C_{N_{\rm source}, \, N_{\rm sink}}(t) = \Big\langle \sum_{\vec{x}} \Tr \left[  \Gamma \, \left(S^R_{(N_{\rm source},N_{\rm sink})} (x,\,0) \right)   \, \bar{\Gamma} \, \left(S^R_{(N_{\rm source},N_{\rm sink})} (x,\,0) \right)  \right]\Big\rangle \,.
\end{equation}
 
The measurement of APE and Wuppertal smeared two-point mesonic correlation functions are performed using the HiRep code~\cite{DelDebbio:2008zf,HiRepSUN,HiRepSpN}.
The APE and Wuppertal smearing step-sizes, $\alpha_\textrm{APE}, \, \varepsilon_{\rm f}, \, \varepsilon_{\rm as}$, and the number of iterations, $ N_{\rm APE}, \, N_{\rm source}, \, N_{\rm sink}$, are parameters that are tuned to optimize the signal,   both by improving the effective mass plateaux,  
as well as the resolution of peaks in the spectral density reconstruction---as we will discuss later. 
For our purposes, for ensembles M1 to M4, we apply Wuppertal smearing step-sizes $\varepsilon_{\rm f} = 0.20$ for the fundamental sector, and $\varepsilon_{\rm as} = 0.12$ for the antisymmetric one. For the ensemble M5, $\varepsilon_{\rm f} = 0.24$ for the fundamental sector and $\varepsilon_{\rm as} = 0.12$ for the antisymmetric one. Typical iteration numbers for sink and sources are $N_{\rm sink}, \, N_{\rm source} \in [0, 80]$. For  APE smearing, we apply $\alpha_\textrm{APE} = 0.4$ and $N_{\rm APE} = 50$. We verified explicitly that the process did not lead to large changes in the mass of the pseudoscalar ground state, and that the convexity of the effective mass plots is not altered, hence excluding the possibility of oversmearing.

\subsection{Generalized Eigenvalue Problem}
\label{Sec:GEVP_method}

The ground state in a given channel, $E_0$,  can be identified by the plateau in effective mass,
at least as long as the ground state is clearly separated in mass from its excited states.  A pragmatic way to isolate excitations involves performing multi-functional fits by minimizing a correlated chi-square functional
\begin{equation}
\label{eq:correlated_chisq_corr}
    \chi^2 = \sum_{t, \, t^{'}} \left( h^{(k)} (t) - C (t) \right) \text{Cov}^{-1}_{t\,t^{'}} [C] \left( h^{(k)} (t^{'}) - C (t^{'}) \right)\,,
\end{equation}
where the fitting function is the periodic version of a multi-exponential fit,
\begin{equation}
    h^{(k)} (t) = \sum_{n = 1}^{k} \mathcal{B}_n \, b(t,E_n)\,,
\end{equation}
while $b(t,E)$ is defined in Eq.~(\ref{eq:basis_functions}). Yet,
 the extraction of such excitations, $E_n$,  is hindered by the increasing number of degrees of freedom it requires, and the signal gets exponentially suppressed with growing $n$.

A variational approach based on solving a \emph{generalized eigenvalue problem} (GEVP) can overcome this problem~\cite{Blossier:2009kd}. 
One starts by defining a matrix-valued correlation function, $\mathcal{C}_{ij}(t)$, encompassing the Euclidean-space two-point correlation functions of a set of operators, ${\cal O}_i$:
\begin{equation}
\label{eq:variational_matrix}
    \mathcal{C}_{ij} (t) = \langle \mathcal{O}_i (t) \bar{\mathcal{O}}_j (0) \rangle, \quad i,j = 0, \dots, N - 1\,.
\end{equation}
Under the assumption that  non-degenerate energy levels exist, they can be ordered with $E_{n} < E_{n+1}$,
where the eigenvalues are labelled as $n = 0, \, \dots, \, N - 1$ In this notation, $E_0$ denotes the ground state energy. The GEVP is defined as
\begin{equation}
\label{eq:GEVP_equation}
\mathcal{C}(t)v_n(t,t_1) = \lambda_n(t,t_1)\mathcal{C}(t_1)v_n(t,t_1)\,, 
\end{equation}
in which the number of new functions, $v_n$ and $\lambda_n$, matches
 the dimension of the variational basis used to build the matrix $\mathcal{C}$. 
 By fixing a reference value, $t_1$, Eq.~(\ref{eq:GEVP_equation}) can be solved as an eigenvalue equation for each lattice-time slice $t>t_1$. 
This procedure results in determining the eigenvalues, $\lambda_n (t,t_1)$.

In the next step, the eigenvalue functions, $\lambda_n (t, t_1)$, are used to define $N$ effective-mass plateaux, $am_n^{\rm eff} (t, t_1)$:
\begin{equation}
    m^{\rm eff}_{n} (t, t_1) = - \dfrac{\partial \log \lambda_n (t,t_1)}{\partial t} = - \dfrac{1}{a} \left\{ \log \left[ \lambda_n (t+a, t_1) \right] - \log \left[ \lambda_n ( t, t_1) \right] \right\}\,,
\end{equation}
or, in the more realistic case of a periodic lattice:
\begin{equation}
    m^{\rm eff}_{n} (t, t_1) =  \dfrac{1}{a} \, \cosh^{-1} \left[ \dfrac{{\lambda_n (t+a, t_1) + \lambda_n (t-a, t_1) }}{2\lambda_n ( t, t_1)} \right] \,.
\end{equation}
These quantities are expected to converge to the energy levels,
\begin{equation}
    aE_n = \lim_{t\to \infty} am_n^{\rm eff} (t, t_1) \quad n = 0, \, \dots, \, N - 1\,.
\end{equation}

In the case of lattice correlation functions, to solve the GEVP in Eq.~(\ref{eq:GEVP_equation}) one focuses on
late-time slices, so that the corrections due to higher-energy excitations, with $E > E_{N-1}$, are suppressed.  As shown in Ref.~\cite{LUSCHER1990222}, for fixed $t_1$ such contributions are expected to take the form 
\begin{equation}
    am^{\rm eff}_n (t, t_1) = aE_n + \mathcal{O}(e^{-\Delta (aE_n) t}), \quad \Delta (aE_n) = \min_{m \neq n} |aE_m - aE_n|\,,
\end{equation}
so that the size of the contaminations depends on the gap between the energy level of the spectrum, and the size of $t$. As discussed in \cite{Blossier:2009kd} there are systematic effects proportional to $\mathcal{A} \cdot \exp(E_N \cdot t_1)$.
On general grounds, then, one expects the quality of the results and signal in each plateau to improve when considering lattices with larger time extents, $N_t$. 
In our study, we build a variational basis by considering different levels of Wuppertal smearing for the interpolating source and sink operators in each mesonic channel. 
For the pseudoscalar, axial-vector, axial-tensor, and scalar channels (and similar for mesons made of ${\rm(as)}$ fermions) we use three different levels of smearing: $N_{\rm source}, \, N_{\rm sink} = 0, \, 40, \, 80$.
Therefore, in this case, the correlation matrix $\mathcal{C}$ is
\begin{equation}
\label{eq:simpler_GEVP_eq}
    \mathcal{C}^i (t) = \begin{pmatrix}
C^{i}_{0, \, 0} (t) & C^{i}_{0, \,  40} (t) & C^{i}_{0, \, 80} (t)\\
C^{i}_{40, \, 0} (t) & C^{i}_{40, \,  40} (t) & C^{i}_{40, \, 80} (t)\\
C^{i}_{80, \, 0} (t) & C^{i}_{80, \,  40} (t) & C^{i}_{80, \, 80} (t)
\end{pmatrix}
\end{equation}
with $C^{i}_{nm} (t) = \langle \mathcal{O}^{i} (t) \bar{\mathcal{O}}^{i} (0) \rangle_{nm}$, where $i = \text{PS}, \, \text{AV}, \, \text{AT}, \, \text{S}, \, \text{ps}, \, \text{av}, \, \text{at}, \, \text{s}$ and $n, \, m$ indicate the level of Wuppertal smearing applied to source and sink operators, $n = N_{\rm source}, \, m = N_{\rm sink}$.

In the case of V and T channels,  the operators transform in the same way under the unbroken symmetry groups, and are expected to mix, and source the same spectrum.  Of course, this argument is no longer true for our measurements on a discretized lattice as the rotational symmetry is broken. Yet, we found no discernible difference between these two channels for given statistical errors. Therefore, we extend the correlation matrix to include the cross-channels $V/T$ and $T/V$, resulting in the following
\begin{equation}
\label{eq:crosschannel_GEVP_eq}
    \mathcal{C}^{V,\,T} (t) = \begin{pmatrix}
C^{V}_{0, \, 0} (t) & C^{V}_{0, \,  40} (t) & C^{V}_{0, \, 80} (t) & C^{V/T}_{0, \, 0} (t) & C^{V/T}_{0, \,  40} (t) & C^{V/T}_{0, \, 80} (t)\\
C^{V}_{40, \, 0} (t) & C^{V}_{40, \,  40} (t) & C^{V}_{40, \, 80} (t) & C^{V/T}_{40, \, 0} (t) & C^{V/T}_{40, \,  40} (t) & C^{V/T}_{40, \, 80} (t)\\
C^{V}_{80, \, 0} (t) & C^{V}_{80, \,  40} (t) & C^{V}_{80, \, 80} (t) & C^{V/T}_{80, \, 0} (t) & C^{V/T}_{80, \,  40} (t) & C^{V/T}_{80, \, 80} (t)\\
C^{T/V}_{0, \, 0} (t) & C^{T/V}_{0, \,  40} (t) & C^{T/V}_{0, \, 80} (t) & C^{T}_{0, \, 0} (t) & C^{T}_{0, \,  40} (t) & C^{T}_{0, \, 80} (t)\\
C^{T/V}_{40, \, 0} (t) & C^{T/V}_{40, \,  40} (t) & C^{T/V}_{40, \, 80} (t) & C^{T}_{40, \, 0} (t) & C^{T}_{40, \,  40} (t) & C^{T}_{40, \, 80} (t)\\
C^{T/V}_{80, \, 0} (t) & C^{T/V}_{80, \,  40} (t) & C^{T/V}_{80, \, 80} (t) & C^{T}_{80, \, 0} (t) & C^{T}_{80, \,  40} (t) & C^{T}_{80, \, 80} (t)\\
\end{pmatrix}
\end{equation}
where the cross-channels correlators are defined as $C^{V/T}_{nm} (t) = \langle \mathcal{O}^{V} (t) \bar{\mathcal{O}}^{T} (0) \rangle_{nm}$, and $C^{T/V}_{nm} (t) = \langle \mathcal{O}^{T} (t) \bar{\mathcal{O}}^{V} (0) \rangle_{nm}$ 
The enlarged variational basis can allow for resolving higher excitations.

\section{Spectral densities}
\label{Sec:spectral_density}

In this section, we introduce and critically appraise the spectral density reconstruction algorithm and its application to meson spectroscopy. We assess its dependence on the energy smearing kernel and on the  APE and Wuppertal smearing present in the input data---the two-point correlation function. To this extent, we implemented this technology in the Python software package \texttt{LSDensities}, and made it publicly available in Ref.~\cite{Forzano:2024}. Specific details of the latter are described in the Appendix. We pay particular attention to estimating systematic effects in the reconstruction procedure, and to optimising parameter choices.

\subsection{The Hansen-Lupo-Tantalo (HLT) method}
\label{Sec:HLT_theory}

Given a generic two-point correlation function, $C(t)$, 
 the spectral density, $\rho(E)$, is defined as follows:
\begin{equation}
\label{eq:inverse_laplace}
C(t) = \int_{E_{\rm min}}^{\infty} dE \, \rho(E) \, b(t,E)\,,
\end{equation}
where $b(t,E)$ has been introduced in Eq.~(\ref{eq:basis_functions}), for  periodic time extent.
This definition reduces to an inverse Laplace transform for infinite time extent,
when one chooses vanishing $E_{\rm min}$. This lower bound in the integration can be chosen between zero and the energy corresponding to the ground state of the theory in the channel defined by 
the interpolating operators yielding $C(t)$. 

The input data takes the form of a set of measurements, labeled by $k = 1,\, \dots,\, N_m$ (where $N_m \leq N_{\textrm{conf}}$ of Tab.~\ref{tab:ensembles}), 
of the given  correlation function, $C_k (t)$. On the lattice, time is discretized,
so that $t=1,\,\cdots,\,t_{\rm max}\leq N_t / 2$. The $t_{\rm max} \times t_{\rm max}$ covariance matrix is
\begin{equation}
\label{Eq:cov}
\hbox{Cov}_{t t{\prime}} [C] \equiv \dfrac{1}{N_m} \sum_{k = 1}^{N_m } \, \Big( C_k (t) - \langle C(t) \rangle \Big) \Big( C_k (t^{\prime}) - \langle C(t^{\prime}) \rangle \Big)\,,
\end{equation}
where $\langle C(t) \rangle$ is the arithmetic average over the available measurements at given $t$.

In the literature, several approaches have been applied to reconstruct spectral densities starting from a finite set of measurements, $C(t)$, which are affected by noise, circumstances under which the inversion of Eq.~\eqref{eq:inverse_laplace} is an ill-posed problem. Among them, the Backus-Gilbert algorithm was originally devised in Ref.~\cite{Backus:1968svk}, and then modified in Ref.~\cite{Hansen:2019idp} to better suit the context of lattice simulations. We refer to this improved version of the Backus-Gilbert algorithm as Hansen-Lupo-Tantalo (HLT) method.

The starting point is the introduction of  smeared  spectral densities\footnote{Smearing of spectral densities should not be confused with (APE and Wuppertal) smearing of two-point correlation functions.}, via the
rewriting
\begin{equation}
\label{eq:smeared_sp_dens}
\rho_\sigma (\omega) \equiv \int_{E_{\rm min}}^{\infty} \, dE \, \Delta_\sigma (E - \omega) \, \rho(E)\,,
\end{equation}
which consists of a convolution of the original spectral density, $\rho(E)$,
with a \textit{smearing kernel}, $\Delta_\sigma (E - \omega)$. 
The parameter $\sigma$ characterizes the \textit{smearing radius} around the point $\omega$. 
 At non-zero smearing radius, the convolution defined in Eq.~\eqref{eq:smeared_sp_dens} is such that the smeared spectral density is always a smooth function.

In a finite volume, the spectral density is a sum of Dirac $\delta$ functions, corresponding to the discrete eigenvalues of the Hamiltonian:
\begin{equation}
\label{eq:unsmeared_spectral_density}
    \rho(E) = \sum_{n} k_n \, \delta (E-\omega_n)\,,
\end{equation}
where the sum runs over the eigenvalues $\omega_n$, and the coefficients, $k_n$, depends on the lattice spatial volume $N_s a$.
The choice of the smearing kernel is guided by the requirement that it is a smooth function,
 approaching a Dirac $\delta$ function as $\sigma \to 0$:
\begin{equation}
\label{eq:constraint_kernel}
\Delta_\sigma (E - \omega) \xrightarrow[\sigma \to 0]{} \delta(E-\omega)\,,
\end{equation}
and hence in this limit, the regulator disappears:
\begin{equation}
\rho_{\sigma}(\omega) \xrightarrow[\sigma \to 0]{} \rho(E=\omega)\,.
\end{equation}

A key idea of the HLT method is to choose and fix a given smearing kernel at the beginning of the procedure. If the chosen smearing kernel can be represented as an infinite sum over the space generated by the basis functions:
\begin{equation}
    \Delta_\sigma(E-\omega) = \sum_{t=1}^{\infty} g_t(\omega, \sigma) \,  b(t,E) ,
\end{equation}
in the HLT method we can look for the coefficients $g_t$ that provide the best approximation at a finite $t_{\rm max} \leq N_t / 2$:
\begin{equation}
\label{eq:reconstructed_kernel}
\bar{\Delta}_\sigma (E - \omega) = \sum_{t=1}^{t_{\rm max}} g_t (\omega, \sigma) \, b(t, E)\,,
\end{equation}
In the HLT procedure, the coefficients $g_t(\omega, \sigma)$ are defined by the minimum value of the following functional:
\begin{equation}
\label{eq:A_functional}
A[\vec{g}] \equiv \int_{E_{\rm min}}^{\infty} dE \,  e^{\alpha E} \, | \bar{\Delta}_\sigma (E - \omega) - \Delta_{\sigma} (E - \omega) |^2,
\end{equation}
which  measures the difference between the reconstructed smearing kernel, $\bar{\Delta}_\sigma (E - \omega)$, and the target kernel, ${\Delta}_\sigma (E - \omega)$. 
The unphysical $\alpha$ parameterises different choices of norm~\cite{DelDebbio:2022qgu}, 
and we shall discuss it later.
If the data are known with infinite precision, the inversion of Eq.~\eqref{eq:A_functional} is enough to provide the best values for the coefficients $g_t$ defining the smearing kernel. In any realistic case, with data affected by uncertainties, the minimization of Eq.~\eqref{eq:A_functional} amounts to the inversion of a highly ill-conditioned matrix. As suggested in Ref.~\cite{Backus:1968svk}, the problem can be regularised by adding a second functional:
\begin{equation}
B[\vec{g}] \equiv \sum_{t, \, t^{\prime} = 1}^{t_{\rm max}} \, g_{t} \, \hbox{Cov}_{t t^{\prime}} [C] \, g_{t^{\prime}}\,,
\end{equation}
where the $t_{\rm max} \times t_{\rm max}$ covariance matrix, $\hbox{Cov}_{t t^{\prime}} [C]$, is taken from the input  data.

With all of the above in place, the third step is to
 define the functional $W[\vec{g}]$~\cite{Hansen:2019idp}:
\begin{equation}
\label{eq:HLT_functional}
W[\vec{g}] \equiv \dfrac{A[\vec{g}]}{A[0]} + \lambda \dfrac{B[\vec{g}]}{B_{\rm norm}(\omega)} \quad , \quad \lambda \in (0,\infty)\,,
\end{equation}
where $B_{\rm norm}(\omega) = C^2(1) / \omega^2$ is written in terms of the normalization of the correlator at the initial time slice $C(t=1)$,  and $\lambda$ we refer to as the trade-off parameter. 
The second part of $W[\vec{g}]$ is called the \emph{statistical error functional}, and it is introduced to regularize 
the problem.

By minimizing $W[\vec{g}]$ for any given value of $\omega$, one can determine a set of coefficients $ \vec{g}(\omega) = (g_1 (\omega, \sigma), \,\cdots, \,g_{t_{\rm max}} (\omega, \sigma))$ which corresponds to the following estimator for the smeared spectral density:

\begin{equation}
\label{Eq:hat}
\hat{\rho}_{\sigma}(\omega) = \sum_{t = 1}^{t_{\rm max}} g_{t} (\omega, \sigma) \, C(t)\,.
\end{equation}

\begin{center}
   \begin{figure}
    \includegraphics[width=0.5\linewidth]{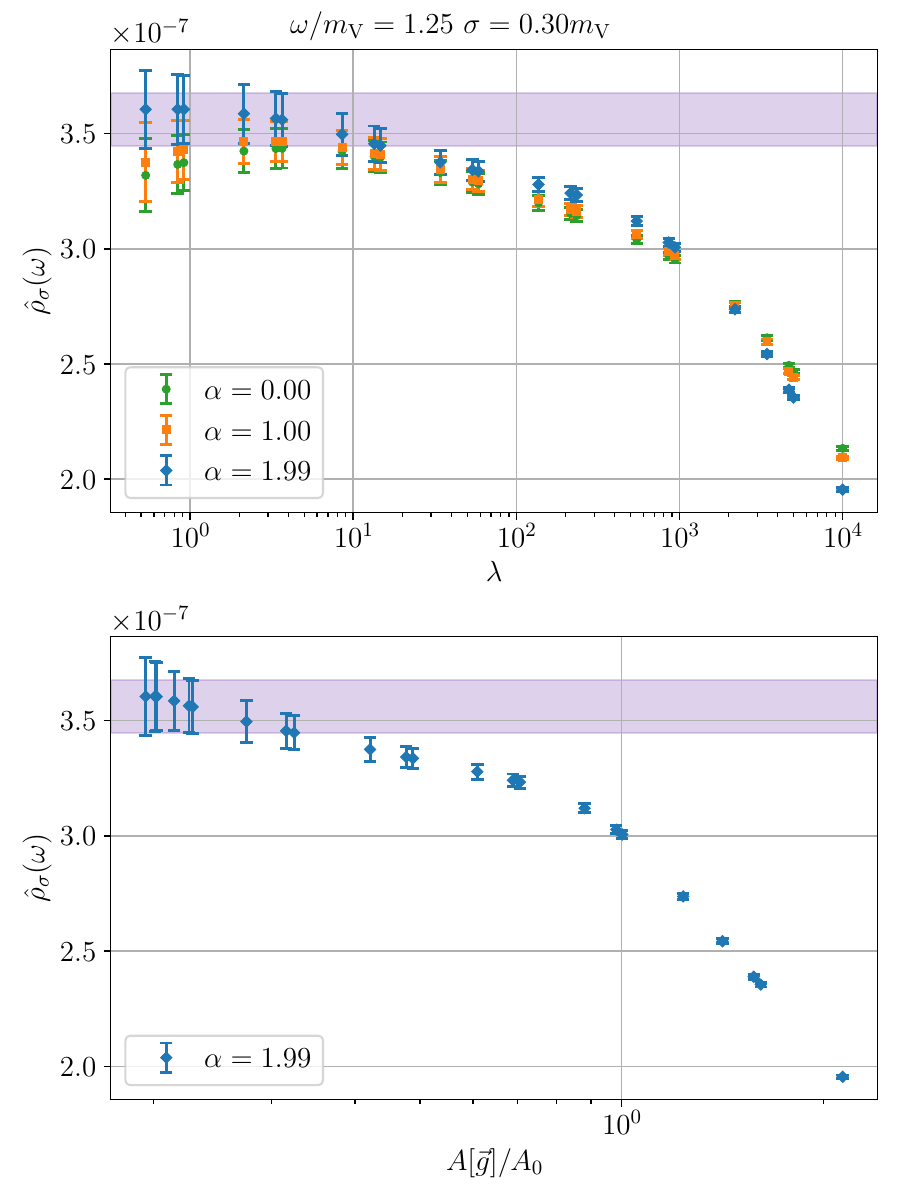}
    \caption{Examples of the plateaux in spectral density reconstruction for a fixed value of energy. In the top panel, the variation of the trade-off parameter $\lambda$ is shown on the x-axis, whereas $A_0 = A[\vec{g} = 0]$ is varied in the bottom panel. The correspondent values of reconstructed spectral densities $\hat{\rho}_{\sigma}(\omega)$ (shown on y-axes of both panels) are within statistical uncertainties of the unphysical parameters $\lambda$ and $\alpha$. Some representative values of $\alpha$ and $\lambda$ are chosen in the figure. The data correspond to the fundamental vector meson channel of ensemble M1, using $t_{\rm max} = N_t / 2$, $\alpha = 0$, $\sigma = 0.30m_{\mathrm{V}}$. 
    \label{fig:plateauscan}}
     \end{figure}
\end{center}

We devote the rest of this subsection to a critical discussion of the various systematics associated with the minimization of $W$.
 The ratio $\sqrt{A[\vec{g}] / A_0}$ provides an indication of the size of the systematic error due to the HLT method 
 for extracting $g_t(\omega, \sigma)$ with finite $t_{\rm max}$ and $N_m$, as this quantity describes the relative deviation between the targeted smearing kernel, $\Delta_{\sigma} (E - \omega)$, and the reconstructed one, $\bar{\Delta}_{\sigma} (E - \omega)$. 
 Conversely, the $B[\vec{g}] / B_{\rm norm}$ part of the functional $W[\vec{g}]$ provides an estimate of the statistical uncertainty for the reconstructed spectral density $\hat{\rho}_\sigma (\omega)$. 
To illustrate this point, 
in Fig.~\ref{fig:plateauscan},  we display the ratio $A[\vec{g}] / A_0$, 
evaluated at the minimum of $W$ for a given choice of $\lambda$ (and $\alpha$).
Systematic effects due to the reconstruction are unsuppressed when $A[\vec{g}] / A_0$ is larger,
which corresponds to choices of the parameters $\lambda$ and $\alpha$ that make the processing of information ineffective. Indeed, in this regime changes of $\lambda$ result in sizeable changes in $\hat{\rho}_\sigma (\omega)$.
In this case, 
 different numerical choices of  norm parameter, $\alpha$, yield results that are not  compatible with one another
  within statistical uncertainties.

Conversely, we can reduce the trade-off parameter, $\lambda$, reducing the size of $A[\vec{g}] / A_0$, and hence yielding smaller systematic effects. This comes at the price of larger statistical uncertainties, as can be seen in Fig.~\ref{fig:plateauscan}. This is expected on the basis of the very definition of  the functional $W[\vec{g}]$ in 
Eq.~(\ref{eq:HLT_functional}), as having a small trade-off parameter, $\lambda$, corresponds 
to minimising predominantly the $A[\vec{g}]/A_0$ functional. Therefore, considering smaller $\lambda$ values corresponds to forcing the systematic error to be smaller, and in principle the reconstruction more accurate,
but at the same time the functional $B[\vec{g}]/B_{\rm norm}$ is poorly constraining the system. 
In order to find a value of lambda that sits on middle grounds, we adopt the procedure described in Ref.~\cite{Bulava:2021fre}, where one looks for small enough values of lambda such that fluctuations due to this unphysical parameter are not relevant compared to the dominating statistical error. An example of this stability analysis is shown in Fig.~\ref{fig:plateauscan}: we can see how values for the prediction exist such that the dependence on unphysical parameters is not significant, and yet the reconstruction is not lost into statistical noise. Additional tunable parameters, such as $\alpha$, can be used with the same criterion to better identify this region in parameter space.

   \begin{figure}[h]
    \includegraphics[width=0.50\linewidth]{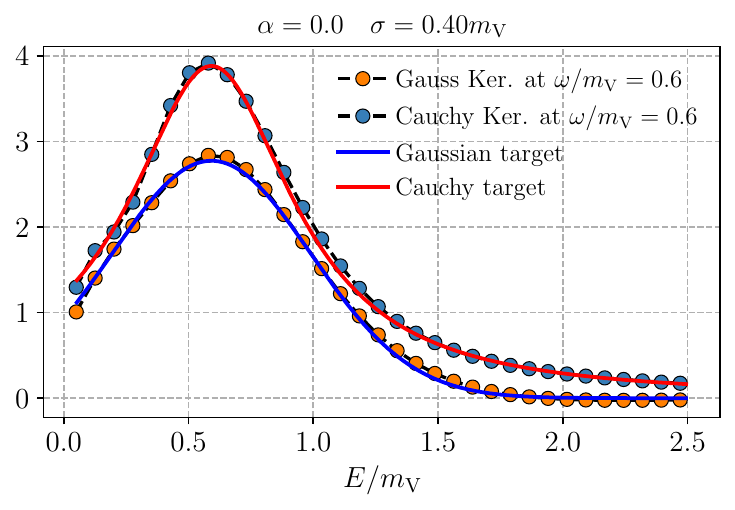}
    \caption{Smearing kernel reconstruction for the vector meson (V) consisting of fermions transforming in the fundamental representation of $Sp(4)$. The data analyzed is taken from ensemble M2, and  $m_V$ has been measured through the GEVP procedure---see Tab.~\ref{table:E2_results_ground}. For both the Gaussian and Cauchy kernels, we superimpose the target kernel, 
    ${\Delta}_\sigma (E - \omega)$,  and the reconstructed one, $\bar{\Delta}_\sigma (E - \omega)$. Parameters used are $t_{\rm max} = N_t / 2$, $\alpha=0$, and $\sigma=0.40\,m_{\text{V}}$.}
    \label{fig:smearingkernel}
     \end{figure}

While the stability analysis supports the idea that the dependence on the bias can be absorbed into statistical fluctuation, we take additional steps to estimate a possible systematic error left in our estimate, by further varying the algorithmic parameters:
\begin{itemize}
    \item The first component of the systematic error is
    \begin{equation}
    \sigma_{1, \,\rm sys}(E) \equiv |\rho_{\lambda_*}(E) - \rho_{\lambda_{*}/10}(E)|\,,
\end{equation}
where $\lambda_*$ is determined by the procedure described above, and $\alpha$ is fixed.
\item The second component of the systematic error is
\begin{equation}
    \sigma_{2, \,\rm sys} (E) = |\rho_{\lambda_*, \alpha}(E) - \rho_{\lambda_{*}, \alpha^{'}}(E)|\,,
\end{equation}
where $\lambda_*$ is determined as above and fixed, where we use different norms, $\alpha \neq \alpha^{'}$.
\end{itemize}
Then, these effects are summed up in quadrature with the statistical error to obtain the error estimate
\begin{equation}
    \sigma^2_{\rho_\sigma} = \sigma_{\rm stat}^2 + \sigma_{1,\, \rm sys}^2 + \sigma_{2,\, \rm sys}^2\,.
\end{equation}

\subsection{Smearing kernel and smearing radius}
\label{Sec:kernels}

Under reasonable conditions, the HLT method is robust, in the sense that physical results 
can be extracted effectively. Given the freedom to choose different smearing kernels, $\Delta_\sigma (E, \omega)$, as the arguments leading to Eq.~(\ref{eq:constraint_kernel}) can be satisfied with a broad class of choices. As a check of the stability of our results, two 
particularly interesting choices are the Gaussian kernels, defined as    
\begin{equation}
\label{eq:gauss_kernel}
\Delta^{(1)}_\sigma(E - \omega) \equiv \frac{1}{Z(\omega)}\exp\left[{\frac{-(E-\omega)^2}{2\sigma^2}}\right] \; ,
\end{equation}
with $Z(\omega) = \int_0^\infty dE \,  e^\frac{-(E-\omega)^2}{2\sigma^2}$, and the
Cauchy (or Breit-Wigner) kernels, written as
\begin{equation}
\label{eq:cauchy_kernel}
    \Delta^{(2)}_{\sigma} (E - \omega) = \dfrac{\sigma}{ (E - \omega)^2 + \sigma^2 }\,.
\end{equation}
Comparing the results obtained with these two choices provides an estimate of the systematic uncertainty for the fits of spectral densities deriving from this source. As expected, we anticipate that no relevant dependence (compared to statistical uncertainties) will be observed in this latter check.

   \begin{figure}[t]
    \includegraphics[width=0.55\linewidth]{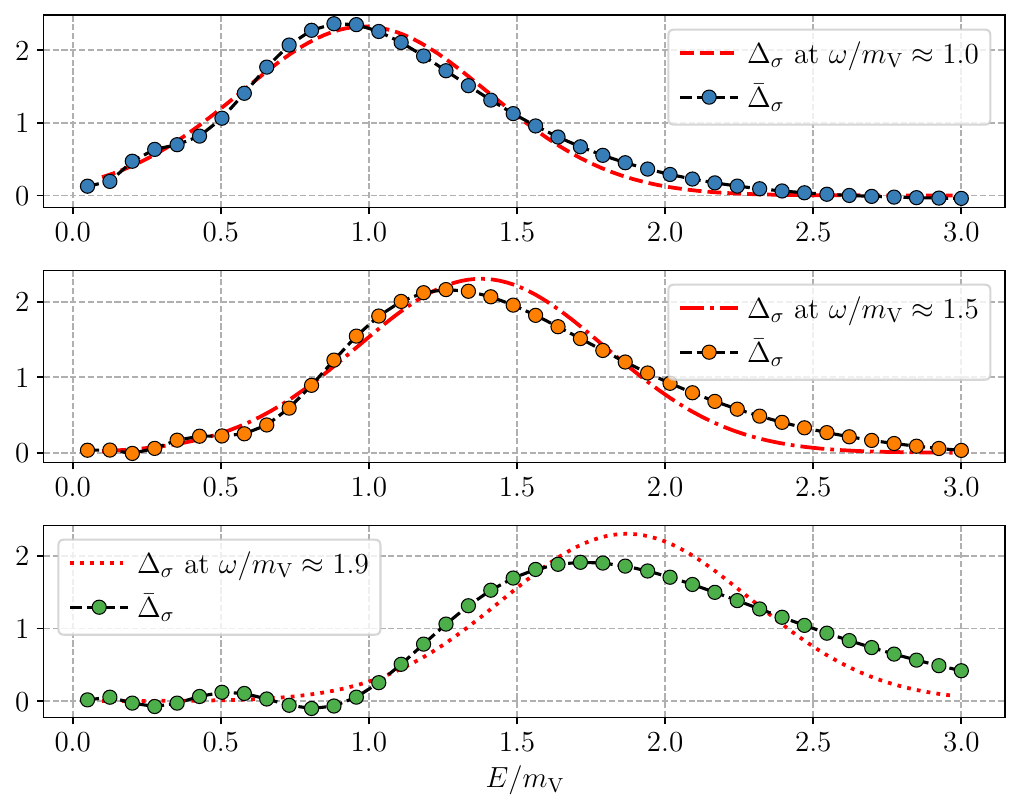}
    \caption{Illustration of kernel reconstruction systematics dependence on energy. The curves are the target Gaussian kernels, $\Delta_\sigma (E - \omega)$, for $\omega/m_{\mathrm{V}} \approx 1.0, \, 1.5, \,1.9$ (top to bottom panels), and $\sigma = 0.28\,m_{\mathrm{V}}$. The points represent the reconstructed kernels, $\bar{\Delta}_\sigma (E - \omega)$. The data is taken from ensemble M2, and $m_{\mathrm{V}}$ is extracted through the GEVP procedure---see Tab.~\ref{table:E2_results_ground}. 
    Other parameters are $t_{\rm max} = N_t / 2$, and $\alpha=0$.}
    \label{fig:energy_worsening}
     \end{figure}

An example of a successful reconstruction of such kernels, $\bar{\Delta}_\sigma (E - \omega)$ (Gaussian and Cauchy cases), compared to the target ones, $\Delta_\sigma (E - \omega)$, is shown in Fig.~\ref{fig:smearingkernel}.
As an additional precaution, we only consider values of $\lambda$ for which $A[\vec{g}]/A_0 < 0.1$, which constrains the size of the systematics. The other parameters are discussed in the caption. Such an outcome corresponds to what we generally find in our analysis. 
As expected, given the finiteness of the lattice and the finite statistics of correlation function measurements, 
the quality of the reconstruction deteriorates at the largest energies considered, as shown in Fig.~\ref{fig:energy_worsening}. The goal for a successful reconstruction is to keep the discrepancy between the targets $\Delta_\sigma (E - \omega)$ and the reconstructed $\bar{\Delta}_\sigma (E - \omega)$ as small as possible for each energy considered.

We find it convenient to express the choice of smearing radius, $\sigma$,
in units of the ground state mass,  $m_0$,  for the mesonic channel of interest,
which can be measured independently, for example by using the GEVP procedure.
The optimal choice of $\sigma$ depends on the quality of the data: the amount of statistics available for the input two-point correlation functions also affects the lower bound for the smearing radius $\sigma$. 
Typically, the values used for our ensembles are in the range $0.18m_0\leq \sigma \leq 0.35m_0$. 
In this interval, the smaller smearing radii are being used to resolve the details of the spectrum
when the discrete energy levels are tightly packed close together.

\begin{center}
   \begin{figure}
    \includegraphics[width=0.50\linewidth]{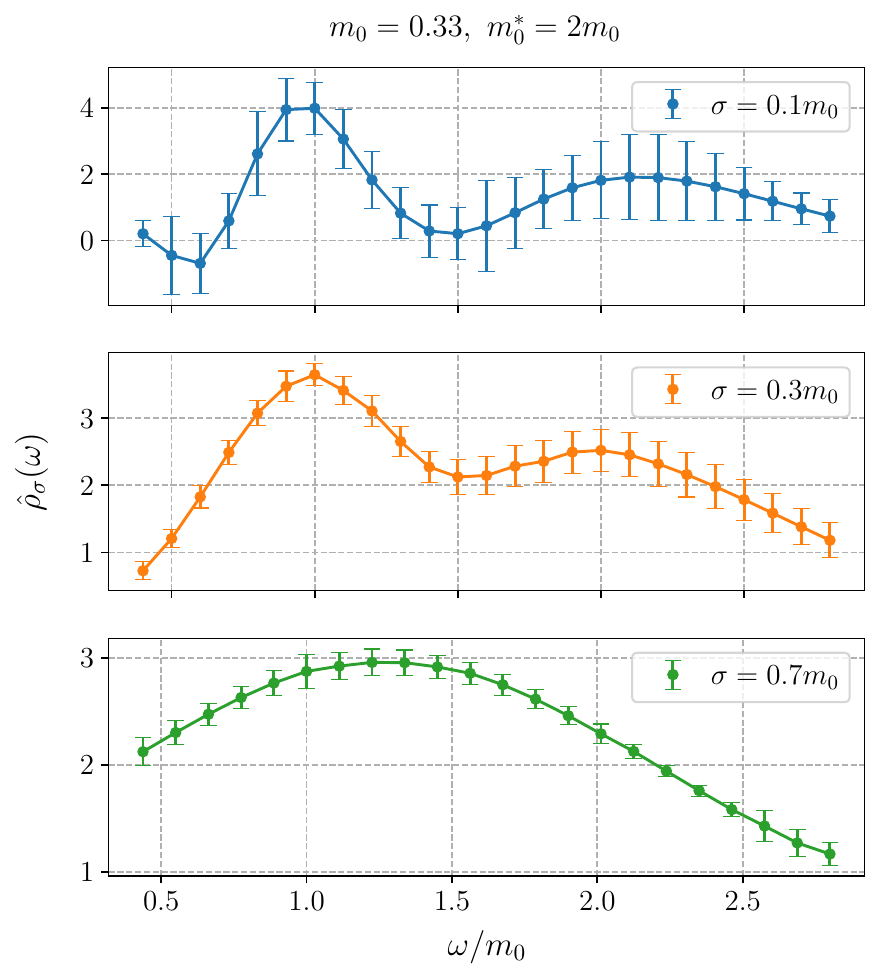}
    \caption{Reconstruction of spectral densities from synthetic 2-point correlation functions,
engineered to have ground state mass $a m_0 = 0.33$, and contain one excited state with $m^{*}_0 = 2m_0$. The number of configurations is $N_{\rm conf} = 1000$, the lattice time extent is $N_t = 96$, and the relative uncertainty is $2\%$. The three panels correspond to Gaussian kernel reconstructions with $\sigma = 0.1m_0, \, 0.3m_0, \, 0.7m_0$
(top to bottom).
    \label{fig:varyingsigma}}
     \end{figure}
\end{center}

Figure~\ref{fig:varyingsigma} displays the reconstruction of a set of synthetic two-point correlation functions,
built to contain a ground state and an excited state.
We have computed $\hat{\rho}_\sigma(E)$ using three illustrative values of the smearing radii ($\sigma = 0.1m_0, \, 0.3m_0, \, 0.7m_0$). For the smallest
smearing radius, $\sigma = 0.1m_0$, as one increases the energy considered, $E/m_0$, the smearing kernels appear to  be poorly reproduced  when  $E/m_0 \gtrapprox 1.0$, so that 
excited states are lost below the level of systematic errors.
On the other hand, with the extreme choice  $\sigma = 0.7m_0$, the two expected peaks merge completely, and the reconstruction keeps statistical and systematic error moderate. By choosing an intermediate value, 
$\sigma = 0.3 m_0$, there is evidence that the reconstruction resolves well the two peaks, and keeps the systematic and statistical uncertainties controllable.

\subsection{Spectral density fits}
\label{Sec:fitting_chisquare}

In this Section, we describe how smeared spectral densities can be used for meson spectroscopy. The method involves minimising the correlated functional, $\chi^2$, defined as in Ref.~\cite{DelDebbio:2022qgu}, along the 
same lines as Eq.~(\ref{eq:correlated_chisq_corr}), as follows: 
\begin{equation}
\label{eq:correlated_chisquare}
    \chi^2 \equiv \sum_{E, \, E^{'}} \left( f^{(k)}_\sigma (E) - \hat{\rho}_\sigma (E) \right) \text{Cov}^{-1}_{E\,E^{'}} [\rho_\sigma] \left( f^{(k)}_\sigma (E^{'}) - \hat{\rho}_\sigma (E^{'}) \right)\,,
\end{equation}
where the covariance matrix, $ \text{Cov}_{E\,E^{'}} $, has been defined in Eq.~(\ref{Eq:cov}), 
and the  fitting function, $f^{(k)}_\sigma (E)$, is defined as a sum of instances of either the Gaussian ($i=1$, Eq.~\eqref{eq:gauss_kernel}) or Cauchy ($i=2$, Eq.~\eqref{eq:cauchy_kernel}) kernel:
\begin{equation}
\label{eq:sum_of_kernels}
    f^{(k)}_\sigma (E) = \sum_{n = 1}^{k} \mathcal{A}_n \, \Delta^{(i)}_\sigma (E - E_n)\,.
\end{equation}
The fit parameters $E_n$ will be identified with the eigenvalues of the finite-volume Hamiltonian.
The freedom in choosing $k$ refers to the fact that \textit{a priori} one does not know how many excited states  the
method can identify; this is fixed \textit{a posteriori}.

\begin{figure}[t!]
\begin{center}
\begin{picture}(150,65)
     \put(-10,0){\includegraphics[width=0.44\linewidth]{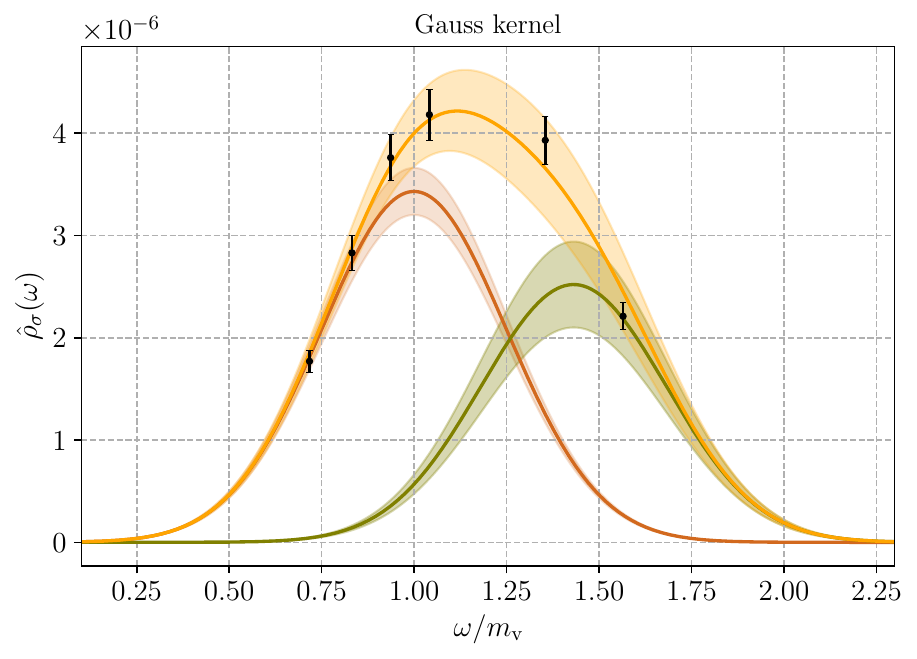}}
     \put(70,0){\includegraphics[width=0.44\linewidth]{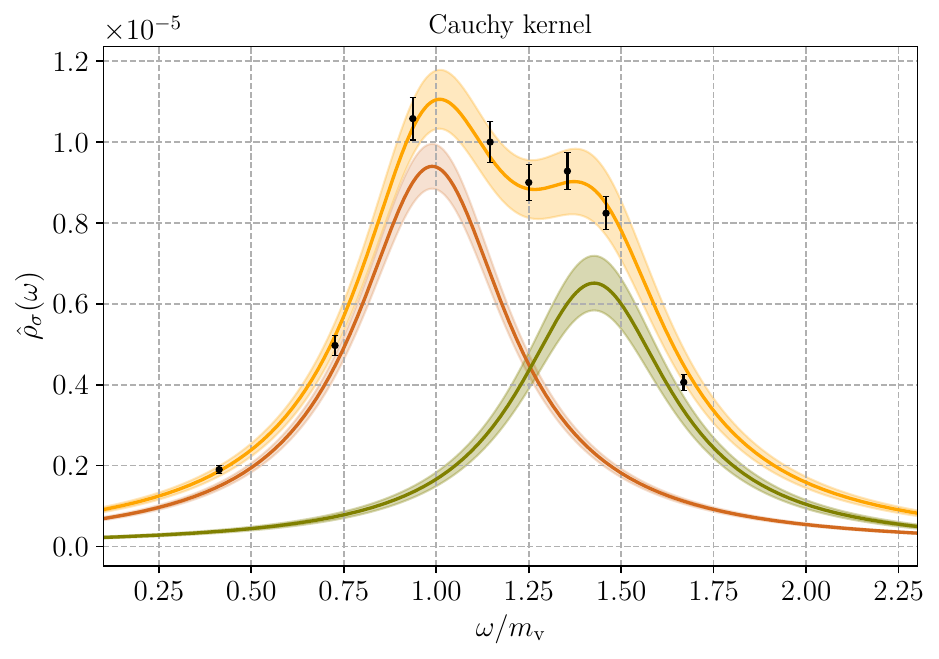}}
\end{picture}
\end{center}
\caption{Spectral density reconstruction in the 
$\rm v$ channel,
in ensemble M2 ($N_t = 64$). The black datapoints are the optimal reconstructed spectral density values $\hat{\rho}_{\sigma} (\omega)$. The yellow line corresponds to the fitted curve $f^{(2)}_{\sigma} (E)$ in Eq.~\eqref{eq:sum_of_kernels}, the red line corresponds to the deriving ground state curve $\Delta_\sigma (E - E_0)$ from the fit, the green line is the first excited state $\Delta_\sigma (E - E_1)$. For the left panel, the reconstruction is done using Gaussian kernels. For the right panel, it is performed using Cauchy kernels. Both the spectral densities have smearing radius $\sigma = 0.25\,m_{\hbox{\small{v}}}$. The Wuppertal smearing applied has $N_{\rm source} = 0$, $N_{\rm sink} = 40$ iterations. For the left panel: 
the fit has reduced ${\chi}^2/N_{\rm d.o.f.} = 1.6$, the energy levels are $aE_0 = 0.6511(32)$ and $aE_1 = 0.932(11)$, and the associated amplitudes have $\mathcal{A}_1 = 3.73(23) \times 10^{-6}$ and $\mathcal{A}_2 = 2.52(52) \times 10^{-6}$, respectively. 
For the right panel: ${\chi}^2/N_{\rm d.o.f.}  = 1.1$,  $aE_0 = 0.6442(37)$ and $aE_1 = 0.929(12)$,  the associated amplitudes are $\mathcal{A}_1 = 2.33(12) \times 10^{-6}$ and $\mathcal{A}_2 = 1.63(16) \times 10^{-6}$. The value of $m_{\hbox{\small{v}}}$ has been found through the GEVP analysis---see Tab.~\ref{table:E2_results_ground}.}
\label{Fig:fit_comparison_gauss_cauchy}
\end{figure}

As anticipated, we exploit the availability of two different and independent smearing kernels as an additional sense check for the results of our spectroscopy study, given that
 the eigenvalues of the finite-volume Hamiltonian do not depend on the kernel. For the determination of the $n^{th}$ energy level,
 we define the systematic error to be
\begin{equation}
\label{eq:sys_err1}
    \sigma_{\rm sys}(aE_n) = |aE_{n, \, {\rm Gauss}} - aE_{n, \, {\rm Cauchy}}|\,.
\end{equation}
In Sec.~\ref{Sec:spectra}, we demonstrate that possible systematics due to the choice of the kernel never lead to fluctuations outside of statistical errors.

We display in Fig.~\ref{Fig:fit_comparison_gauss_cauchy} an example of the comparison between the results of the reconstruction process performed using Gauss and Cauchy kernels. The plots show qualitative and quantitative differences in the spectral shapes, due to the use of different choices of $\Delta_\sigma$ functions. Nonetheless, the position of the corresponding (Gaussian or Cauchy) peaks are compatible within statistical uncertainties.

\begin{figure}[t!]
\begin{center}
\begin{picture}(0,125)
     \put(-70,65){\includegraphics[width=0.74\linewidth]{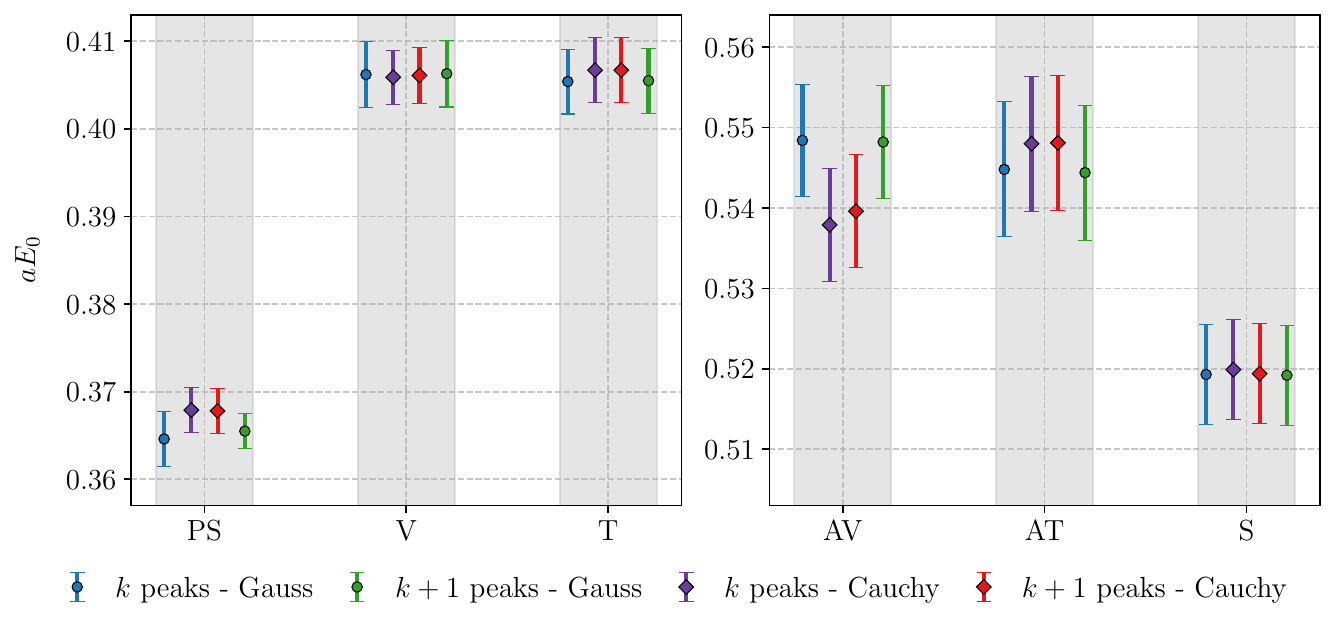}}
     \put(-70,0){\includegraphics[width=0.74\linewidth]{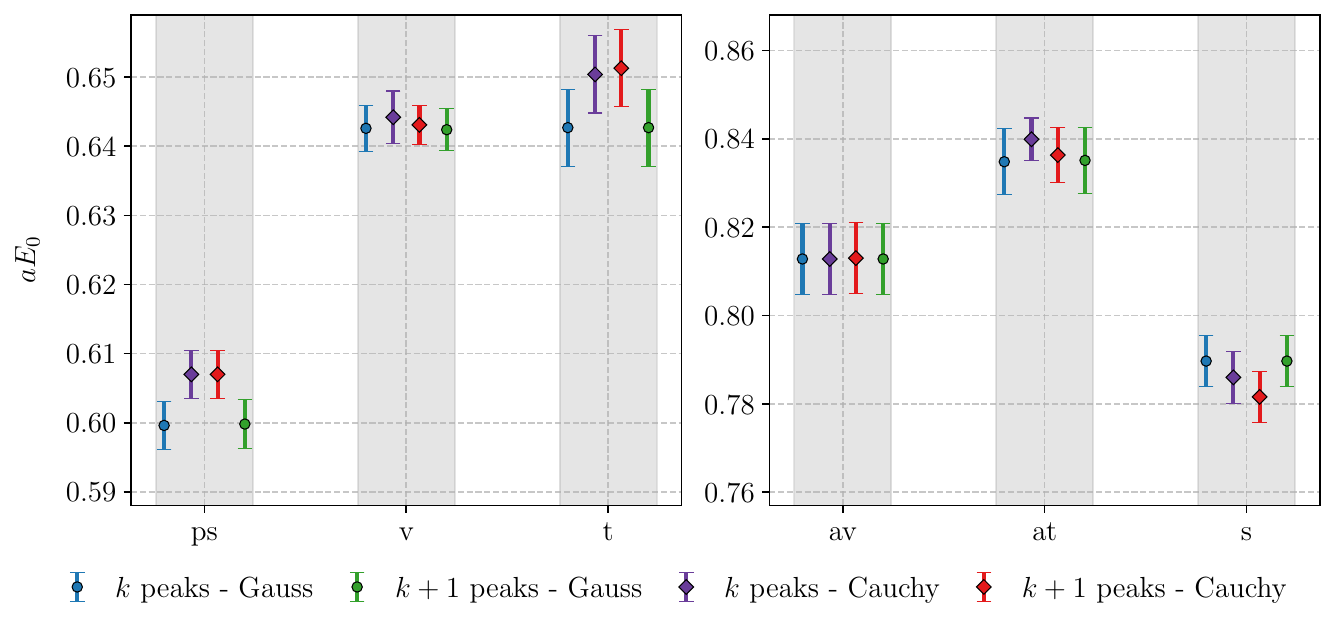}}
\end{picture}
\end{center}
\caption{Assessment of the fitting systematic error in the ground state energy level, for ensemble M2. For each channel, the results obtained by fitting using $k$ and $k+1$ peaks, and Gaussian and Cauchy kernels, are compared. Further details are presented in Tab.~\ref{table:E2_results_ground}. The offset between datapoints in the same mesonic channel is set for visual clarity.}
\label{Fig:systematics_spectrum_nt64_fund_ground}
\end{figure}

\begin{figure}[t!]
\begin{center}
\begin{picture}(0,125)
     \put(-70,65){\includegraphics[width=0.74\linewidth]{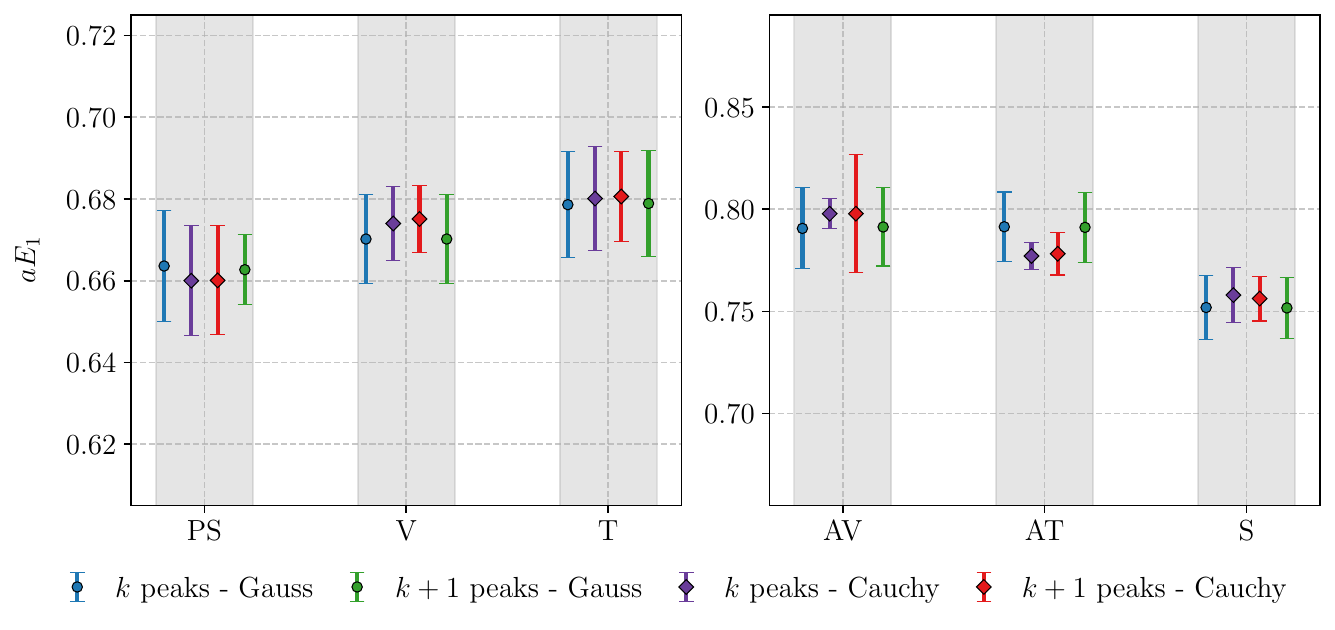}}
     \put(-70,0){\includegraphics[width=0.74\linewidth]{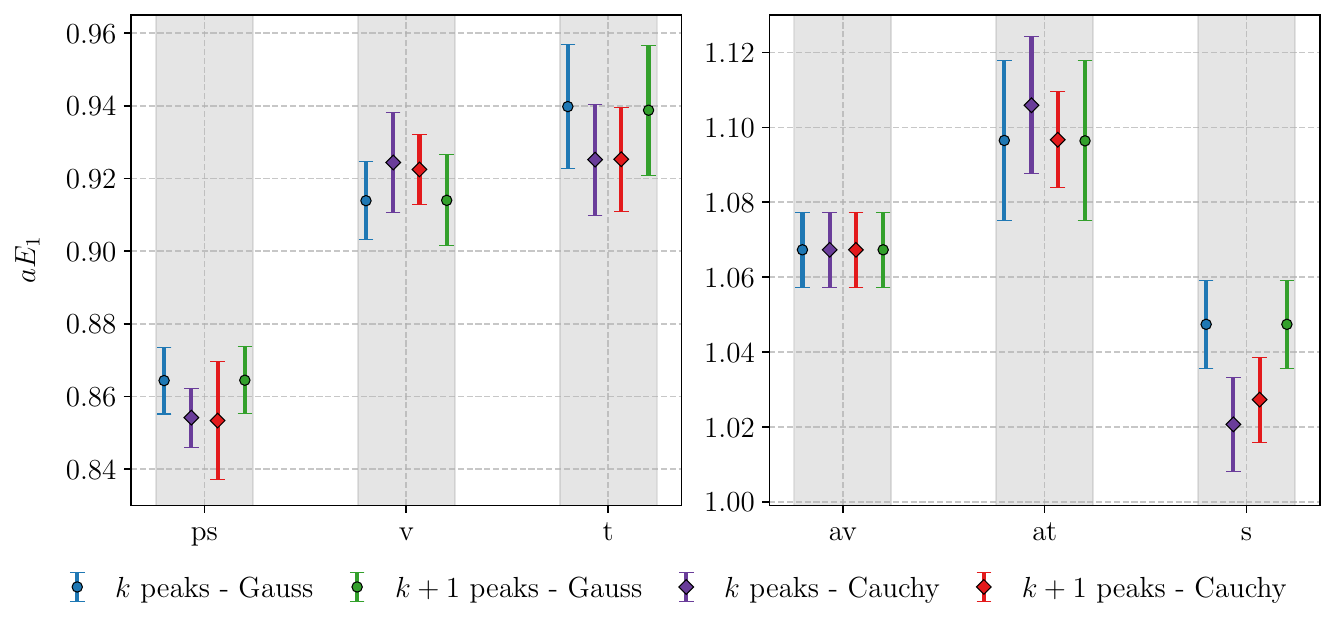}}
\end{picture}
\end{center}
\caption{Assessment of the fitting systematic error in the first excited energy level, for ensemble M2. For each channel, the results obtained by fitting using $k$ and $k+1$ peaks, and Gaussian and Cauchy kernels, are compared. Further details are presented in Tab.~\ref{table:E2_results_first}. The offset between datapoints in the same mesonic channel is set for visual clarity.}
\label{Fig:systematics_spectrum_nt64_fund_first}
\end{figure}

Given the finiteness of the smearing radius, and the deterioration of the reconstruction at high energies, one will only be able to access a certain number of states in the mesonic spectrum. Even then, fit results can be contaminated by additional states that are not accounted for in the model. In order to ensure this does not hinder our prediction for the first $k$ states, we can repeat the fit by adding a state in the model--performing a reconstruction with $k+1$ (Gaussian or Cauchy) peaks in Eq.~(\ref{eq:correlated_chisquare}), to check the presence of excited states contamination. If the targetted states are stable under this procedure, and the chi-square does not improve significantly, we consider the estimate reliable.  
As we will show, our measurements appear free from this effect. 
An analysis of the systematics due to different kernels, and contaminations from excited states, is shown in Figs~\ref{Fig:systematics_spectrum_nt64_fund_ground} and \ref{Fig:systematics_spectrum_nt64_fund_first}, for the ground states and first excited states, respectively, in a selection of ensembles and meson channels.

\subsection{On the effect of APE and Wuppertal smearing on the spectral densities}
\label{Sec:varying_smearings}

In analogy with what is done for the correlation functions in Eq.~(\ref{eq:correlator_matrix_elements}), one can re-express the spectral density 
\begin{equation}
\label{eq:rewriting_spec_dens}
\rho_\sigma (E) = \sum_n \dfrac{\langle 0 | \mathcal{O}(0) | n \rangle \langle n |  \bar{\mathcal{O}}(0) | 0 \rangle }{2E_n} \, \Delta_\sigma \Big( E-E_n (N_s a)\Big)\,,
\end{equation}
using  the matrix elements $\langle 0 | \mathcal{O}(0) | n \rangle$ and $\langle n |  \bar{\mathcal{O}}(0) | 0 \rangle$, in agreement with the  functional form in Eq.~(\ref{eq:sum_of_kernels}). 
The matrix elements in Eq.~(\ref{eq:rewriting_spec_dens})  determine the relative contribution of each energy level to the smeared spectral density.

A major obstacle in performing spectroscopy of lattice gauge theories consists in the difficulties of disentangling their spectrum, because the energies contributing to Eq.~\eqref{eq:rewriting_spec_dens} can be very close to each other. Moreover, certain states can have large contributions, at the risk of obfuscating others.
This phenomenon may result in considerable discrepancies in the matrix elements corresponding to different states, $|n\rangle$, posing a challenge to applications in spectroscopy. 

The same potential obstruction would affect the direct study of two-point correlation functions, as it results in a distortion of the  large-$t$ behavior of two-point correlation function in Eq.~(\ref{eq:correlator_matrix_elements}), and hence the quality of multi-exponential fits of effective mass plateaux deteriorates.
As discussed in Sec.~\ref{Sec:correlation_functions}, this difficulty is addressed by 
introducing an appropriately tuned combination of APE and Wuppertal smearing in the extraction of the correlation functions. By doing so, one improves the overlap of the states of interest with the interpolating operator, therefore reducing the importance of other, undesired states.

\begin{figure}[t!]
\begin{center}
\begin{picture}(140,57)
     \put(-10,5){\includegraphics[width=0.35\linewidth]{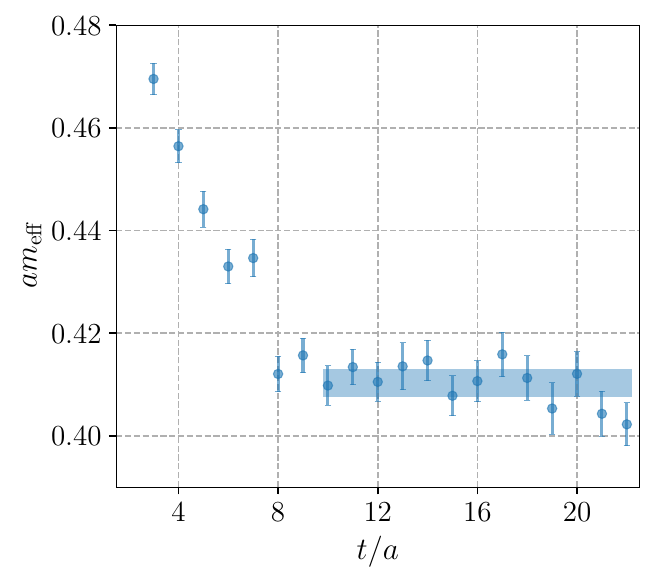}}
     \put(55,7){\includegraphics[width=0.42\linewidth]{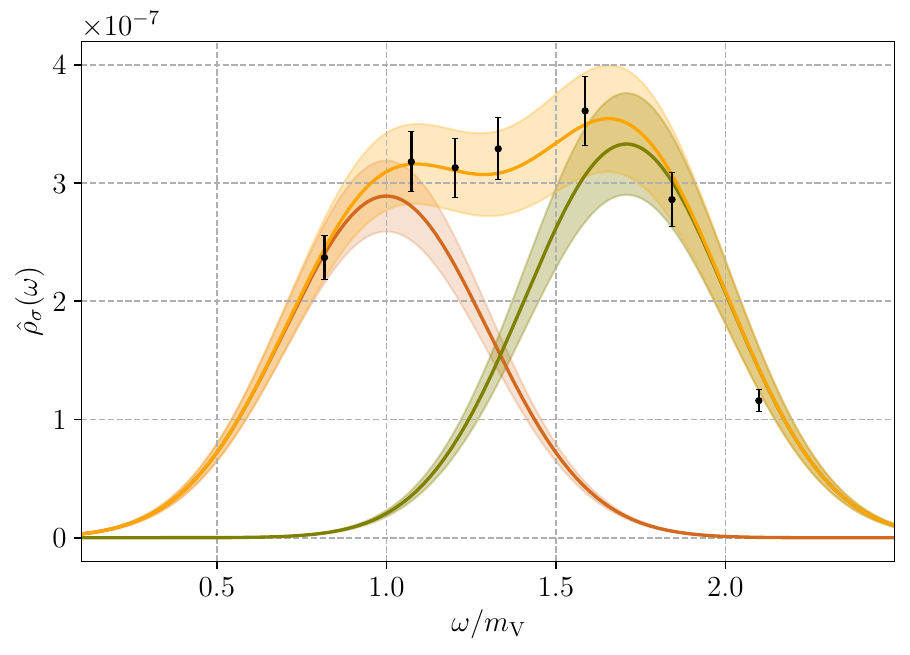}}
\end{picture}
\end{center}
\caption{Example of the high quality plateau and fit obtained with an appropriate application of APE and Wuppertal smearing.
For ensemble M1, the correlation function in the V channel is obtained by applying APE smearing with $N_{\rm APE} = 50$ steps (APE smearing step-size $\alpha_\textrm{APE} = 0.4$), and Wuppertal smearing with $N_{\rm source} = 80$ source smearing steps and $N_{\rm sink} = 40$ sink smearing steps (Wuppertal smearing step size $\varepsilon_{\rm f} = 0.12$). The effective mass is displayed in the left panel. Applying the GEVP procedure, we find
$m_{\mathrm{V}} = 0.4098(25)$,  $m^{*}_{\mathrm{V}} = 0.700(26)$.
The right panel displays the reconstructed spectral density with Gaussian kernel, for $\sigma = 0.30\,m_{V}$, including two states. The black datapoints are the optimal reconstructed spectral density values $\hat{\rho}_{\sigma} (\omega)$. The yellow line corresponds to the fitted curve $f^{(2)}_{\sigma} (E)$ in Eq.~\eqref{eq:sum_of_kernels}, the red line corresponds to the deriving ground state curve $\Delta_\sigma (E - E_0)$ from the fit, the green line is the first excited state $\Delta_\sigma (E - E_1)$.  The fit results are $\mathcal{A}_1 = 2.89(30) \times 10^{-7}$, $\mathcal{A}_2 = 3.33(37) \times 10^{-7}$, $a E_0 = 0.4139(49)$, and $aE_1 = 0.702(19)$. The reduced chi square is ${\chi}^2/N_{\rm d.o.f.} = 0.15$.}
\label{Fig:fit_N80_N40}
\end{figure}

    \begin{table}[b]
    \begin{tabular}{ |c|c|c|c|c|c|c|c| }
     \hline \hline
     Case & $\alpha_\textrm{APE}$& $\varepsilon_{\rm f}$ & $N_{\rm source}$ & $N_{\rm sink}$ & $\mathcal{A}_2/\mathcal{A}_1$ & $aE_0$ & $aE_1$ \\
     \hline
     A & 0.4 & 0.12 & 80 & 20  & 1.32(19) & 0.4144(50) & 0.692(27) \\
     B & 0.4 & 0.12 & 80 &  40  & 1.15(11) & 0.4139(49) & 0.702(19) \\
     C & 0.4 & 0.12 & 80 &  80  & 0.75(15) & 0.4131(52) & 0.699(22) \\
     D & 0.4 & 0.12 & 40 &  80  & 1.24(18) & 0.4132(43) & 0.694(27) \\
     E & 0.4 & 0.12 & 20 &  80  & 1.80(28) & 0.4148(51) & 0.714(23) \\
     F & 0.4 & 0.24 & 90 & 30 & 1.01(20) & 0.4148(52) & 0.698(19) \\
     G & 0.4 & 0.4 & 170  & 170 & 0.63(11) & 0.4113(82) & 0.717(33) \\
     H & 0.4 & 0.05 & 20 & 20 & 2.28(27) & 0.4136(74) & 0.705(32) \\
     I & 0.0 & 0.12 & 80  & 40 & 1.27(11) & 0.4154(73) & 0.698(32) \\
     \hline \hline
    \end{tabular}
    \caption{\label{table:smearing_levels} 
    Illustration of the effectiveness of APE and Wuppertal smearing in spectroscopy analysis.
For ensemble M1, the correlation function in the V channel is obtained by applying
different levels of sink and source Wuppertal smearing and APE smearing, and we demonstrate the
effect on the output results, $\mathcal{A}_2/\mathcal{A}_1$, $aE_0$, and $aE_1$.}
    \end{table}

An example of the result of the optimization of the smearing parameters is shown in Fig.~\ref{Fig:fit_N80_N40}.
The right panel displays the effect of applying a level of smearing that optimizes the GEVP extraction of the ground state,
chosen so that the effective mass plateaux are clearly discernible (left panel).
The ground state has comparable amplitude with the first excited state(s).
To be more explicit, in Tab.~\ref{table:smearing_levels} we report 
various choices of the smearing parameters and they affect the output results, for the same channel and ensemble as in 
Fig.~\ref{Fig:fit_N80_N40}.
Cases A to F in Tab.~\ref{table:smearing_levels} all correspond to reasonable levels of smearing: the relative amplitudes of the peaks corresponding to the ground and first excited states are comparable and yield to accurate 
spectroscopy results. 
Conversely, if one applies too much smearing, such as case G in Tab.~\ref{table:smearing_levels}, or too little, as in H, the relative difference between the amplitudes increases, which results in less precise determinations of the energy levels. 
We verified that the amplitudes scale as expected from Eq.~(\ref{eq:rewriting_spec_dens}). 
By applying smearing to the operators increases the overlap with the ground state and decreases the overlap with the excited states. 
The ratio $\mathcal{A}_2 / \mathcal{A}_1$ is smaller for larger smearing.
For the same reason, and as expected from Eq.~(\ref{eq:correlator_matrix_elements}), another effect of smearing is the appearance of longer effective-mass plateaux for the ground state.

\begin{figure}[t!]
\begin{center}
\begin{picture}(140,115)
     \put(-10,65){\includegraphics[width=0.35\linewidth]{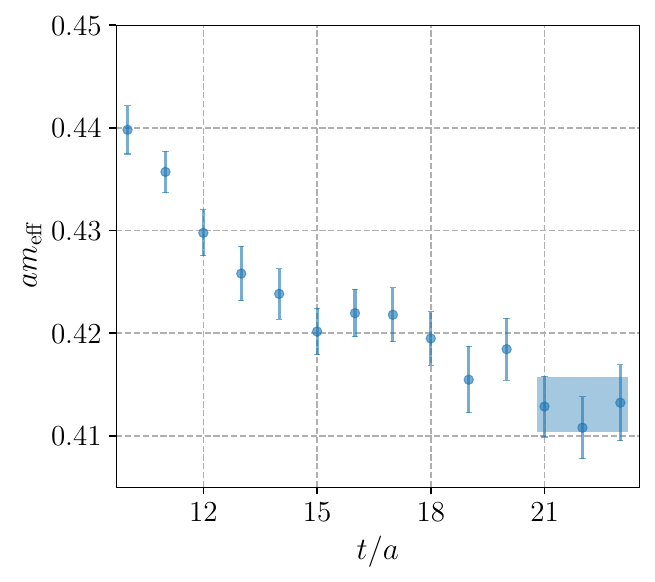}}
     \put(55,67){\includegraphics[width=0.42\linewidth]{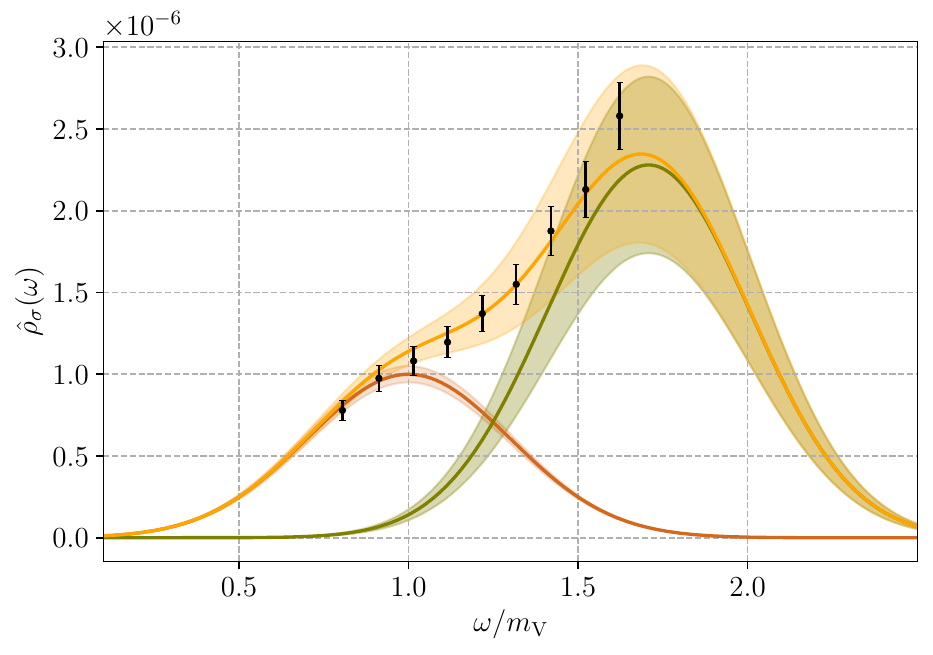}}
     \put(-10,0){\includegraphics[width=0.35\linewidth]{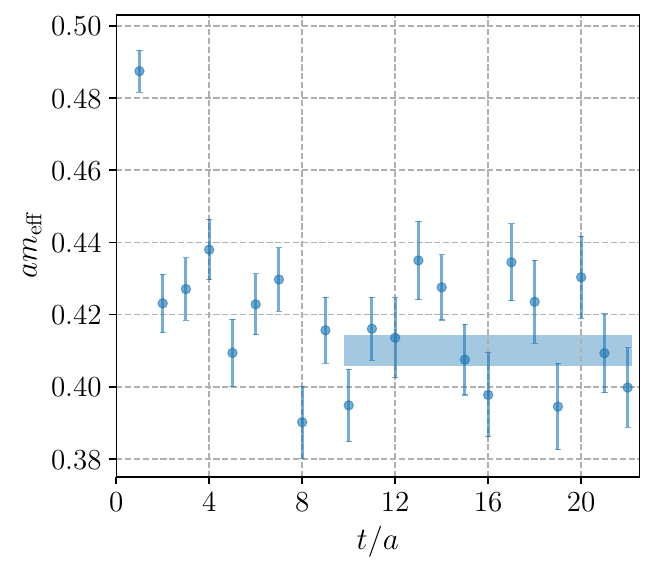}}
     \put(55,2){\includegraphics[width=0.42\linewidth]{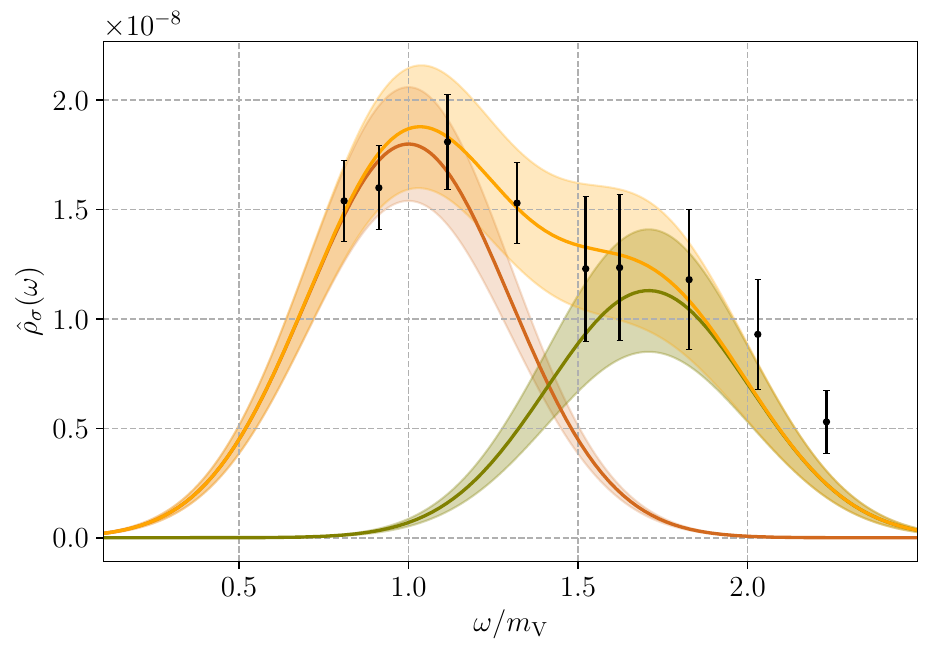}}
\end{picture}
\end{center}
\caption{
The V channel in ensemble M1, 
for which the GEVP procedure yields $m_{\mathrm{V}} = 0.4098(25)$ and  $m^{*}_{\mathrm{V}} = 0.700(26)$.
Top panels: effective mass and spectral density, for $\sigma = 0.30\,m_{\mathrm{V}}$,  with APE smearing with $N_{\rm APE} = 50$ steps (APE smearing step-size $\alpha_\textrm{APE} = 0.4$), and Wuppertal smearing with $N_{\rm source} = 20$ source steps and $N_{\rm sink} = 20$ sink  steps (Wuppertal step size $\varepsilon_{\rm f} = 0.05$). An example of an optimal amount of (APE and Wuppertal smearing) in this case is shown in Fig.~\ref{Fig:fit_N80_N40}. The black datapoints are the optimal reconstructed spectral density values $\hat{\rho}_{\sigma} (\omega)$. The yellow line corresponds to the fitted curve $f^{(2)}_{\sigma} (E)$ in Eq.~\eqref{eq:sum_of_kernels}, the red line corresponds to the deriving ground state curve $\Delta_\sigma (E - E_0)$ from the fit, the green line is the first excited state $\Delta_\sigma (E - E_1)$. Fit results are $\mathcal{A}_1 = 1.00(5) \times 10^{-6}$,  $\mathcal{A}_2 = 2.28(54) \times 10^{-6}$, 
$aE_0 = 0.4166(74)$, $aE_1 = 0.705(32)$, and reduced  ${\chi}^2/N_{\rm d.o.f.} = 0.9$. 
Bottom panels: same as top, except that Wuppertal smearing has $N_{\rm source} = 170$ source steps and $N_{\rm sink} = 170$ sink steps (Wuppertal step size $\varepsilon_{\rm f} = 0.90$). Fit results are 
$\mathcal{A}_1 = 1.80(26) \times 10^{-8}$, $\mathcal{A}_2 = 1.13(28) \times 10^{-8}$, $aE_0 = 0.4113(82)$,
$aE_1 = 0.717(33)$, and reduced ${\chi}^2/N_{\rm d.o.f.} = 0.8$.}
\label{Fig:wrong_smearings}
\end{figure}

\begin{figure}[h]
\begin{center}
\begin{picture}(140,65)
     \put(-10,0){\includegraphics[width=0.35\linewidth]{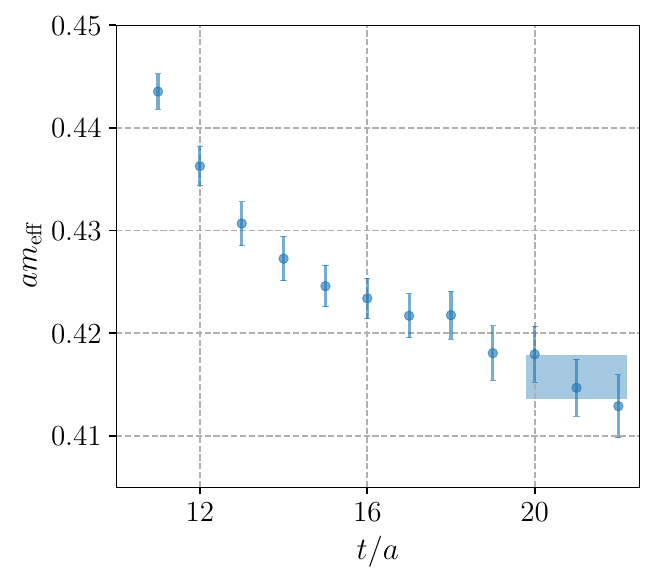}}
     \put(55,2){\includegraphics[width=0.42\linewidth]{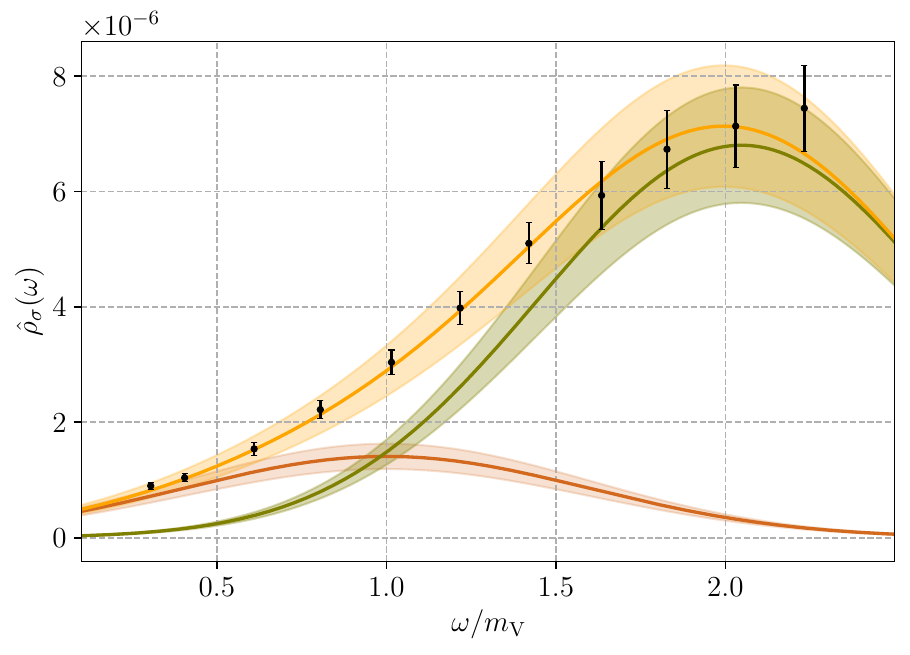}}
\end{picture}
\end{center}
\caption{
The V channel in ensemble M1, 
for which the GEVP procedure yields $m_{\mathrm{V}} = 0.4098(25)$ and  $m^{*}_{\mathrm{V}} = 0.700(26)$.
With no APE nor Wuppertal smearing  applied, 
 spectral density reconstruction uses $\sigma = 0.60\,m_{\mathrm{V}}$. An example of an optimal amount of (APE and Wuppertal smearing) in this case is shown in Fig.~\ref{Fig:fit_N80_N40}. The same optimal case without applying APE smearing is shown in Fig.~\ref{Fig:smearing_NOAPE}. The black datapoints are the optimal reconstructed spectral density values $\hat{\rho}_{\sigma} (\omega)$. The yellow line corresponds to the fitted curve $f^{(2)}_{\sigma} (E)$ in Eq.~\eqref{eq:sum_of_kernels}, the red line corresponds to the deriving ground state curve $\Delta_\sigma (E - E_0)$ from the fit, the green line is the first excited state $\Delta_\sigma (E - E_1)$. Fit results are $\mathcal{A}_1 = 1.41(22) \times 10^{-6}$, 
 $\mathcal{A}_2 = 6(1) \times 10^{-6}$, $aE_0 = 0.413(11)$,  $aE_1 = 0.839(77)$, and reduced 
 ${\chi}^2/N_{\rm d.o.f.} = 0.80$.}
\label{Fig:no_smearing}
\end{figure}

\begin{figure}[b!]
\begin{center}
\begin{picture}(140,70)
     \put(-10,0){\includegraphics[width=0.37\linewidth]{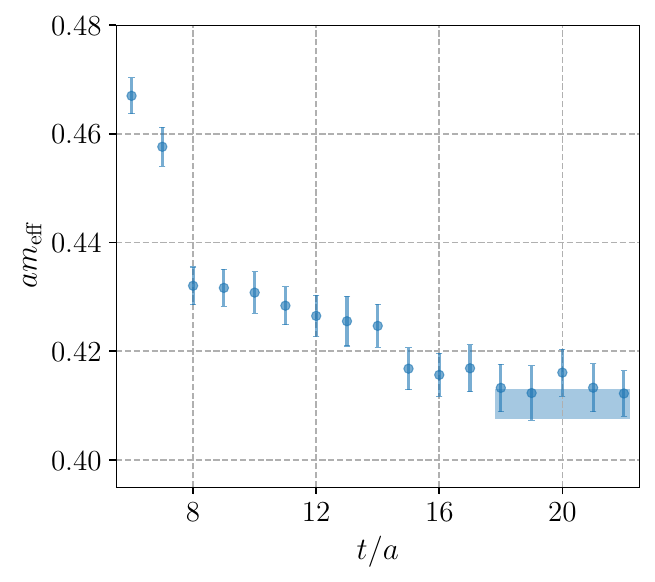}}
     \put(55,2){\includegraphics[width=0.47\linewidth]{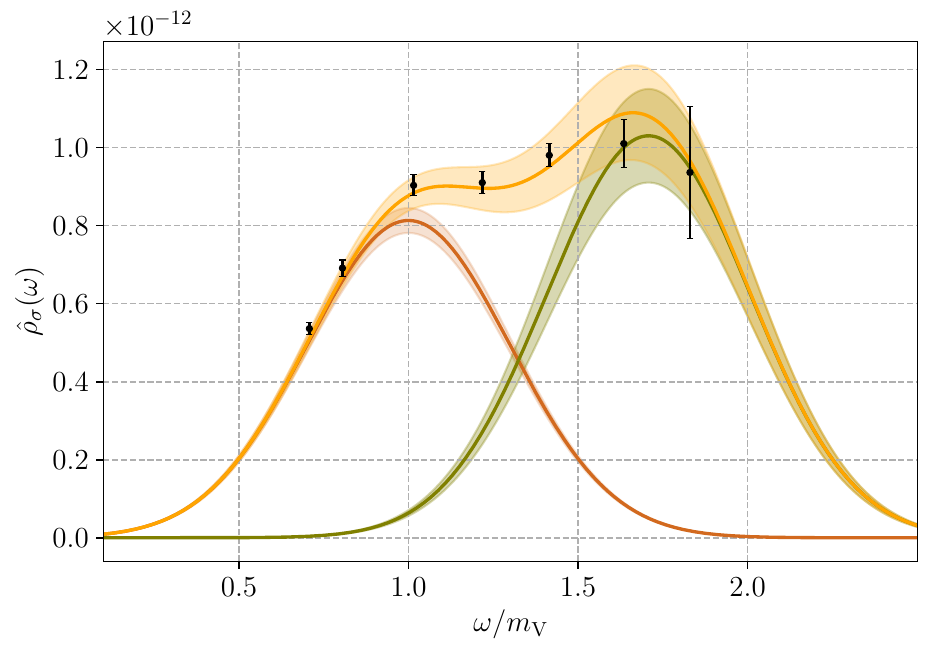}}
\end{picture}
\end{center}
\caption{
The V channel in ensemble M1, 
for which the GEVP procedure yields $m_{\mathrm{V}} = 0.4098(25)$ and  $m^{*}_{\mathrm{V}} = 0.700(26)$.
 No APE smearing has been applied, while Wuppertal smearing has $N_{\rm source} = 80$ source steps and $N_{\rm sink} = 40$ sink steps (Wuppertal step size $\varepsilon_{\rm f} = 0.12$). An example of an optimal amount of (APE and Wuppertal smearing) in this case is shown in Fig.~\ref{Fig:fit_N80_N40}.
The spectral density assumes $\sigma = 0.30\,m_{\mathrm{V}}$. The black datapoints are the optimal reconstructed spectral density values $\hat{\rho}_{\sigma} (\omega)$. The yellow line corresponds to the fitted curve $f^{(2)}_{\sigma} (E)$ in Eq.~\eqref{eq:sum_of_kernels}, the red line corresponds to the deriving ground state curve $\Delta_\sigma (E - E_0)$ from the fit, the green line is the first excited state $\Delta_\sigma (E - E_1)$.
 The fit results are $\mathcal{A}_1 = 8.13(32) \times 10^{-13}$, $\mathcal{A}_2 = 1.03(12) \times 10^{-12}$,
  $aE_0 = 0.4154(73)$, $aE_1 = 0.698(32)$, and reduced ${\chi}^2/N_{\rm d.o.f.} = 1.3$.}
\label{Fig:smearing_NOAPE}
\end{figure}

For pedagogical purposes, we find it useful to display, in Fig.~\ref{Fig:wrong_smearings},  also two cases in which the choice of smearing parameters leads away from optimal results, in cases G and H of Tab.~\ref{table:smearing_levels}.
In producing the top panels, only a tiny amount of Wuppertal smearing is applied to the two-point correlation functions, resulting in very short (or absent) effective mass plateaux and poor resolution of the spectral density. 
A similar difficulty emerges if one applies too large amounts of Wuppertal smearing, as depicted in the bottom
 panels of Fig.~\ref{Fig:wrong_smearings}: the plateau practically disappears from the effective mass plot,
 which appears to be dominated by uncontrolled systematics, 
and the resolution of the spectral density deteriorates.

To make the point clearer, in Fig.~\ref{Fig:no_smearing} we depict the case in which no APE nor Wuppertal smearing has been applied.  In this case, the effective-mass plateau is short and might lead to arguable determinations of effective mass fits. This is a 
particularly effective illustration of why the use of point sources may be problematic. In such a case, to perform a reliable reconstruction it is necessary to use a larger spectral density smearing radius, $\sigma$, compared to all the cases considered above. Moreover, the difference in peak heights is substantial. All these factors lead to a deterioration of the signal.

 In Fig.~\ref{Fig:smearing_NOAPE}, the same reconstruction as in Fig.~\ref{Fig:fit_N80_N40} is considered, but with no APE smearing applied. 
By comparing the effective mass plateaux in the
left panels of Figs.~\ref{Fig:fit_N80_N40} and~\ref{Fig:smearing_NOAPE} (cases B and I), one
sees how  APE smearing affects the effective mass plots and spectroscopy.
 As APE smearing averages out the ultraviolet fluctuations in gauge links,
 without it, the spectral density fits deteriorate at large values of the energy, as illustrated 
 by the right panels of the figures.
 Hence, APE smearing of two-point correlation functions results in a widening of  the energy window available to
  spectral reconstruction.

\section{Spectral energy density reconstruction and lattice time extent }
\label{Sec:varyingT}

       \begin{figure}[t]
        \includegraphics[width=0.6\linewidth]{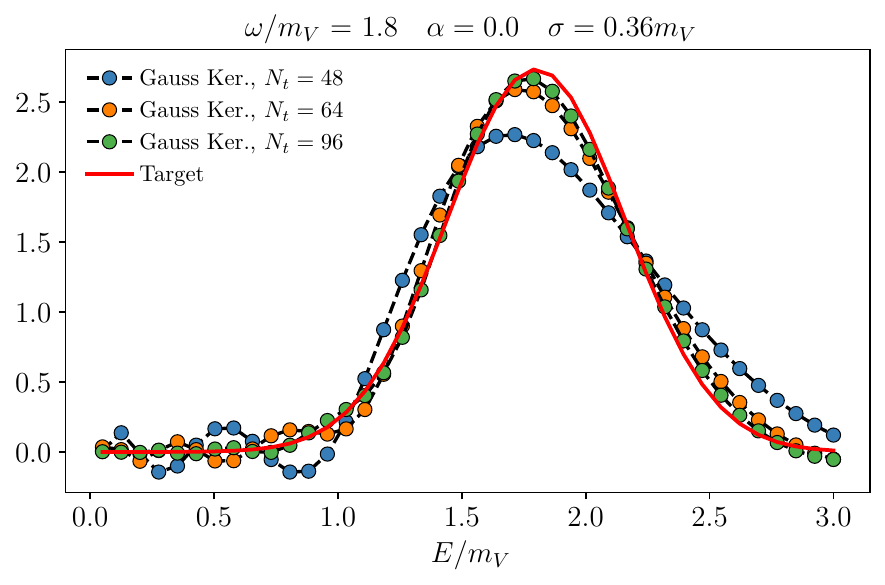}
        \caption{Examples of reconstructed Gaussian kernels at large energy, for $\omega/m_{\mathrm{V}}=1.8$, using different lattice temporal extents, $N_t = 48$, $N_t = 64$, and $N_t = 96$ cases, in ensembles M1, M2, and M3, respectively. The value of $m_V$ is the ground state for the V channel in ensemble M3, extracted through the GEVP process---see Tab.~\ref{table:E3_results_ground}.}
        \label{fig:increasing_nt_kernels}
         \end{figure}

\begin{figure}[t]
\begin{picture}(120,180)
     \put(0,130){\includegraphics[width=0.52\linewidth]{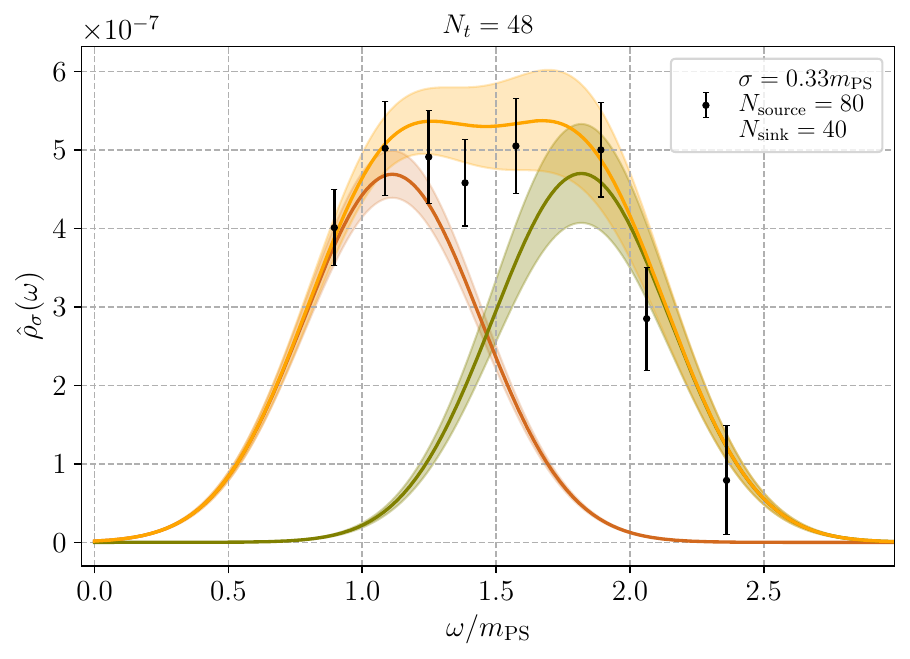}}
     \put(0,65){\includegraphics[width=0.52\linewidth]{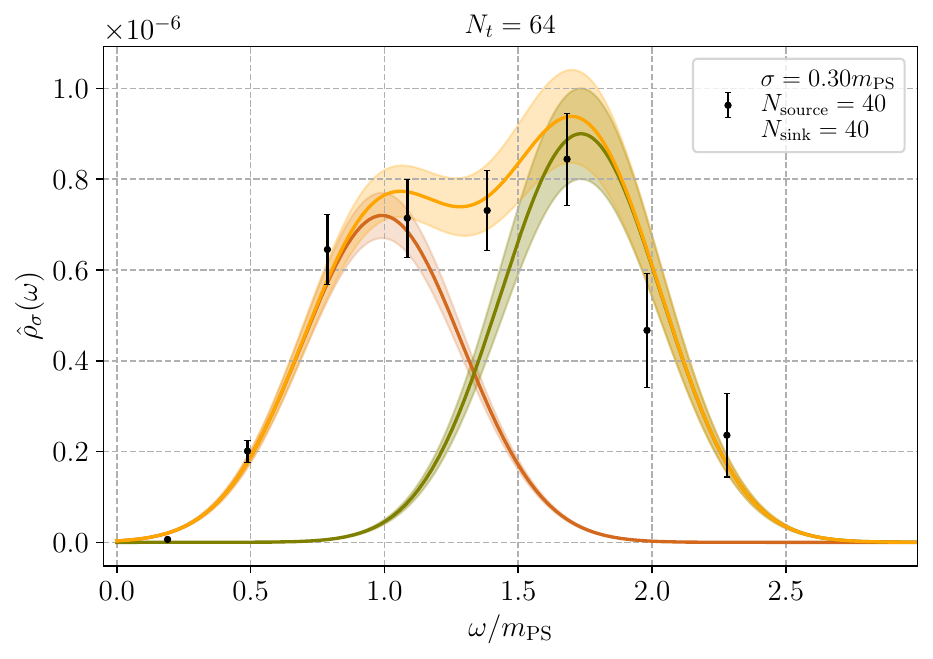}}
     \put(0, 0){\includegraphics[width=0.52\linewidth]{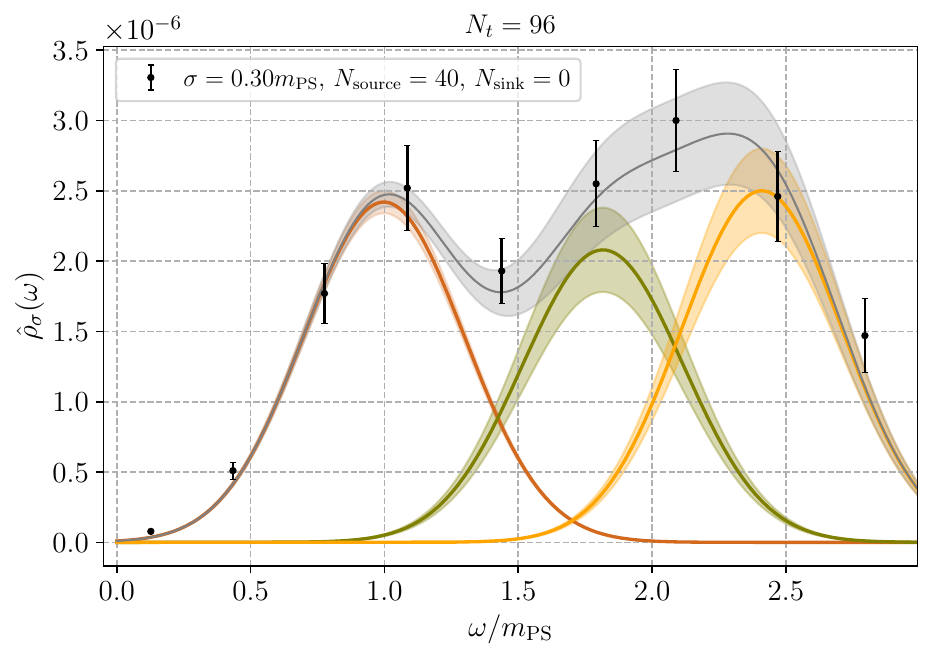}}
\end{picture}
\caption{Reconstructed spectral density in the PS meson channel, with Gaussian kernel. 
Correlation functions are obtained for APE smearing with $N_{\rm APE} = 50$ steps and step-size $\alpha_{\textrm{APE}} = 0.4$. The ground state mass, $m_{\rm PS}$, is obtained for the GEVP of the longer time extent---see Tab.~\ref{table:E3_results_ground}. In all the three panels, the black datapoints are the optimal reconstructed spectral density values $\hat{\rho}_{\sigma} (\omega)$.  Top panel:  M1 ensemble ($N_t = 48$), 
Wuppertal smearing step size $\varepsilon_{\rm f} = 0.18$, fit results $aE_0 = 0.3685(21)$, $aE_1 = 0.649(20)$ and reduced ${\chi}^2/N_{\rm d.o.f} = 0.2$. 
The yellow line corresponds to the fitted curve $f^{(2)}_{\sigma} (E)$ in Eq.~\eqref{eq:sum_of_kernels}, the red line corresponds to the deriving ground state curve $\Delta_\sigma (E - E_0)$ from the fit, the green line is the first excited state $\Delta_\sigma (E - E_1)$.
Middle panel:  M2 ensemble ($N_t = 64$), Wuppertal smearing step size $\varepsilon_{\rm as} = 0.20$, fit results
 $aE_0 = 0.3652(18)$, $aE_1 = 0.659(18)$, and reduced  ${\chi}^2/N_{\rm d.o.f.} = 0.9$. The yellow line corresponds to the fitted curve $f^{(2)}_{\sigma} (E)$ in Eq.~\eqref{eq:sum_of_kernels}, the red line corresponds to the deriving ground state curve $\Delta_\sigma (E - E_0)$ from the fit, the green line is the first excited state $\Delta_\sigma (E - E_1)$.
Bottom panel:  M3 ensemble ($N_t = 96$), Wuppertal smearing step size $\varepsilon_{\rm as} = 0.20$, fit results $aE_0 = 0.3678(10)$, $aE_1 = 0.6693(92)$, $aE_2 = 0.888(24)$, and reduced  ${\chi}^2/N_{\rm d.o.f.} = 0.5$. The grey line corresponds to the fitted curve $f^{(3)}_{\sigma} (E)$ in Eq.~\eqref{eq:sum_of_kernels}, the red line corresponds to the deriving ground state curve $\Delta_\sigma (E - E_0)$ from the fit, the green line is the first excited state $\Delta_\sigma (E - E_1)$, and the yellow line is the second excited state $\Delta_\sigma (E - E_2)$.}
\label{Fig:improvement_fit_nt}
\end{figure}

   \begin{figure}[t]
    \includegraphics[width=0.9\linewidth]{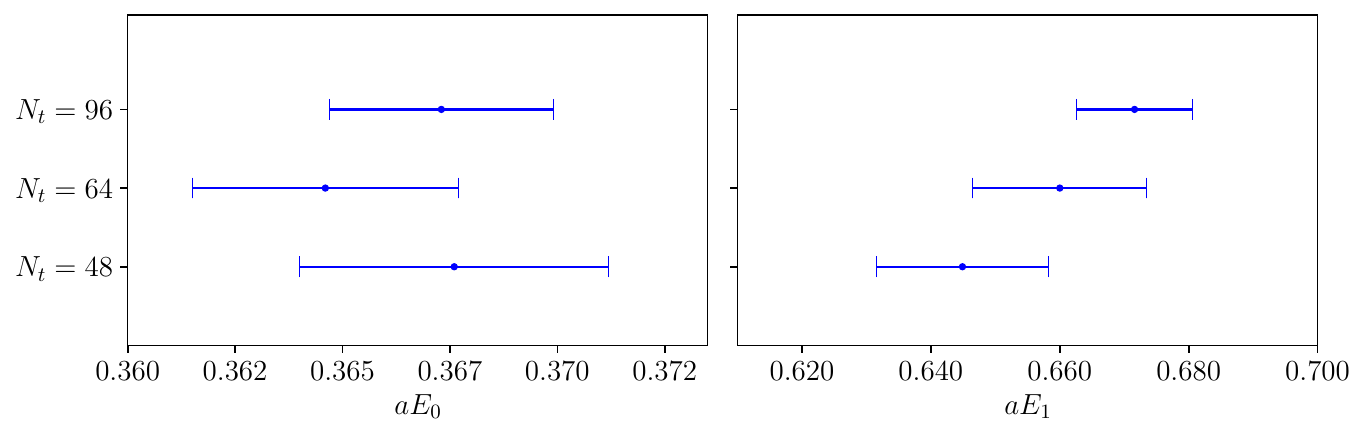}
    \caption{Spectroscopy results corresponding to Fig.~\ref{Fig:improvement_fit_nt} for ground and first excited states. The central value of the corresponding energy levels corresponds to the datapoints, whereas the width is the uncertainty in their determination. Off-sets between the different datapoints are kept for visual clarity.
    \label{fig:varyingsigma2}}
     \end{figure}

The reconstructed smearing kernel, $\bar{\Delta}_\sigma (E,\omega)$, can be expressed
 in terms of the function $b(t,E)$---see Eq.~(\ref{eq:reconstructed_kernel}) in Sec.~\ref{Sec:HLT_theory}.
 This finite sum over $t$ has a precision that is limited by the lattice extent, 
 as the largest possible choice of  $t = t_{\rm max}$ is bound by the constraint $t_{\rm max} \leq N_t / 2$. 

At a vanishing value of the trade-off parameter $\lambda$, this sum is known to converge to the true kernel in the limit of infinite time extents~\cite{Hansen:2019idp}. At a non-zero $\lambda$, one expects the reconstruction to improve, because the basis of functions generating the kernel is larger, and one can afford smaller values of $\lambda$, thus reducing the bias.
To quantify this effect, we consider lattices with different extents, $N_t$, in the temporal direction, while keeping all other lattice parameters fixed. Ensembles M1, M2, and M3, have time extents $N_t = 48, \, 64$ and $96$, respectively, but all other lattice parameters are common. We then compare the spectral reconstruction, focusing our attention on the region 
of parameter space at large energy.

The first result of this comparative study is shown in  Fig.~\ref{fig:increasing_nt_kernels}, which displays
the reconstructed Gaussian kernel, $\bar{\Delta}_\sigma (E, \omega)$,
 with $\omega/m_{\rm V}=1.8$, for the three aforementioned ensembles.
The choice of  $t_{\rm max}$ is the maximum possible value, $t_{\rm max} = N_t / 2$.
The reconstruction obtained with the shorter available time extent, $N_t = 48$, 
leads to non-negligible deviations from the target kernel, $\Delta_\sigma (E, \omega)$, and hence higher systematic uncertainties, both in the region away from the maximum of the kernel, but also in the central region.
These indications are compatible with the results shown in Fig.~(\ref{fig:energy_worsening}).
Conversely, the longest available time extent, $N_t = 96$, leads to reconstructed kernels that have visibly smaller deviations from the exact one.

By increasing the time extent of the lattice
one can perform accurate spectral density fits and spectroscopy, reaching progressively higher energies. 
Hence, it becomes possible to extend the number of measurable excited states. 
We illustrate this phenomenon in Fig.~\ref{Fig:improvement_fit_nt}, for the $({\rm PS}$) meson channel and the three time extents $N_t = 48, 64$ and $96$. 
A stabilization of the spectral reconstruction with respect to the algorithmic parameters, described in Sec.~\ref{Sec:HLT_theory}, can be obtained at higher energies, which become accessible by extending the time extent. As a consequence, we are able to obtain a second peak structure, which encodes information about excited states. 
Moreover, as will be discussed in Sec.~\ref{Sec:spectra}, the positions of the first
 two peaks agree, within statistical errors, and do not show a significant dependence on $N_t$. 
As a consequence, we can expect our fits to the smeared spectral densities to improve, possibly capturing more states. We illustrate this phenomenon in Fig.~\ref{fig:varyingsigma2}. The three ensembles considered, M1, M2, and M3,  respectively, yield the difference 
 $[aE_n - \sigma_{aE_n}, aE_n + \sigma_{aE_n}]$, where $\sigma_{aE_n}$ is the  uncertainty in 
 the determination of the $n^{th}$ energy level.

\begin{figure}[t]
\includegraphics[width=0.6\linewidth]{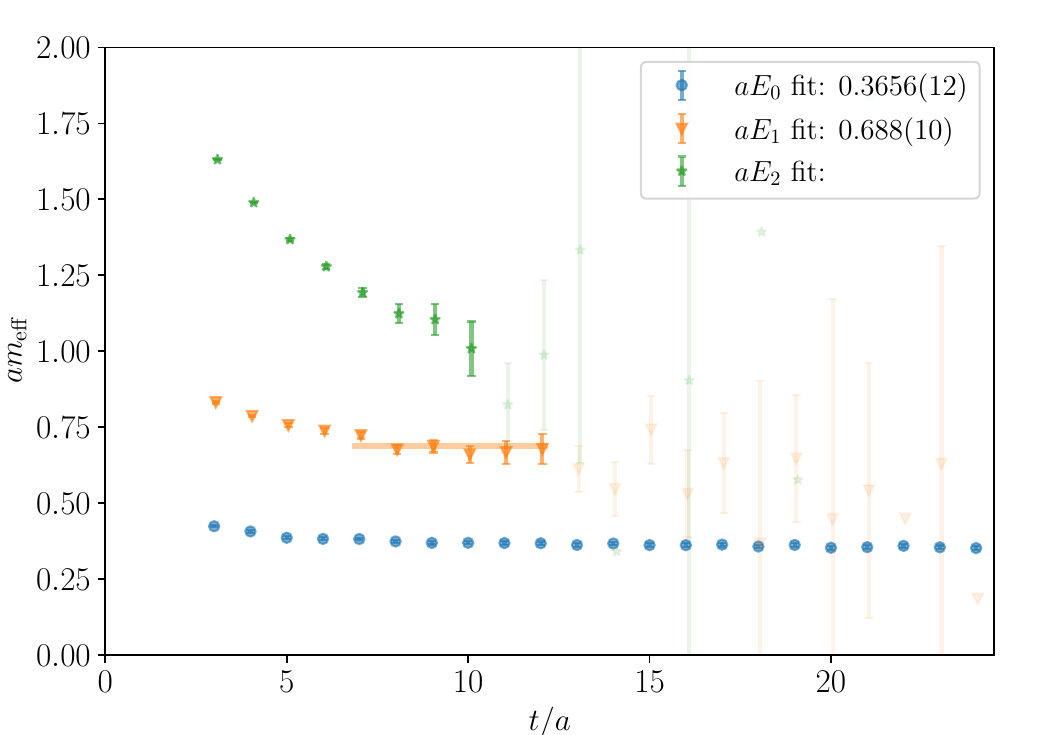}
\caption{Plateau in the effective mass, $am_{\rm eff}$, for the {\rm PS} channel in ensemble M2 ($N_t = 64$). 
The plateaux are identified via the GEVP process, by using a basis of nine correlators as shown in Eq.~\eqref{eq:simpler_GEVP_eq}, obtained by varying $N = 0$, $40$, $80$ sink and source Wuppertal smearing steps
 (keeping fixed $\varepsilon_{\rm as} = 0.10$, $\alpha_\textrm{APE} = 0.4$, and $N_{\rm APE} = 50$).}
\label{fig:nt64_plateau_PS_fund}
 \end{figure}

\begin{center}
\begin{figure}[t]
\includegraphics[width=0.6\linewidth]{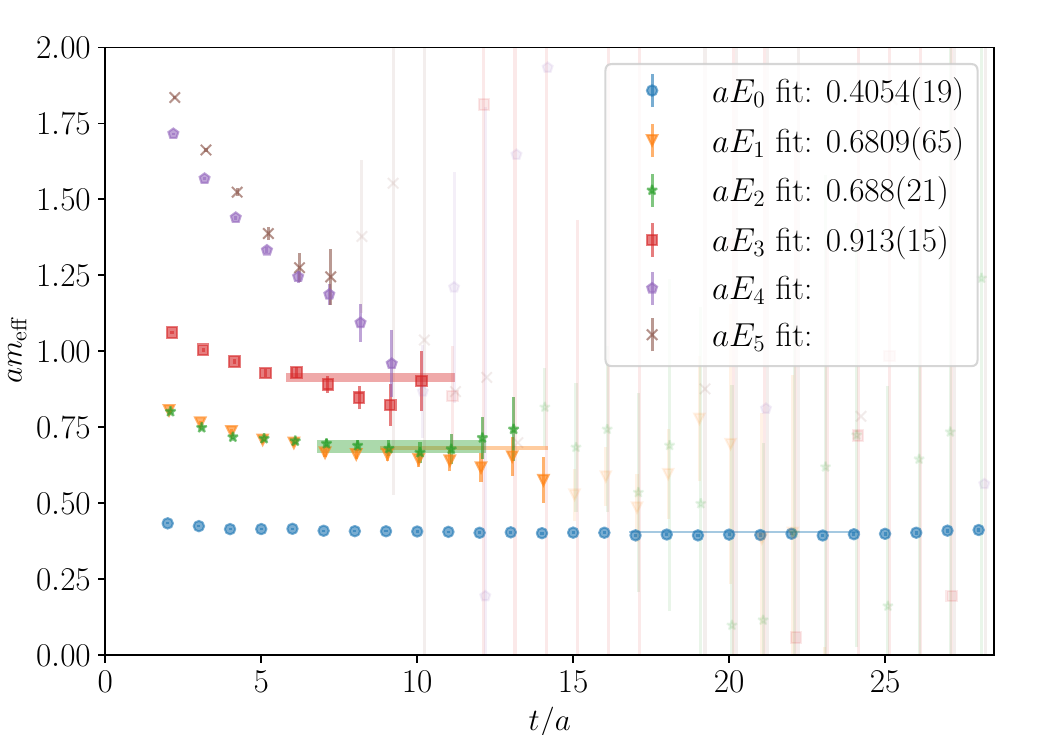}
\caption{Plateaux in effective mass, $am_{\rm eff}$, for the combined V and T channels, in ensemble M2 
($N_t = 64$). The plateaux are identified via the GEVP process, by using a basis of thirty-six correlators as shown in Eq.~\eqref{eq:crosschannel_GEVP_eq}: besides varying the smearing levels, $N = 0$, $40$, $80$, of both sink and source Wuppertal smearing (and keeping fixed $\varepsilon_{\rm f} = 0.2$, $\alpha_\textrm{APE} = 0.4$, and $N_{\rm APE} = 50$), we combine the channels with the cross channels.
In the case of degeneracy in the plateau fitting (within statistical uncertainties), we choose the energy level with a smaller statistical error as a representative value for the considered energy level. This is the case in this example, where $aE_1$ has been chosen over $aE_2$ (table~\ref{table:E2_results_first}) as an estimation for the first excited state.
\label{fig:nt64_plateau_VT_fund}}
 \end{figure}
\end{center}

\section{Numerical results and comparisons with GEVP }
\label{Sec:spectra}

   \begin{figure}[t]
    \includegraphics[width=0.4\linewidth]{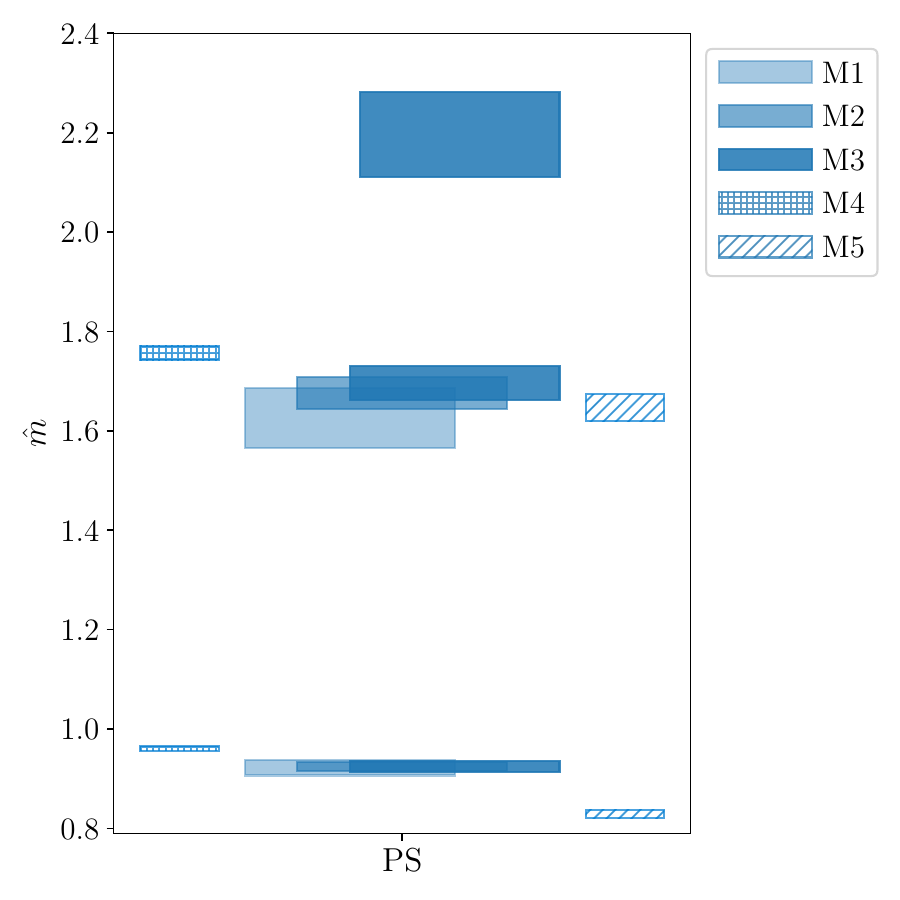}
    \caption{The spectrum, obtained through spectral density fits, for the pseudoscalar operators made of fundamental representation fermions for all the ensembles studied. Tower of masses, $\hat{m} \equiv w_0 \cdot m$ (in Wilson flow units), corresponding to ground, first and (possibly) second excited states is shown. The vertical midpoint of each colorblock is the numerical result, whereas the heights are comprehensive of statistical and systematic errors summed in quadrature. Horizontal offsets distinguish different ensembles. Different shadings of the same colors differentiate ensembles that differ in time extents ($N_t = 48, \, 64, \, 96$ for ensembles M1, M2 and M3), whereas hatched patterns, with no fill, are used to indicate ensembles that differ also in bare parameters (ensembles M4 and M5). As expected, progressively smaller uncertainties are obtained in the results for M1, M2 and M3.
    \label{fig:final_spectrum_PS}}
     \end{figure}

   \begin{figure}[t]
    \includegraphics[width=0.8\linewidth]{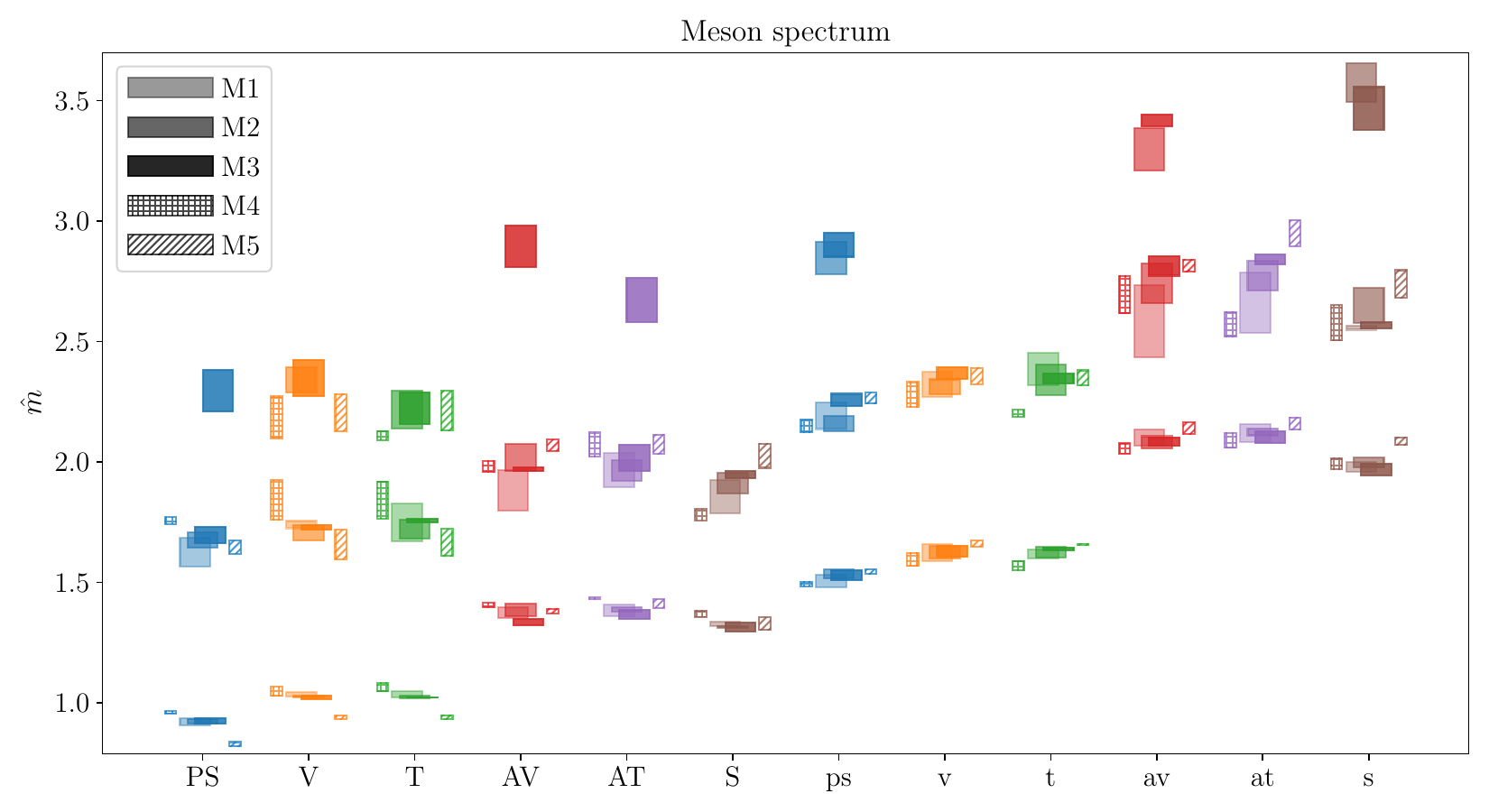}
    \caption{Meson spectrum in all the ensembles (summarised in Tab.~\ref{tab:ensembles}), for both fundamental and antisymmetric representation fermions, found through spectral densities fitting analysis. For each channel, a tower of masses, $\hat{m} \equiv w_0 \cdot m$ (in Wilson flow units), corresponding to ground, first and (where available) second excited states is shown. The vertical midpoint of each colorblock is the numerical result, whereas the heights are comprehensive of statistical and systematic errors summed in quadrature. Horizontal offsets distinguish different ensembles. Different shadings of the same colors differentiate ensembles that differ in time extents ($N_t = 48, \, 64$ and $96$ for ensembles M1, M2 and M3), whereas no filling color and patterns are used to indicate ensembles that differ also in bare parameters (ensembles M4 and M5). Progressively smaller uncertainties are obtained in the results for M1, M2 and M3. Six colors distinguish the different meson channels, and the colors match where two different representations--fundamental and antisymmetric--are used to study meson operators built with the same gamma-matrix structure.
    \label{fig:final_spectrum}}
     \end{figure}

In this section, we display and discuss our spectroscopy results, obtained with the HLT method. 
Firstly, we showcase how our results match the expectations from the GEVP ones.
Figures~\ref{fig:nt64_plateau_PS_fund} and \ref{fig:nt64_plateau_VT_fund} display two representative examples of GEVP computations. The former case is obtained as described in Eq.~(\ref{eq:simpler_GEVP_eq}), by using a variational basis of nine elements, whereas the latter combines the V and T channels, according to Eq.~(\ref{eq:crosschannel_GEVP_eq}).
As shown in the figures, typically the signal-to-noise ratio is good enough to find both ground state and first excited state even for the smaller variational basis. By adding the cross channel one gains access to the second excited state, $E_2$.

In Fig.~\ref{fig:final_spectrum_PS}, the spectrum for the fundamental pseudoscalar meson is shown for all the ensembles of Tab.~\ref{tab:ensembles}. 
The figure shows masses normalized in Wilson flow units, $\hat{m} \equiv w_0 \cdot m$. Different shadings and small horizontal offsets have been applied to distinguish the ensembles M1, M2 and M3, whereas larger offsets and patterns display the ensembles M4 and M5. The vertical extent of each colorblock represents the sum in quadrature of statistical errors and systematic effects. The latter ones are given by the excited states contaminations artifacts and by evaluating methodological effects using different smearing kernels, as the maximal difference between lattice results in the $k$ and $k+1$ peaks Gauss and Cauchy fits, $k$-G, $(k+1)$-G and $k$-C, $(k+1)$-C. In Fig.~\ref{fig:final_spectrum}, the same logic is applied to the whole meson spectrum of the theory, and six different colors have been used for different meson channels, and the matching colors have been used for the same meson channel in different fermion representations--fundamental and antisymmetric.

The numerical results for the ground states, first excited states, and (where available) second excited states obtained are reported in Tabs.~\ref{table:E1_results_ground} to \ref{table:E5_results_second}, for all the ensembles M1 to M5.  We write the masses in units of the lattice spacing, $a$, but these can be converted to Wilson flow units with the numerical results reported in Tab.~\ref{tab:ensembles}. 
The tables display the results for all the twelve meson channels of interest, together with details of the smearing level,
and the number of functions, $k$, included in Eq.~(\ref{eq:sum_of_kernels}). 
We tabulate the results obtained with five alternative analyses of the same correlation functions: two measurements obtained from spectral density reconstruction with Gaussian kernels, with $k$ or $k+1$ functions, 
two measurements obtained with Cauchy kernels, with $k$ or $k+1$ functions, and one obtained with the GEVP analysis. We also report the value of the smearing radius in the Gaussian and Cauchy cases, respectively.

As described in Sec.~\ref{Sec:varyingT}, we expect to find smaller uncertainties, and achieve more precise estimates 
of the energy levels, $aE_n$, by considering longer time extents, $N_t$. This is confirmed by inspecting the results in ensembles M1, M2, and M3, which are characterized by the same lattice bare parameters and lattice spatial extent, $N_s$, while only the time extent, $N_t$, is varied. This is also shown in Fig.~\ref{fig:final_spectrum_PS}, where the pseudoscalar fundamental channel is shown as a representative case of the data shown in Fig.~\ref{fig:final_spectrum}: the numerical results for the masses concerning ensembles M1, M2, M3 progressively present smaller uncertainties, as expected, and their values are compatible. We consider and show also the cases where the bare fundamental fermion masses, $am_0^{\rm f}$, are varied, in ensembles M4 and M5, for intermediate lattice time extent, $N_t = 64$.

For each energy level, $aE_n$, we report the fits obtained with $k$ and $k+1$ peaks,  both for Gaussian and Cauchy kernels, used in Eq.~(\ref{eq:correlated_chisquare}). If contaminations from further excited states are negligible, one expects compatible results, for both kernels, in going from $k$ to $k+1$; this is confirmed by the numerical results tabulated. The number of peaks optimizes the fitting procedure, $k$, is reported in all cases. The number of iterations for Wuppertal operator smearing, at source and sink, $N_{\rm source}$, and $N_{\rm sink}$, used for performing the spectral density fits, is reported in the tables as well.

The usage of multiple smearing kernels is an additional safety check against systematic effects that can potentially be overlooked in the reconstruction. Where the latter are absent, the spectroscopy should be untainted when adopting different kernels. This is confirmed by our numerical results. The fitted energy levels, $aE_n$,  are compatible with one another, within statistical uncertainties. In summary, as anticipated in Sec.~\ref{Sec:fitting_chisquare}, and shown in an example in Figs.~\ref{Fig:systematics_spectrum_nt64_fund_ground} and~\ref{Fig:systematics_spectrum_nt64_fund_first}, all the known sources of systematic uncertainty in the spectral density reconstruction 
 appear to be under control.

Comparison between results obtained by using spectral densities and GEVP show agreement in correspondence of all mesonic channels of the spectrum of our theory. The robustness of the results is reassuring and in line with one of the purposes of this paper, which is to show that measurements obtained with the former, novel, method match those of the latter, well established, one. 
Focussing on the spectrum, in our studies the ground state predictions from spectral densities tend to have larger uncertainties. This observation agrees with the findings of Ref.~\cite{DelDebbio:2022qgu}.

We also find a general trend towards improvement in the extraction of excited state masses, for which the quality of results is competitive with the GEVP ones.
This is noteworthy, for the quantity of information given as input for spectral density measurements is smaller than what used in the GEVP analysis. The simple case of Eq.~(\ref{eq:simpler_GEVP_eq}) makes use of nine two-point correlation functions in the GEVP analysis, and the cross-channel in Eq.~(\ref{eq:crosschannel_GEVP_eq}) uses thirty-six of them; by comparison, in both cases, the spectral density method uses only a single measurement. 
This finding  suggests an opportunity for possible gains in efficiency in future  large-scale studies
of multiple excited states.

\section{Summary and outlook } 
\label{Sec:outlook}

The first purpose of this paper is to report on software development and testing related to a new analysis package that implements the HLT method as a spectral density reconstruction tool for two-point functions obtained on the lattice.  We performed a systematic study of the method itself, in order to optimize the parameter choices entering the spectral reconstruction. We provide details, and numerical examples, in the main body of the paper. Our main findings can be summarized as follows. 

\begin{itemize}
\item For a given smearing kernel, $\Delta_{\sigma}(E,\,\omega)$, and a given input set of  2-point functions, $C_{ij}(t)$,
spectral density reconstruction requires minimizing a functional, $W[\vec{g}]$, defined in Eq.~(\ref{eq:HLT_functional}).
This procedure depends on two unphysical parameters, $\lambda$ and $\alpha$. We identify a range of $\lambda\sim {\cal O}(10)$ for which the smeared spectral densities do not depend on the parameters $\lambda$ and $\alpha$ within the statistical error, which is nonetheless not substantial.

\item In this work, we were able to afford values of the smearing radius, $\sigma$, in the range $0.18\,m_0 \leq \sigma
\leq 0.35\,m_0$, where $m_0$ is the mass of the ground state appearing 
in the two-point correlation functions used as input data.

\item Under the conditions identified at the previous points, our fit results are independent of the specific choice of smearing kernel. We illustrate this point by repeating our analysis with both Gaussian and Cauchy kernel, and demonstrating that the results of the two processes are compatible with one another.

\item APE and Wuppertal smearings are essential in the production of correlation functions to be analyzed. APE smearing is necessary in order to explore the high-energy behavior of the spectral density. Wuppertal smearing must be tuned so that contributions to the spectral density of all states of interest have comparable amplitudes.

\item A critical quantity for spectral density reconstruction, in particular in reference to the identification of excited states, is the time extent of the lattice, $N_t$. We illustrate its impact on the results of the physical analysis, by comparing lattices with $N_t=48$, $N_t=64$, and $N_t=96$, while keeping all other parameters fixed.
For the longest time extent, it is possible to reconstruct reliably the ground state, first and second excited states.

\end{itemize}

A concept that is worth discussing when testing a method for spectroscopy is the one of near-degenerate states. The GEVP offers a theoretically clean way to identify near-degeneracies due to the orthogonality of the eigenvectors that would be associated to similar (within noise) eigenvalues. The task would require, however, an adequately large basis of operators, and sufficient statistics. A similar statement holds for spectroscopy performed by fits of spectral densities. In this case, one would require enough energy resolution to distinguish two energy levels. Again, this is only possible with enough statistics; a large operator basis provides benefits here by allowing combined fits to be performed. Importantly, we have demonstrated that the smearing radius of the spectral density does not need to be smaller than the spacing between of two neighbouring states for the fit to separate them, see e.g. Fig~\ref{Fig:fit_comparison_gauss_cauchy}.

The second purpose of this document is to report on progress 
in the study of meson spectroscopy in the $Sp(4)$ lattice gauge theory coupled to $N_f=2$ Dirac fermions 
transforming according to the fundamental representation of the gauge group, and $n_f=3$ on the two-index antisymmetric one. This theory is a prominent candidate for new physics in the context
of composite Higgs models implementing top partial compositeness~\cite{Barnard:2013zea}.
To this purpose, we generated, using the (R)HMC algorithm, five new gauge ensembles. 
Allowing for thermalization, we selected configurations to retain
 in the ensembles so that
the plaquette does not show significant indications of autocorrelation.
We applied the Wilson flow as a smoothening process and measured the topological charge of the configurations, to
demonstrate the absence of topological freezing. Some modest level of autocorrelation is visible in the topology, but we do not expect this phenomenon to affect meson spectroscopy studied in this work.

We then measured the two-point correlation functions involving twelve flavored meson operators.
We applied both the HLT spectral density method, and the GEVP analysis, and reported the resulting mass spectra in Tabs.~\ref{table:E1_results_ground} to \ref{table:E5_results_second}.
We expressed the masses in units of the lattice spacing, $a$, and verified that the results were independent of the
analysis process applied, within statistical uncertainties.
Besides the fact that the analysis involves new ensembles obtained for different choices of lattice parameters, the main advance with respect to the literature in this theory is that we measured excited-state masses,
including both first and second excited states. Figure~\ref{fig:final_spectrum} is a way to visualize the results: we display the masses, $\hat{m}$, expressed in units of the Wilson flow scale, $w_0$, of all the states we could identify in the spectral density analysis, and for all the available channels and ensembles.

The work  reported in this paper  sets the stage for a plethora of future studies, both within 
the ongoing program of systematic exploration of $Sp(2N)$ gauge theories, but also in contexts of more general interest.
Firstly, the new ensembles can be analyzed in other physical channels, looking at correlation functions involving 
the singlet meson operators and the chimera baryon operators. Such measurements are currently underway.
Second, having put in place all the necessary technology, it is possible to generate and analyze new ensembles
with different couplings, $\beta$, and masses, $a m^{\mathrm{f}}$ and $a m^{\rm as}$, of the fermions.
By combining the results within the framework of  Wilson chiral perturbation theory~\cite{Sheikholeslami:1985ij,Rupak:2002sm} (see also Refs~\cite{Sharpe:1998xm}, and, in the context of improvement: Refs~\cite{Symanzik:1983dc,Luscher:1996sc}, as well as, for baryon chiral perturbation theory in QCD: Refs. ~\cite{Jenkins:1990jv,Bernard:1995dp,Beane:2003xv}), one can perform preliminary extrapolations towards the continuum and chiral limits.
Doing so will allow us to classify the properties of the theory that are of relevance to model building considerations in the CHM context, and to compare the spectroscopy results with those descending from other, complementary approaches~\cite{Bizot:2016zyu, Erdmenger:2020lvq, Erdmenger:2020flu, Elander:2020nyd,Elander:2021bmt,
Erdmenger:2023hkl,Erdmenger:2024dxf}.

The third further direction of future development reverts back to the original motivation of spectral density studies: the reconstruction of off-shell properties of correlation functions. One wants to put in place the technology needed to measure not just the position of the poles in the two-point functions, but also their residues, which are related to the decay constants and the overlap functions between operators and individual states.
Generally, one would like  to reconstruct the whole complex-space structure of the correlators,
because these enter observable quantities of theoretical and 
phenomenological relevance, as discussed in the introduction.

Last but not least, we wrote the main body of this paper by providing complete technical details, because our systematic study  
has general implications for any lattice gauge theory, including also QCD, the theory of strong nuclear interactions.
We expect the progress we reported here to be useful in a broader context, as spectral densities 
have broad ranges of phenomenological applications. 
We made all our new software and numerical results available as open source, so that users can download them, use them, and modify them as appropriate to their own physics goals.
Given the potential that spectral density analysis has for lattice field theory, we envision this work to contribute towards the general goal of the lattice community to develop and master such novel techniques, and turn them into high-precision analysis tools.

\begin{acknowledgments}

We would like to thank Giacomo Cacciapaglia, Thomas Flacke, Anna Hasenfratz, Chulwoo Jung,  Gabriele Ferretti, and Sarada Rajeev, 
for useful discussions during the ``PNU Workshop on Composite Higgs: Lattice study and all”, at Haeundae, Busan, in February 2024, where preliminary results of this study were presented. We also thank Nazario Tantalo for the discussions.

The work of E.B. and B.L. is supported in part by the EPSRC ExCALIBUR programme ExaTEPP (project EP/X017168/1). The work of E.B. has also been supported by the UKRI Science and Technology Facilities Council (STFC) Research Software Engineering Fellowship EP/V052489/1. The work of E.B., B.L., and M.P. has been supported in part by the STFC Consolidated Grant No. ST/X000648/1.

The work of N.F. has been supported by the STFC Consolidated Grant No. ST/X508834/1.

A.L. is funded in part by l’Agence Nationale de la Recherche (ANR), under grant ANR-22-CE31-0011.

The work of D.K.H. was supported by Basic Science Research Program through the National Research Foundation of Korea (NRF) funded by the Ministry of Education (NRF-2017R1D1A1B06033701). The work of D.K.H. was further supported by the National Research Foundation of Korea (NRF) grant funded by the Korea government (MSIT) (2021R1A4A5031460). 

L.D.D. and R.C.H. are supported by the UK Science and Technology Facility Council (STFC) grant ST/P000630/1.
The work of L.D.D. was supported by the ExaTEPP project EP/X01696X/1.

The work of J.W.L. was supported by IBS under the project code, IBS-R018-D1. 

The work of H.H. and C.J.D.L. is supported by the Taiwanese MoST grant 109-2112-M-009-006-MY3 and NSTC grant 112-2112-M-A49-021-MY3. The work of C.J.D.L. is also supported by Grants No. 112-2639-M-002-006-ASP and No. 113-2119-M-007-013-.

The work of B.L. and M.P. has been supported in part by the STFC  Consolidated Grant No. ST/T000813/1.
B.L., M.P. and L.D.D. received funding from the European Research Council (ERC) under the European Union’s Horizon 2020 research and innovation program under Grant Agreement No.~813942.

The work of D.V. is supported by STFC under Consolidated Grant No. ST/X000680/1.

The work of F.Z. is supported by the STFC Grant No.~ST/X000648/1.

Numerical simulations have been performed on the DiRAC Extreme Scaling service at the University of Edinburgh, and on the DiRAC Data Intensive service at Leicester.
The DiRAC Extreme Scaling service is operated by the Edinburgh Parallel Computing Centre on behalf of the STFC DiRAC HPC Facility (www.dirac.ac.uk). This equipment was funded by BEIS capital funding via STFC capital grant ST/R00238X/1 and STFC DiRAC Operations grant ST/R001006/1. DiRAC is part of the National e-Infrastructure. Spectral density measurements have been performed by using the \texttt{LSDensities} software package in Ref.~\cite{Forzano:2024}.

\vspace{1.0cm}
{\bf Research Data Access Statement}---The analysis code and data generated for this manuscript can be downloaded from Refs.~\cite{analysis_release} and~\cite{data_release}, respectively. 
\vspace{1.0cm}

{\bf Open Access Statement}---For the purpose of open access, the authors have applied a Creative Commons 
Attribution (CC BY) license to any Author Accepted Manuscript version arising.

\end{acknowledgments}

\appendix

\section{More about spectral density reconstruction}
\label{Sec:spec_dens_formulas}

In this appendix, we show details of our implementation for the HLT method. 
As described in Sec.~\ref{Sec:HLT_theory},  the functional $W[\vec{g}]$ in Eq.~(\ref{eq:HLT_functional}) consists 
of two parts: the systematic-error functional $A[\vec{g}] / A_0$ and the statistical-error functional $B / B_{\rm norm}$. 
Minimisation of $W[\vec{g}]$ is tantamount to solving the linear system
\begin{equation}
\vec{g} = \left( \mathtt{S} + \frac{\lambda \, A_0}{ B_{\rm norm}}\; \mathtt{B}  \right)^{-1} \vec{\mathtt{f}} \,.
\end{equation}
where $B_{\rm norm}$ is defined below Eq.~(\ref{eq:HLT_functional}). 
The matrix $\mathtt{S}$  descends from minimising in $A[\vec{g}] / A_0$  the contribution quadratic in $\vec{g}$. 
Its matrix elements can be always expressed as follows
\begin{equation}
\begin{split}
    \mathtt{S}_{tr} = \int_{E_{\rm min}}^{\infty}& dE e^{\alpha E} \, b(t+1,E) \, b(r+1, E) = \\ 
    &= \frac{e^{E_{\rm min}(\alpha -r -t -2)}}{t+r+2-\alpha} + \frac{e^{E_{\rm min}(\alpha +r +t +2 -2(N_t a))}}{2(N_t a) -t-r-2-\alpha} + \frac{e^{E_{\rm min}(\alpha +r -t -(N_t a))}}{(N_t a) +t-r-\alpha} + \frac{e^{E_{\rm min}(\alpha -r +t -(N_t a))}}{(N_t a) -t+r-\alpha}
\end{split}
\end{equation}
where $b(t, E)$ are the functions defined in Eq.~(\ref{eq:basis_functions}), where $\mathtt{S}$ is a $t_{\rm max} \times t_{\rm max}$ matrix. \\
The error functional $B[\vec{g}] / B_{\rm norm}$ contributes another functional quadratic part in $\vec{g}$, 
minimized by
\begin{equation}
    \texttt{B}_{tr} = \text{Cov}_{tr} \; .
\end{equation}
The vector $\vec{f}$  appears due to the minimization of terms linear in $\vec{g}$ inside $A[\vec{g}] / A_0$.
Its entries are defined to coincide with $f_{t+1}$, where
\begin{equation}
\label{eq:linear_part_HLT}
f_t(\omega) = \int_{E_{\rm min}}^\infty dE \; \Delta_\sigma(E,\omega) \, e^{\alpha E} \, b(t, E)\,.
\end{equation}

For a Gaussian kernel, $A_0 (\omega)$ and the entries $f_{t+1} (\omega)$ from Eq.~(\ref{eq:linear_part_HLT}) are
calculable:
\begin{equation}
    A_0(\omega) \equiv A[0](\omega) = \int_{E_{\rm min}}^{\infty} dE \, e^{\alpha E} \left[ \Delta^{(1)}_\sigma(E,\omega) \right]^2 = \frac{\hbox{Erf}\left(\frac{\omega-E_{\rm min}+\alpha\sigma^2/2}{\sigma}\right)+1}{\sigma \sqrt{\pi }  \left(\hbox{Erf}\left(\frac{\omega}{\sqrt{2} \sigma}\right)+1\right)^2} \; e^{\alpha \omega + \alpha^2 \sigma^2 / 4}\,,
\end{equation}
and 
\begin{equation}
\tilde{f}_t(\omega) = \int_{E_{\rm min}}^\infty dE \; \Delta^{(1)}_\sigma(E,\omega) \, e^{(-t+\alpha) E}  = \dfrac{e^{\frac{\sigma^2}{2}(\alpha-t)^2} e^{(\alpha-t)\omega} \, \left\{ 1 - \hbox{Erf} \left[ \frac{E_{\rm min}-\omega + \sigma^2(t-\alpha) }{\sqrt{2} \sigma}\right]  \right\}}{1+ \hbox{Erf} \left[ \frac{\omega}{\sigma \sqrt{2}} \right]}\,,
\end{equation}
respectively, with
\begin{equation}
    f_t(\omega) = \tilde{f}_t(\omega) + \tilde{f}_{T-t}(\omega) \,.
\end{equation}
For the Cauchy kernel, the integrals have to be computed numerically.

\section{Tables}
\label{sec:tables}
\newpage
\begin{table}[]
    \centering
    \begin{tabular}{ |c |c|c|c|c|c|c|c|c|c|c| }
     \hline  \hline
     $C$ & $k$ & $N_{\rm source}$ & $N_{\rm sink}$ & $aE_0$ $k$-G & $aE_0$ ($k+1$)-G & $aE_0$ $k$-C & $aE_0$ ($k+1$)-C & $am_C$ & $\sigma_{G} / m_C$ & $\sigma_C / m_C$ \\
     \hline
     PS & 2 & 80 & 40 & 0.3682(20) & 0.3695(27) & 0.3676(33)    & 0.3670(37) & 0.3678(17) & 0.33 & 0.32 \\
     V & 2 & 80 & 40 & 0.4092(58) & 0.4137(80) &  0.4080(24)  & 0.4078(27) & 0.4098(25) & 0.30 & 0.22  \\
     T & 2  & 80 & 40 & 0.4087(26) & 0.4094(58) &  0.4035(56)  & 0.4095(34) & 0.4098(25) & 0.30 & 0.30 \\
     AV & 2 & 80 & 40 & 0.5545(72) & 0.5492(93) &  0.5466(84)  & 0.5552(95) & 0.5485(81) & 0.20 & 0.18 \\
     AT & 2 & 80 & 40 & 0.5518(85) & 0.5495(77) &  0.5459(74)  & 0.5454(88) & 0.5514(73)  & 0.18 & 0.20 \\
     S & 2 & 80 & 40 & 0.5287(84) & 0.5272(99) & 0.5250(70) & 0.5292(80) & 0.5241(64)   & 0.20 & 0.20\\
      \hline
     ps & 2 & 80 & 40 & 0.5999(31) & 0.6009(29) & 0.5997(34)    &  0.6010(31) & 0.60161(91)  &0.18 & 0.20 \\
     v & 2 & 80 & 40 & 0.6457(45) & 0.6452(38) &  0.6547(30)  & 0.6492(27) & 0.6503(13)  & 0.20 & 0.26\\
     t & 2 & 80 & 40 & 0.6476(38) & 0.6479(33) &  0.6552(36)  & 0.6582(70) & 0.6503(13)  & 0.30 & 0.26 \\
     av & 2 & 80 & 40 & 0.8375(71) & 0.8370(93) &  0.8335(72)  & 0.8290(62) & 0.8299(81)  & 0.20 & 0.18 \\
     at & 2 & 80 & 40 & 0.8516(86) & 0.8567(98) &  0.8500(71)  & 0.8533(92) & 0.8408(87)  & 0.18 & 0.18 \\
     s & 2 & 80 & 40 & 0.7994(84) & 0.7998(83) & 0.7953(79) & 0.7929(86) & 0.7957(83)  & 0.18 & 0.20 \\

     \hline  \hline
    \end{tabular}
    \caption{\label{table:E1_results_ground} Numerical results for the ground state mass in ensemble M1. $k$-G stands for $k$-Gauss fit, ($k+1$)-G is $k+1$-Gauss fit, $k$-C stands for $k$-Cauchy function fit, ($k+1$)-G $k+1$-Cauchy function fit, $C$ indicates the mesonic channel considered, $am_C$ is the result of the GEVP analysis, $\sigma_G$ is the energy smearing radius used for the Gaussian fits, $\sigma_C$ for the Cauchy fit.}
\end{table}

    \begin{table}[]    
    \centering
    \begin{tabular}{ |c |c|c|c|c|c|c|c|c|c|c| }
     \hline  \hline
     $C$ & $k$ & $N_{\rm source}$ & $N_{\rm sink}$ & $aE_1$ $k$-G & $aE_1$ ($k+1$)-G & $aE_1$ $k$-C & $aE_1$ ($k+1$)-C & $am_C$ & $\sigma_{G} / m_C$ & $\sigma_C / m_C$ \\
     \hline
     PS & 2 & 80 & 40 & 0.650(19) & 0.652(25) & 0.653(30) & 0.648(26)  & 0.661(25)  & 0.33 & 0.32 \\
     V & 2 & 80 & 40 & 0.703(20) & 0.698(27) & 0.699(20) & 0.692(12) & 0.700(26)  & 0.30 & 0.22 \\
     T & 2 & 80 & 40 & 0.707(13) & 0.696(23) & 0.693(22) & 0.696(31) & 0.700(26)  & 0.30 & 0.30 \\
     AV & 2 & 80 & 40 & 0.761(32) & 0.753(33) & 0.726(24) & 0.722(31) & 0.743(44)  & 0.20 & 0.18 \\
     AT & 2 & 80 & 40 & 0.790(42) & 0.790(42) & 0.766(32) & 0.784(17) & 0.768(47)  & 0.18 & 0.20 \\
     S & 2 & 80 & 40 & 0.750(23) & 0.746(29) & 0.755(24) & 0.742(18) & 0.748(14)  & 0.20 & 0.20 \\
      \hline
     ps  & 2 & 80 & 40 & 0.880(22) & 0.876(30) & 0.865(30) & 0.878(35) & 0.891(19)  & 0.18 & 0.20 \\
     v & 2 & 80 & 40 & 0.931(23) & 0.923(19)  & 0.940(22) & 0.932(20) & 0.955(11)  & 0.20 & 0.26 \\
     t & 2 & 80 & 40 & 0.960(13) & 0.966(18) & 0.949(23) & 0.967(22) & 0.955(11)  & 0.30 & 0.26 \\
     av & 2 & 80 & 40 & 1.040(31) & 1.046(33) & 1.028(40) & 1.018(43) & 1.063(42)  & 0.20 & 0.18 \\
     at & 2 & 80 & 40 & 1.062(35) & 1.067(35) & 1.061(30) & 1.056(34) & -  & 0.18 & 0.18 \\
     s & 2 & 80 & 40 & 1.036(22) & 1.037(21) & 1.024(18) & 1.039(28) & 1.052(19)  & 0.18 & 0.20 \\
     \hline  \hline
    \end{tabular}
    \caption{\label{table:E1_results_first} Numerical results for the first excited  state mass in ensemble M1. $k$-G stands for $k$-Gauss fit, ($k+1$)-G is $k+1$-Gauss fit, $k$-C stands for $k$-Cauchy function fit, ($k+1$)-G $k+1$-Cauchy function fit, $C$ indicates the mesonic channel considered, $am_C$ is the result of the GEVP analysis, $\sigma_G$ is the energy smearing radius used for the Gaussian fits, $\sigma_C$ for the Cauchy fit.}
    \end{table}

    \begin{table}[]    
    \centering
    \begin{tabular}{ |c |c|c|c|c|c|c|c|c|c|c| }
     \hline  \hline
     $C$ & $k$ & $N_{\rm source}$ & $N_{\rm sink}$ & $aE_0$ $k$-G & $aE_0$ ($k+1$)-G & $aE_0$ $k$-C & $aE_0$ ($k+1$)-C & $am_C$ & $\sigma_{G} / m_C$ & $\sigma_C / m_C$ \\
     \hline
     PS & 2 & 80 & 40 & 0.3633(31) & 0.3653(12) & 0.3642(13) & 0.3624(25)  & 0.3656(12) & 0.35 & 0.30 \\
     V & 3 & 0 & 40 & 0.4081(31) & 0.4034(33) & 0.4040(23) & 0.4050(20) & 0.4054(19) & 0.28 & 0.33\\
     T & 3 & 0 & 40 & 0.4022(35) & 0.4043(36) & 0.4053(24) & 0.4043(30) & 0.4054(19) & 0.23 & 0.23\\
     AV & 2 & 80 & 40 & 0.5484(81) & 0.5464(83) & 0.5453(92) & 0.5462(90) & 0.5423(90) & 0.30 & 0.18 \\
     AT & 2 & 80 & 40 & 0.5522(74) & 0.5513(84) & 0.5474(66) & 0.5428(68) & 0.5477(84) & 0.30 & 0.20 \\
     S & 2 & 0 & 40 & 0.5173(75) & 0.5201(75) & 0.5163(93) & 0.5155(99) & 0.5222(78) & 0.30 & 0.20 \\
      \hline
     ps  & 3 & 0 & 40 & 0.6021(11) & 0.6047(11) & 0.6067(17) & 0.6024(15) & 0.6007(11) & 0.18 & 0.18 \\
     v &  2 & 40 & 80 & 0.6510(31) & 0.6523(43) & 0.6443(38) & 0.6453(29) & 0.6473(12) & 0.20 & 0.20 \\
     t & 2 & 40 & 40 & 0.6443(38) & 0.6468(24) & 0.6500(33) & 0.6499(27) & 0.6473(12) & 0.20 & 0.20 \\
     av & 3 & 0 & 40 & 0.8212(81) & 0.8186(76) & 0.8216(92) & 0.829(11) & 0.821(11) & 0.18 & 0.20 \\
     at & 2 & 0 & 40 & 0.832(13) & 0.835(11) & 0.843(17) & 0.836(17) & 0.834(15) & 0.18 & 0.23 \\
     s & 3 & 80 & 40 & 0.7863(88) & 0.7882(95) & 0.7872(98) & 0.7913(90) & 0.7820(96) & 0.18 & 0.18\\
     \hline  \hline
    \end{tabular}
    \caption{\label{table:E2_results_ground} Numerical results for the ground state mass in ensemble M2.  $k$-G stands for $k$-Gauss fit, ($k+1$)-G is $k+1$-Gauss fit, $k$-C stands for $k$-Cauchy function fit, ($k+1$)-G $k+1$-Cauchy function fit, $C$ indicates the mesonic channel considered, $am_C$ is the result of the GEVP analysis, $\sigma_G$ is the energy smearing radius used for the Gaussian fits, $\sigma_C$ for the Cauchy fit.}
    \end{table}

    \begin{table}[]    
    \centering
    \begin{tabular}{ |c |c|c|c|c|c|c|c|c|c|c| }
     \hline  \hline
     $C$ & $k$ & $N_{\rm source}$ & $N_{\rm sink}$ & $aE_1$ $k$-G & $aE_1$ ($k+1$)-G & $aE_1$ $k$-C & $aE_1$ ($k+1$)-C & $am_C$ & $\sigma_{G} / m_C$ & $\sigma_C / m_C$ \\
     \hline
     PS & 2 & 80 & 40 & 0.658(24) & 0.663(18) & 0.671(11) & 0.678(12)  & 0.688(10) & 0.35 & 0.30 \\
     V & 3 & 0 & 40 & 0.676(12) & 0.6712(94) & 0.670(13) & 0.669(10) & 0.6809(65) & 0.28 & 0.33 \\
     T & 3 & 0 & 40 & 0.678(11) & 0.670(13) & 0.6783(99) & 0.679(11) & 0.6809(65) & 0.23  & 0.23 \\
     AV & 2 & 80 & 40 & 0.792(24) & 0.770(23) & 0.767(23) & 0.742(23) & 0.772(30) & 0.30 & 0.18 \\
     AT & 2 & 80 & 40 & 0.802(31) & 0.798(34) & 0.824(30) & 0.804(29) & 0.789(39) & 0.30 & 0.20 \\
     S & 2 & 0 & 40 & 0.749(28) & 0.754(21) & 0.761(21) & 0.763(23) & 0.783(25) & 0.30 & 0.20\\
      \hline
     ps & 3 & 0 & 40 & 0.870(17) & 0.874(20) & 0.864(19) & 0.864(16) & 0.895(16) & 0.18 & 0.18 \\
     v & 2 & 40 & 80 & 0.933(12) & 0.929(12) & 0.928(11) & 0.9361(83) & 0.9392(67) & 0.20 & 0.20 \\
     t & 2 & 40 & 40 & 0.927(22) & 0.9333(91) & 0.9432(65) & 0.9397(71) & 0.9392(67) & 0.20 & 0.20 \\
     av & 3 & 0 & 40 & 1.070(18) & 1.073(15) & 1.079(16) & 1.081(20) & 1.087(14) & 0.18 & 0.20 \\
     at & 2 & 0  & 40 & 1.102(21) & 1.101(22) & 1.102(23) & 1.088(22) & 1.076(28) & 0.18 & 0.23 \\
     s & 3 & 80 & 40 & 1.036(22) & 1.035(21) & 1.058(12) & 1.059(14) & 1.053(12) & 0.18 & 0.18 \\
     \hline  \hline
    \end{tabular}
    \caption{\label{table:E2_results_first} Numerical results for the first excited  state mass in ensemble M2.  $k$-G stands for $k$-Gauss fit, ($k+1$)-G is $k+1$-Gauss fit, $k$-C stands for $k$-Cauchy function fit, ($k+1$)-G $k+1$-Cauchy function fit, $C$ indicates the mesonic channel considered, $am_C$ is the result of the GEVP analysis, $\sigma_G$ is the energy smearing radius used for the Gaussian fits, $\sigma_C$ for the Cauchy fit.}
    \end{table}

    \begin{table}[]    
    \centering
    \begin{tabular}{ |c |c|c|c|c|c|c|c|c|c|c| }
     \hline  \hline
     $C$ & $k$ & $N_{\rm source}$ & $N_{\rm sink}$ & $aE_2$ $k$-G & $aE_2$ ($k+1$)-G & $aE_2$ $k$-C & $aE_2$ ($k+1$)-C & $am_C$ & $\sigma_{G} / m_C$ & $\sigma_C / m_C$ \\
     \hline
      PS & 2 & 80 & 40 & - & - & - & - & - & 0.35 & 0.30 \\
     V & 3 & 0 & 40 & 0.923(19) & 0.908(21) & 0.906(18) & 0.906(21) & 0.913(15) & 0.28 & 0.33 \\
     T & 3 & 0 & 40 & 0.879(31) & 0.875(29) & 0.902(18) & 0.899(13) & 0.913(15) & 0.23  & 0.23 \\
     AV & 2 & 80 & 40 & - & - & - & - & - & 0.30 & 0.18 \\
     AT & 2 & 80 & 40 & - & - & - & - & - & 0.30 & 0.20 \\
     S & 2 & 0 & 40 & - & - & - & - & - & 0.30 & 0.20 \\
      \hline
     ps & 3 & 0 & 40 & 1.136(41) & 1.143(44) & 1.131(39) & 1.131(35) & - & 0.18 & 0.18 \\
     v & 2 & 40 & 80 & - & - & - & - & 1.027(11) & 0.20 & 0.20 \\
     t & 2 & 40 & 40 & - & - & - & - & 1.027(11) & 0.20 & 0.20 \\
     av & 3 & 0 & 40 & 1.347(36) & 1.336(33) & 1.336(32) & 1.349(41) & - & 0.18 & 0.20 \\
     at & 2 & 0 & 40 & - & - & - & - & - & 0.18 & 0.23 \\
     s & 3 & 80 & 40 & 1.388(37) & 1.405(41) & 1.396(38) & 1.393(40) & - & 0.18 & 0.18 \\
     \hline  \hline
    \end{tabular}
    \caption{\label{table:E2_results_second} Numerical results for the second excited state mass in ensemble M2.  $k$-G stands for $k$-Gauss fit, ($k+1$)-G is $k+1$-Gauss fit, $k$-C stands for $k$-Cauchy function fit, ($k+1$)-G $k+1$-Cauchy function fit, $C$ indicates the mesonic channel considered, $am_C$ is the result of the GEVP analysis, $\sigma_G$ is the energy smearing radius used for the Gaussian fits, $\sigma_C$ for the Cauchy fit.}
    \end{table}

    \begin{table}[]    
    \centering
    \begin{tabular}{ |c |c|c|c|c|c|c|c|c|c|c| }
     \hline \hline
     $C$ & $k$ & $N_{\rm source}$ & $N_{\rm sink}$ & $aE_0$ $k$-G & $aE_0$ ($k+1$)-G & $aE_0$ $k$-C & $aE_0$ ($k+1$)-C & $am_C$ & $\sigma_{G} / m_C$ & $\sigma_C / m_C$ \\
     \hline
     PS & 3 & 40 & 0 & 0.3677(11) & 0.36591(81) & 0.36733(91) & 0.36692(80) & 0.36657(82) & 0.30 & 0.27 \\
     V & 3 & 40 & 40 & 0.4105(14) & 0.4101(18) & 0.4079(16) & 0.4091(10) & 0.4083(12) & 0.28 & 0.25 \\
     T & 3 & 0 & 40 & 0.4071(22) & 0.4073(20) & 0.4082(11) & 0.4092(16) & 0.4083(12) & 0.33 & 0.28 \\
     AV & 3 & 40 & 0 & 0.5363(54) & 0.5374(53) & 0.5384(48) & 0.5361(73) & 0.5351(64) & 0.28 & 0.32 \\
     AT & 3 & 40 & 40 & 0.5459(56) & 0.5458(66) & 0.5501(53) & 0.5508(50) & 0.5487(46) & 0.30 & 0.18 \\
     S  & 2 & 40 & 40 & 0.5152(32) & 0.5130(37) & 0.5177(54) & 0.5167(29) &  0.5166(44) & 0.30 & 0.24 \\
      \hline
     ps  & 3 & 0 & 40 & 0.60183(90) & 0.60142(80) & 0.60183(61) & 0.60183(52) & 0.60132(57) & 0.23 & 0.22 \\
     v &  2 & 0 & 40 & 0.6480(20) & 0.6491(14) & 0.6499(18) & 0.6502(19) & 0.6491(16) & 0.24 & 0.25 \\
     t & 2 & 0 & 40 & 0.6481(16) & 0.6481(18) & 0.6517(22) & 0.6520(21) & 0.6491(16) & 0.28 & 0.28 \\
     av & 3 & 0 & 40 & 0.8361(57) & 0.8358(62) & 0.8348(47) & 0.8371(60) & 0.8348(64) & 0.18 & 0.25 \\
     at & 2 & 0 & 40 & 0.8451(64) & 0.8401(80) & 0.8393(81) & 0.8398(73) & 0.8431(73) & 0.25 & 0.25 \\
     s & 3 & 0 & 40 & 0.798(13) & 0.800(15) & 0.797(12) & 0.796(12) &  0.789(13) & 0.23 & 0.23 \\

     \hline  \hline
    \end{tabular}
    \caption{\label{table:E3_results_ground} Numerical results for the ground state mass in ensemble M3.  $k$-G stands for $k$-Gauss fit, ($k+1$)-G is $k+1$-Gauss fit, $k$-C stands for $k$-Cauchy function fit, ($k+1$)-G $k+1$-Cauchy function fit, $C$ indicates the mesonic channel considered, $am_C$ is the result of the GEVP analysis, $\sigma_G$ is the energy smearing radius used for the Gaussian fits, $\sigma_C$ for the Cauchy fit.}
    \end{table}

    \begin{table}[]    
    \centering
    \begin{tabular}{ |c |c|c|c|c|c|c|c|c|c|c| }
     \hline  \hline
     $C$ & $k$ & $N_{\rm source}$ & $N_{\rm sink}$ & $aE_1$ $k$-G & $aE_1$ ($k+1$)-G & $aE_1$ $k$-C & $aE_1$ ($k+1$)-C & $am_C$ & $\sigma_{G} / m_C$ & $\sigma_C / m_C$ \\
     \hline
     PS & 3 & 40  & 0 & 0.6694(93) & 0.6672(97) & 0.6786(85) & 0.6798(82) & 0.6851(88) & 0.30 & 0.27 \\
     V & 3 & 40 & 40 & 0.6855(92) & 0.6757(98) & 0.6741(13) & 0.6781(15) & 0.6818(99) & 0.28 & 0.25 \\
     T & 3 & 0 & 40 & 0.6881(12) & 0.6847(98) & 0.6901(11) & 0.6911(14) & 0.6818(99) & 0.33 & 0.28 \\
     AV & 3 & 40 & 0 & 0.791(18) & 0.787(17) & 0.793(19) & 0.794(18) & 0.783(13) & 0.28 & 0.32 \\
     AT & 3 & 40 & 40 & 0.783(22) & 0.778(24) & 0.777(27) & 0.775(29) & 0.795(24) & 0.30 & 0.18 \\
     S  & 2 & 40 & 40 & 0.778(23) & 0.784(32) & 0.778(30) & 0.778(38) & 0.792(34) & 0.30 & 0.24 \\
      \hline
     ps  & 3 & 0 & 40 & 0.908(13) & 0.905(15) & 0.907(14) & 0.906(16) & 0.9116(95) & 0.23 & 0.22 \\
     v & 2 & 0 & 40 & 0.9332(90) & 0.928(12) & 0.9332(60) & 0.9327(75) & 0.9378(70) & 0.24 & 0.25 \\
     t & 2 & 0 & 40 & 0.9332(85) & 0.9317(90) & 0.9427(81) & 0.9417(68) & 0.9378(70) & 0.28 & 0.28 \\
     av &  3 & 0 & 40 & 1.102(15) & 1.098(15) & 1.098(14) & 1.097(17) &  1.109(16) & 0.18 & 0.25 \\
     at & 2 & 0 & 40 & 1.112(23) & 1.112(21) & 1.118(16) & 1.118(16) & 1.114(21) & 0.25 & 0.25 \\
     s & 3 & 0 & 40 & 1.033(20) & 1.027(20) & 1.041(33) & 1.042(18) &  1.044(47) & 0.23 & 0.23 \\

     \hline  \hline
    \end{tabular}
    \caption{\label{table:E3_results_first} Numerical results for the first excited  state mass in ensemble M3.  $k$-G stands for $k$-Gauss fit, ($k+1$)-G is $k+1$-Gauss fit, $k$-C stands for $k$-Cauchy function fit, ($k+1$)-G $k+1$-Cauchy function fit, $C$ indicates the mesonic channel considered, $am_C$ is the result of the GEVP analysis, $\sigma_G$ is the energy smearing radius used for the Gaussian fits, $\sigma_C$ for the Cauchy fit.}
    \end{table}

    \begin{table}[]    
    \centering
    \begin{tabular}{ |c |c|c|c|c|c|c|c|c|c|c| }
     \hline  \hline
     $C$ & $k$ & $N_{\rm source}$ & $N_{\rm sink}$ & $aE_2$ $k$-G & $aE_2$ ($k+1$)-G & $aE_2$ $k$-C & $aE_2$ ($k+1$)-C & $am_C$ & $\sigma_{G} / m_C$ & $\sigma_C / m_C$ \\
     \hline
     PS & 3 & 40  & 0 & 0.889(25) & 0.877(22) & 0.860(22) & 0.855(23) & - & 0.30 & 0.27 \\
     V & 3 & 40  & 40 & 0.923(20) & 0.908(22) & 0.922(18) & 0.910(19) & 0.913(19) & 0.28 & 0.25 \\
     T & 3 & 0  & 40 & 0.901(18) & 0.899(20) & 0.920(22) & 0.918(20) & 0.913(19) & 0.33 & 0.28 \\
     AV & 3 & 40 & 0 & 1.140(30) & 1.111(33) & 1.131(36) & 1.138(33) & - & 0.28 & 0.32 \\
     AT & 3 & 40 & 40 & 1.050(38) & 1.043(35) & 1.020(32) & 1.018(34) & - & 0.30 & 0.18 \\
     S   & 2 & 40 & 40 & - & - & - & - & - & 0.30 & 0.24 \\
      \hline
     ps  & 3 & 0 & 40 & 1.154(39) & 1.161(42) & 1.156(38) & 1.170(40) & - & 0.23 & 0.22 \\
     v & 2 & 0 & 40 & - & - & - & - & 1.0190(68) & 0.24 & 0.25 \\
     t & 2 & 0 & 40 & - & - & - & - & 1.0190(68) & 0.28 & 0.28 \\
     av & 3 & 0 & 40 & 1.373(32) & 1.371(32) & 1.381(28) & 1.375(28) & - & 0.18 & 0.25 \\
     at & 2 & 0 & 40 & - & - & - & - & - & 0.25 & 0.25 \\
     s & 3 & 0 & 40 & 1.340(32) & 1.338(32) & 1.353(32) & 1.351(33) & - & 0.23 & 0.23 \\

     \hline  \hline
    \end{tabular}
    \caption{\label{table:E3_results_second} Numerical results for the second excited state mass in ensemble M3.  $k$-G stands for $k$-Gauss fit, ($k+1$)-G is $k+1$-Gauss fit, $k$-C stands for $k$-Cauchy function fit, ($k+1$)-G $k+1$-Cauchy function fit, $C$ indicates the mesonic channel considered, $am_C$ is the result of the GEVP analysis, $\sigma_G$ is the energy smearing radius used for the Gaussian fits, $\sigma_C$ for the Cauchy fit.}
    \end{table}

    \begin{table}[]    
    \centering
    \begin{tabular}{ |c |c|c|c|c|c|c|c|c|c|c| }
     \hline  \hline
     $C$ & $k$ & $N_{\rm source}$ & $N_{\rm sink}$ & $aE_0$ $k$-G & $aE_0$ ($k+1$)-G & $aE_0$ $k$-C & $aE_0$ ($k+1$)-C & $am_C$ & $\sigma_{G} / m_C$ & $\sigma_C / m_C$ \\
     \hline
     PS & 2 & 0 & 40 &  0.4079(31) & 0.4082(35) &  0.4086(24) & 0.4087(23) & 0.4095(12) & 0.30 & 0.30 \\
     V & 3 & 40 & 40 & 0.4466(30) & 0.4468(32) &  0.4459(33) & 0.4458(34) & 0.4483(17) & 0.30 & 0.27 \\
     T & 3 & 0 & 40 & 0.4472(22) & 0.4474(21) & 0.4487(20) & 0.4490(22) & 0.4483(17) & 0.25 & 0.25 \\
     AV & 2 & 80 & 40 & 0.5973(85) & 0.5976(87) &  0.5973(85) & 0.5996(85) & 0.6015(89) & 0.25 & 0.25 \\
     AT & 2 & 80 & 40 & 0.6027(90) & 0.6020(86) &  0.6187(85) & 0.6137(84) & 0.6140(96) & 0.25 & 0.25 \\
     S  & 2 & 0 & 40 & 0.590(11) & 0.589(11) &  0.576(13) & 0.573(14) & 0.579(11) & 0.25 & 0.30 \\
      \hline
     ps  & 2 & 0 & 40 & 0.6290(13) & 0.6292(14) &  0.6282(12) & 0.6283(15) & 0.62808(95) & 0.24 & 0.24 \\
     v &  2 & 0 & 40 & 0.6691(33) & 0.6689(35) &  0.6721(33) & 0.6724(34) & 0.6715(16) & 0.23 & 0.20 \\
     t & 2 & 0 & 40 & 0.6691(31) & 0.6694(34) &  0.6688(33) & 0.6687(32) & 0.6715(16) & 0.23 & 0.25 \\
     av & 2 & 40 & 40 & 0.8746(55) & 0.8747(57) & 0.8741(48) & 0.8788(55) & 0.8792(56) & 0.20 & 0.20 \\
     at & 2 & 80 & 40 & 0.8744(82) & 0.8764(82) & 0.8764(52) & 0.8754(62) & 0.8796(67) & 0.25 & 0.25 \\
     s & 2 & 40 & 40 & 0.8481(100) & 0.8482(97) & 0.8581(88) & 0.8571(83) & 0.8539(91)  & 0.24 & 0.20 \\
     \hline  \hline
    \end{tabular}
    \caption{\label{table:E4_results_ground} Numerical results for the ground state mass in ensemble M4.  $k$-G stands for $k$-Gauss fit, ($k+1$)-G is $k+1$-Gauss fit, $k$-C stands for $k$-Cauchy function fit, ($k+1$)-G $k+1$-Cauchy function fit, $C$ indicates the mesonic channel considered, $am_C$ is the result of the GEVP analysis, $\sigma_G$ is the energy smearing radius used for the Gaussian fits, $\sigma_C$ for the Cauchy fit.}
    \end{table}

    \begin{table}[]    
    \centering
    \begin{tabular}{ |c |c|c|c|c|c|c|c|c|c|c| }
     \hline \hline
     $C$ & $k$ & $N_{\rm source}$ & $N_{\rm sink}$ & $aE_1$ $k$-G & $aE_1$ ($k+1$)-G & $aE_1$ $k$-C & $aE_1$ ($k+1$)-C & $am_C$ & $\sigma_{G} / m_C$ & $\sigma_C / m_C$ \\
      \hline 
     PS & 2 & 0 & 40 & 0.7401(86) & 0.7371(87) & 0.7351(77)  & 0.7351(73) & 0.7347(87) & 0.30 & 0.30 \\
     V & 3 & 40 & 40 & 0.783(16) & 0.781(15) & 0.763(13) & 0.767(15) & 0.759(16) & 0.30 & 0.27 \\
     T & 3 & 0 & 40 & 0.787(25) & 0.788(25) &  0.753(18) & 0.759(19) & 0.759(16) & 0.25 & 0.25 \\
     AV & 2 & 80 & 40 & 0.822(24) & 0.819(22) &  0.832(29) & 0.830(28) & 0.808(17) & 0.25 & 0.25 \\
     AT & 2 & 80 & 40 & 0.872(14) & 0.868(14) &  0.853(14) & 0.863(14) & 0.856(15) & 0.25 & 0.25 \\
     S  & 2 & 0 & 40 & 0.764(23) & 0.765(24) & 0.754(19) & 0.756(20) & 0.751(19) & 0.25 & 0.30 \\
      \hline
     ps  & 2 & 0 & 40 & 0.922(13) & 0.928(14) &  0.938(12) & 0.922(15) & 0.932(14) & 0.25 & 0.24 \\
     v & 2 & 0 & 40 & 0.947(21) & 0.944(21) &  0.943(24) & 0.942(23) & 0.960(20) & 0.23 & 0.20 \\
     t & 2 & 0 & 40 & 0.952(26) & 0.950(25) &  0.956(23) & 0.951(22) & 0.960(20) & 0.23 & 0.25 \\
     av & 2 & 40 & 40 & 1.119(14) & 1.120(13) &  1.129(16) & 1.122(15) & 1.118(15) & 0.20 & 0.20 \\
     at & 2 & 80 & 40 & 1.103(21) & 1.105(20) & 1.145(20) & 1.135(23) & 1.127(21) & 0.25 & 0.25 \\
     s & 2 & 40 & 40 & 1.101(20) & 1.103(22) & 1.118(18) & 1.107(18) & 1.109(19) & 0.24 & 0.20 \\

      \hline \hline
    \end{tabular}
    \caption{\label{table:E4_results_first} Numerical results for the first excited state mass in ensemble M4. $k$-G stands for $k$-Gauss fit, ($k+1$)-G is $k+1$-Gauss fit, $k$-C stands for $k$-Cauchy function fit, ($k+1$)-G $k+1$-Cauchy function fit, $C$ indicates the mesonic channel considered, $am_C$ is the result of the GEVP analysis, $\sigma_G$ is the energy smearing radius used for the Gaussian fits, $\sigma_C$ for the Cauchy fit.}
    \end{table}

    \begin{table}[]    
    \centering
    \begin{tabular}{ |c |c|c|c|c|c|c|c|c|c|c| }
     \hline \hline
     $C$ & $k$ & $N_{\rm source}$ & $N_{\rm sink}$ & $aE_2$ $k$-G & $aE_2$ ($k+1$)-G & $aE_2$ $k$-C & $aE_2$ ($k+1$)-C & $am_C$ & $\sigma_{G} / m_C$ & $\sigma_C / m_C$ \\
      \hline
     PS & 2 & 0 & 40 & - & - & - & - & - & 0.30 & 0.30 \\
     V  & 3 & 40 & 40 & 0.949(33) & 0.953(32) & 0.954(31) & 0.959(34) & - & 0.30 & 0.27 \\
     T  & 3 & 0 & 40 & 0.944(31) & 0.949(31) & 0.969(31) & 0.963(31) & - & 0.25 & 0.25 \\
     AV & 2 & 80 & 40 & - & - & - & - & - & 0.25 & 0.25 \\
     AT & 2 & 80 & 40 & - & - & - & - & - & 0.25 & 0.25 \\
     S  & 2 & 0 & 40 & - & - & - & - & - & 0.30 & 0.30 \\
      \hline
     ps & 2 & 0 & 40 & - & - & - & - & - & 0.25 & 0.24 \\
     v  & 2 & 0 & 40 & - & - & - & - & - & 0.23 & 0.20 \\
     t  & 2 & 40 & 40 & - & - & - & - & - & 0.23 & 0.25 \\
     av & 2 & 40 & 40 & - & - & - & - & - & 0.20 & 0.20 \\
     at & 2 & 80 & 40 & - & - & - & - & - & 0.25 & 0.25 \\
     s  & 2 & 40 & 40 & - & - & - & - & - & 0.24 & 0.20 \\
      \hline \hline
    \end{tabular}
    \caption{\label{table:E4_results_second} Numerical results for the second excited state mass in ensemble M4.  $k$-G stands for $k$-Gauss fit, ($k+1$)-G is $k+1$-Gauss fit, $k$-C stands for $k$-Cauchy function fit, ($k+1$)-G $k+1$-Cauchy function fit, $C$ indicates the mesonic channel considered, $am_C$ is the result of the GEVP analysis, $\sigma_G$ is the energy smearing radius used for the Gaussian fits, $\sigma_C$ for the Cauchy fit.}
    \end{table}

    \begin{table}[]    
    \centering
    \begin{tabular}{ |c |c|c|c|c|c|c|c|c|c|c| }
     \hline  \hline
     $C$ & $k$ & $N_{\rm source}$ & $N_{\rm sink}$ & $aE_0$ $k$-G & $aE_0$ ($k+1$)-G & $aE_0$ $k$-C & $aE_0$ ($k+1$)-C & $am_C$ & $\sigma_{G} / m_C$ & $\sigma_C / m_C$ \\
      \hline
     PS & 2 & 40 & 40 & 0.3123(13) & 0.3113(14) & 0.3120(11) & 0.3125(17) & 0.31025(64) & 0.25 & 0.25 \\
     V  & 3 & 40 & 40 & 0.3523(22) & 0.3502(21) & 0.3509(25) & 0.3513(24) & 0.3515(13)  & 0.30 & 0.30 \\
     T  & 3 & 40 & 40 & 0.3520(25) & 0.3472(24) & 0.3479(29) & 0.3523(28) & 0.3515(13)  & 0.30 & 0.30 \\
     AV & 2 & 40 & 40 & 0.5142(32) & 0.5133(31) & 0.5140(30) & 0.5131(29) & 0.5121(30) & 0.25 & 0.20 \\
     AT & 2 & 40 & 40 & 0.5211(43) & 0.5203(33) & 0.5231(53) & 0.5215(33) & 0.5201(34)  & 0.20 & 0.20 \\
     S  & 2 & 40 & 40 & 0.4888(36) & 0.4878(32) & 0.4900(31) & 0.4909(38) & 0.4898(30)  & 0.25 & 0.25 \\
      \hline 
     ps & 2 & 40 & 40 & 0.57923(40) & 0.57933(44) & 0.57905(42) & 0.57913(47) & 0.57953(42)  & 0.20 & 0.20 \\
     v  & 2 & 40 & 40 & 0.6222(33) & 0.6231(33) & 0.6212(35) & 0.6210(33) & 0.6222(11)  & 0.20 & 0.25 \\
     t  & 2 & 40 & 40 & 0.6242(31) & 0.6223(37) & 0.6223(32) & 0.6222(38) & 0.6222(11)  & 0.20 & 0.25 \\
     av & 2 & 40 & 40 & 0.7994(62) & 0.7990(67) & 0.7964(54) & 0.7964(42) & 0.7993(42)  & 0.20 & 0.20 \\
     at & 2 & 40 & 40 & 0.8093(45) & 0.8078(70) & 0.8072(60) & 0.8073(53) & 0.8105(49)  & 0.20 & 0.20 \\
     s  & 2 & 40 & 40 & 0.7645(53) & 0.7661(65) & 0.7670(52) & 0.7658(51) & 0.7684(31)  & 0.20 & 0.20 \\
      \hline \hline
    \end{tabular}
    \caption{\label{table:E5_results_ground} Numerical results for the ground state mass in ensemble M5.  $k$-G stands for $k$-Gauss fit, ($k+1$)-G is $k+1$-Gauss fit, $k$-C stands for $k$-Cauchy function fit, ($k+1$)-G $k+1$-Cauchy function fit, $C$ indicates the mesonic channel considered, $am_C$ is the result of the GEVP analysis, $\sigma_G$ is the energy smearing radius used for the Gaussian fits, $\sigma_C$ for the Cauchy fit.}
    \end{table}

    \begin{table}[]    
    \centering
    \begin{tabular}{ |c |c|c|c|c|c|c|c|c|c|c| }
      \hline \hline
     $C$ & $k$ & $N_{\rm source}$ & $N_{\rm sink}$ & $aE_1$ $k$-G & $aE_1$ ($k+1$)-G & $aE_1$ $k$-C & $aE_1$ ($k+1$)-C & $am_C$ & $\sigma_{G} / m_C$ & $\sigma_C / m_C$ \\
     \hline
     PS & 2 & 40 & 40 & 0.615(14) & 0.620(15) & 0.625(13) & 0.628(17) & 0.627(15)  & 0.25 & 0.25 \\
     V  & 3 & 40 & 40 & 0.622(17) & 0.612(18) & 0.619(15) & 0.624(20) & 0.620(15) & 0.30 & 0.30 \\
     T  & 3 & 40 & 40 & 0.625(20) & 0.615(20) & 0.614(18) & 0.622(18) & 0.620(15)  & 0.30 & 0.30 \\
     AV & 2 & 40 & 40 & 0.780(13) & 0.772(13) & 0.770(15) & 0.771(14) & 0.7743(81) & 0.25 & 0.20 \\
     AT & 2 & 40 & 40 & 0.780(17) & 0.776(11) & 0.772(14) & 0.774(14) & 0.7788(92) & 0.20 & 0.20 \\
     S  & 2 & 40 & 40 & 0.741(13) & 0.738(11) & 0.735(15) & 0.740(16) & 0.7391(94) & 0.25 & 0.25 \\
      \hline 
     ps & 2 & 40 & 40 & 0.8783(94) & 0.8763(74) & 0.8803(100) & 0.8763(72) & 0.8798(77)  & 0.20 & 0.20 \\
     v  & 2 & 40 & 40 & 0.8811(100) & 0.8903(100) & 0.8862(100) & 0.8843(96) & 0.8909(86) & 0.20 & 0.25 \\
     t  & 2 & 40 & 40 & 0.8851(99) & 0.8853(93) & 0.8822(83) & 0.8863(85) & 0.8909(86)  & 0.20 & 0.25 \\
     av & 2 & 40 & 40 & 1.013(21) & 1.003(20) & 1.023(23) & 1.016(22) & 1.008(21) & 0.20 & 0.20 \\
     at & 2 & 40 & 40 & 1.043(17) & 1.041(15) & 1.034(14) & 1.030(13) & 1.034(15) & 0.20 & 0.20 \\
     s  & 2 & 40 & 40 & 1.025(20) & 1.016(19) & 1.004(17) & 1.012(19) & 1.001(16) & 0.20 & 0.20 \\
 \hline \hline
\end{tabular}
\caption{\label{table:E5_results_first} Numerical results for the first excited state mass in ensemble M5.  $k$-G stands for $k$-Gauss fit, ($k+1$)-G is $k+1$-Gauss fit, $k$-C stands for $k$-Cauchy function fit, ($k+1$)-G $k+1$-Cauchy function fit, $C$ indicates the mesonic channel considered, $am_C$ is the result of the GEVP analysis, $\sigma_G$ is the energy smearing radius used for the Gaussian fits, $\sigma_C$ for the Cauchy fit.}
\end{table}

\begin{table}[]    
    \centering
    \begin{tabular}{ |c |c|c|c|c|c|c|c|c|c|c| }
     \hline \hline
     $C$ & $k$ & $N_{\rm source}$ & $N_{\rm sink}$ & $aE_2$ $k$-G & $aE_2$ ($k+1$)-G & $aE_2$ $k$-C & $aE_2$ ($k+1$)-C & $am_C$ & $\sigma_{G} / m_C$ & $\sigma_C / m_C$ \\
     \hline
     PS & 2 & 40 & 40 & - & - & - & - & - & 0.25 & 0.25 \\
     V  & 3 & 40 & 40 & 0.833(20) & 0.843(24) & 0.831(19) & 0.838(25) & 0.839(22) & 0.30 & 0.30 \\
     T  & 3 & 40 & 40 & 0.832(24) & 0.840(27) & 0.841(24) & 0.833(21) & 0.839(22) & 0.30 & 0.30 \\
     AV & 2 & 40 & 40 & - & - & - & - & - & 0.25 & 0.20 \\
     AT & 2 & 40 & 40 & - & - & - & - & - & 0.20 & 0.20 \\
     S  & 2 & 40 & 40 & - & - & - & - & - & 0.25 & 0.25 \\
      \hline
     ps & 2 & 40 & 40 & - & - & - & - & - & 0.20 & 0.20 \\
     v  & 2 & 40 & 40 & - & - & - & - & 1.084(51) & 0.20 & 0.25 \\
     t  & 2 & 40 & 40 & - & - & - & - & 1.084(51) & 0.20 & 0.25 \\
     av & 2 & 40 & 40 & - & - & - & - & - & 0.20 & 0.20 \\
     at & 2 & 40 & 40 & - & - & - & - & - & 0.20 & 0.20 \\
     s  & 2 & 40 & 40 & - & - & - & - & - & 0.20 & 0.20 \\
 \hline \hline
\end{tabular}
\caption{\label{table:E5_results_second} Numerical results for the second excited  state mass in ensemble M5.  $k$-G stands for $k$-Gauss fit, ($k+1$)-G is $k+1$-Gauss fit, $k$-C stands for $k$-Cauchy function fit, ($k+1$)-G $k+1$-Cauchy function fit, $C$ indicates the mesonic channel considered, $am_C$ is the result of the GEVP analysis, $\sigma_G$ is the energy smearing radius used for the Gaussian fits, $\sigma_C$ for the Cauchy fit.}
\end{table}

\bibliographystyle{JHEP} 
\bibliography{ref}

\end{document}